\documentclass[aps,prd,preprint,groupedaddress,showpacs,showkeys,nofootinbib,eqsecnum]{revtex4-1}

\usepackage{graphicx}
\usepackage{amsmath}
\usepackage{setspace}
\usepackage{xspace}
\usepackage{hyperref}
\usepackage{color}
\usepackage{slashed}

\newcommand{\ord}{{\mathcal O}}

\newcommand{\be}{\begin{equation}}
\newcommand{\ee}{\end{equation}}
\newcommand{\bea}{\begin{eqnarray}}
\newcommand{\eea}{\end{eqnarray}}
\newcommand{\Appendix}[1]%
    {%
     \section{#1}%
      }

\newcommand{\eq}[1]{Eq.~\eqref{#1}}
\newcommand{\eqs}[2]{Eqs.~\eqref{#1} and \eqref{#2}}

\newcommand{\eqss}[2]{Eqs.~\eqref{#1}-\eqref{#2}}

\newcommand{\fig}[1]{Fig.~\ref{#1}}
\newcommand{\figs}[2]{Figs.~\ref{#1} and~\ref{#2}}
\newcommand{\app}[1]{Appendix~\ref{#1}}
\newcommand{\apps}[2]{Appendices~\ref{#1} and~\ref{#2}}
\newcommand{\Sec}[1]{Sec.~\ref{#1}}
\newcommand{\tab}[1]{Table~\ref{#1}}

\newcommand{\rcite}[1]{Ref.~\cite{#1}}
\newcommand{\rcites}[1]{Refs.~\cite{#1}}

\newcommand{\als}{\alpha_{s}}
\newcommand{\alh}{\alpha_{\rm h}}
\newcommand{\alus}{\alpha_{\rm us}}
\newcommand{\nuus}{\nu_{\rm us}}
\newcommand{\nuh}{\nu_{\rm h}}
\newcommand{\US}{\rm us}
\newcommand{\lQ}{\Lambda_{\rm QCD}}

\newcommand{\nn}{\nonumber}

\newcommand{\dsl}{\slashed{D}}
\newcommand{\MS}{\overline{\rm MS}}
\newcommand{\RS}{\rm RS}

\newcommand{\RG}{{\rm RG}}

\newcommand{\zero}{{(0)}}
\newcommand{\one}{{(1)}}
\newcommand{\two}{{(2)}}

\newcommand{\bfnabla}{{\boldsymbol \nabla}}
\newcommand{\bfsigma}{{\boldsymbol \sigma}}
\newcommand{\lla}{\langle\!\langle}
\newcommand{\rra}{\rangle\!\rangle}
     
\allowdisplaybreaks[3]

\begin{document}

\preprint{TUM-1163/18}


\title{P-wave heavy quarkonium spectrum with next-to-next-to-next-to-leading logarithmic accuracy}

\author{Clara Peset}
\email{clara.peset@tum.de}
\affiliation{Physik Department T31, James-Franck-Stra\ss e 1, Technische Universit\"at M\"unchen,
85748 Garching, Germany}
\author{Antonio Pineda}
\email{pineda@ifae.es}
\author{Jorge Segovia}
\email{jsegovia@ifae.es}
\affiliation{Grup de F\'\i sica Te\`orica, Dept. F\'\i sica and IFAE-BIST, Universitat Aut\`onoma de Barcelona,\\ 
E-08193 Bellaterra (Barcelona), Spain}

\date{\today}

\begin{abstract}
We compute the heavy quarkonium mass of $l\not= 0$ (angular momentum) states, with otherwise arbitrary quantum numbers, with next-next-to-next-to-leading logarithmic (N$^3$LL) accuracy. This constitutes the first observable in heavy quarkonium 
for which two orders of the weak-coupling expansion sensitive to the ultrasoft scale are known and the resummation of 
ultrasoft logarithms is made. We also obtain, for the first time, resummed N$^3$LL expressions for the different fine and hyperfine energy splittings of these states, which are not sensitive to the ultrasoft scale but still require resummation of (hard) logarithms.  We do this analysis for the equal and non-equal mass cases. We also study an alternative computational scheme that treats the static potential exactly. We then perform a comprehensive phenomenological analysis: we apply these results to the $n=2$, $l=1$ bottomonium, $B_c$ and charmonium systems and study their convergence. 
\end{abstract}


\maketitle

\begin{spacing}{1}
\tableofcontents
\end{spacing}


\section{Introduction}

The heavy quarkonium mass has been computed with increasing accuracy in the limit of very large mass (i.e. in the strict weak-coupling approximation) over the years. If $n$ represents the principal quantum number and $l$ the orbital angular momentum, in this paper we exclusively consider non S-wave states (i.e. those states with $l \not=0$). Typically, we will use the notation ``P-wave'' to refer to non S-wave states (unless explicitly stated otherwise). 
The heavy quarkonium mass of the P-wave states has been computed in \rcite{Billoire:1979ih} to next-to-leading order (NLO), in \rcite{Pineda:1997hz} to NNLO, in \rcite{Brambilla:1999xj} the $\ln \als$ term of the N$^3$LO, in \rcites{Kiyo:2013aea,Kiyo:2014uca} with N$^3$LO accuracy for the equal mass case and in \rcite{Peset:2015vvi} for the non-equal mass case. For the $n=2$ and $l=1$ fine splitting in the equal mass case, the N$^3$LO expression was obtained in \rcite{Brambilla:2004wu} and the hyperfine in \rcite{Titard:1994id} (for arbitrary quantum numbers and equal masses). 

Once the spectrum has been obtained with N$^3$LO accuracy one can move to the next step: the computation of the heavy quarkonium mass with N$^3$LL accuracy by the resummation of the large logarithms. This is one of the main purposes of this paper, and we achieve this goal for 
arbitrary P-wave states. Most of the necessary ingredients are already available in the literature. The ultrasoft renormalization group (RG) 
analysis of the potentials relevant for the P-wave states were obtained with N$^3$LL accuracy in \rcite{Pineda:2011aw}. These results, together with the detailed computations in \rcite{Peset:2015vvi}, allow us to obtain the mass of the excited states with N$^3$LL accuracy. We also achieve this precision for the fine and hyperfine P-wave splittings for the first time. Crucial to obtain this last result is the knowledge of the potential to N$^3$LO, of the structure of the potential in terms of Wilson loops, and the confirmation that no ultrasoft effects enter at this order. The above results are obtained using the effective field theory (EFT) named potential nonrelativistic QCD (pNRQCD) \cite{Pineda:1997bj,Brambilla:1999xf} (for reviews see \cite{Brambilla:2004jw,Pineda:2011dg}).

\medskip

The cancellation of the leading renormalon of the pole mass and the static potential, first found in \rcite{Pineda:1998id}, and later in \cite{Hoang:1998nz,Beneke:1998rk}, led to the realization \cite{Beneke:1998rk} that using threshold masses \cite{Bigi:1994em,Beneke:1998rk,Pineda:2001zq,Hoang:2009yr,Brambilla:2017hcq} (which explicitly implement the cancellation of the renormalon in heavy quarkonium observables) improves the convergence of the perturbative series. This makes these very precise computations useful not only for academical purposes but also for phenomenological applications. The applicability of a weak-coupling analysis to the first P-wave heavy quarkonium excited state ($n=2$, $l=1$) is an open issue. Originally, they were studied in \rcites{Brambilla:2001fw,Brambilla:2001qk,Recksiegel:2002za,Recksiegel:2003fm}, where the outcome of the analysis was qualitatively positive. These analyses had NNLO accuracy and used the Upsilon counting \cite{Hoang:1998ng}, which effectively introduces the cancellation of renormalon but does not use threshold masses. An analysis of the fine splittings, which are directly renormalon free, was done in \rcite{Brambilla:2004wu}.
Beyond NNLO there is only a preliminary phenomenological analysis in N$^3$LO using the Upsilon counting \cite{Kiyo:2013aea} and the more recent analysis \cite{Mateu:2017hlz}. 

On the phenomenological side, one of the purposes of this paper is to study the P-wave states of heavy quarkonium for bottomonium, charmonium and $B_c$ (but specially bottomonium) to clarify if a weak-coupling description for them is appropriate, and, if so, to which extent. The threshold mass we will use is the RS' mass \cite{Pineda:2001zq}.
We also want to quantify the impact of the resummation of logarithms in the heavy quarkonium spectrum: for the first time we have two terms of the weak-coupling expansion that depend on the ultrasoft logarithmic resummation. 

Besides the aforementioned phenomenological analysis performed at strict weak coupling, we also study the convergence of an alternative computational scheme that reorganizes the perturbative expansion of the weak-coupling computation. This scheme is characterized by solving the Schr\"odinger equation including the static potential exactly (to the order it is known). This incorporates formally subleading terms in the leading order (LO) solution. 
On the other hand the relativistic corrections to the spectrum are included perturbatively. This working scheme performs a partial resummation of higher order effects. This may accelerate the convergence of the perturbative series. This is indeed the effect seen in (most of) the cases where it has been applied (spectrum and decays) \cite{Recksiegel:2002za,Recksiegel:2003fm,Kiyo:2010jm,Pineda:2013lta} (the acceleration is somewhat more marginal in the analysis in \rcite{Peset:2018ria}). This scheme naturally leads to the organization of the computation in powers of $v$, the relative velocity of the heavy quark in the bound state.

\subsection{pNRQCD}
Integrating out the soft modes in NRQCD \cite{Caswell:1985ui,Bodwin:1994jh}
we obtain the EFT named pNRQCD \cite{Pineda:1997bj}.
The most general pNRQCD Lagrangian 
compatible with the symmetries of QCD that can be constructed
with a singlet and an octet (quarkonium) fields, as well as an ultrasoft gluon field to NLO in the 
multipole expansion has the form~\cite{Pineda:1997bj,Brambilla:1999xf}
\begin{align}
 & \!\!\!\!\!
{\cal L}_{\rm pNRQCD} = \!\! \int \!\! d^3{\bf r} \; {\rm Tr} \,  
\Biggl\{ {\rm S}^\dagger \left( i\partial_0 
- h_s({\bf r}, {\bf p}, {\bf P}_{\bf R}, {\bf S}_1,{\bf S}_2) \right) {\rm S} 
+ {\rm O}^\dagger \left( iD_0 
- h_o({\bf r}, {\bf p}, {\bf P}_{\bf R}, {\bf S}_1,{\bf S}_2) \right) {\rm O} \Biggr\}
\nn
\\
 &\qquad\qquad 
+ V_A ( r) {\rm Tr} \left\{  {\rm O}^\dagger {\bf r} \cdot g{\bf E} \,{\rm S}
+ {\rm S}^\dagger {\bf r} \cdot g{\bf E} \,{\rm O} \right\} 
+ \frac{V_B (r)}{ 2} {\rm Tr} \left\{  {\rm O}^\dagger {\bf r} \cdot g{\bf E} \, {\rm O} 
+ {\rm O}^\dagger {\rm O} {\bf r} \cdot g{\bf E}  \right\}  
\nn
\\
 &\qquad\qquad 
- \frac{1}{ 4} G_{\mu \nu}^{a} G^{\mu \nu \, a} 
+  \sum_{i=1}^{n_f} \bar q_i \, i \dsl \, q_i 
\,,
\label{Lpnrqcd}
\\
 &
\nn 
\\
 &
h_s({\bf r}, {\bf p}, {\bf P}_{\bf R}, {\bf S}_1,{\bf S}_2) = 
 \frac{{\bf p}^2 }{ 2\, m_{ r}}
+ 
\frac{{\bf P}_{\bf R}^2 }{ 2\, M} + 
V_s({\bf r}, {\bf p}, {\bf P}_{\bf R}, {\bf S}_1,{\bf S}_2), 
\\
 & 
h_o({\bf r}, {\bf p}, {\bf P}_{\bf R}, {\bf S}_1,{\bf S}_2) = 
\frac{{\bf p}^2 }{ 2\, m_{ r}}
+ 
\frac{{\bf P}_{\bf R}^2 }{ 2\,M}  + 
V_o({\bf r}, {\bf p}, {\bf P}_{\bf R}, {\bf S}_1,{\bf S}_2), 
\\
&
\nn
\\
& V_s = 
V^{(0)} + \frac{V^{(1,0)} }{ m_1}+\frac{V^{(0,1)}}{ m_2}
+ \frac{V^{(2,0)}}{ m_1^2}+ \frac{V^{(0,2)}}{ m_2^2}+\frac{V^{(1,1)}}{ m_1m_2}+\cdots,
\label{V1ovm2}
\\
& V_o = 
V^{(0)}_o + \frac{V^{(1,0)}_o }{ m_1}+\frac{V^{(0,1)}_o}{ m_2}
+ \frac{V^{(2,0)}_o }{ m_1^2}+ \frac{V^{(0,2)}_o}{ m_2^2}+\frac{V^{(1,1)}_o }{ m_1m_2}+\cdots,
\end{align}
where $iD_0 {\rm O} \equiv i \partial_0 {\rm O} - g [A_0({\bf R},t),{\rm O}]$, 
${\bf P}_{\bf R} = -i{\bfnabla}_{\bf R}$ for the singlet,  
${\bf P}_{\bf R} = -i{\bf D}_{\bf R}$ for the octet (where the covariant derivative is in the adjoint representation), 
${\bf p} = -i\bfnabla_{\bf r}$,
\begin{align}
m_{r} = \frac{m_1 m_2}{m_1+m_2}
\end{align}
and $M = m_1+m_2$. 
We adopt the color normalization  
\begin{align}
{\rm S} = { S\, 1\!\!{\rm l}_c / \sqrt{N_c}} \,, \quad\quad\quad 
{\rm O} = O^a { {\rm T}^a / \sqrt{T_F}}\,,
\label{SSOO}
\end{align}
for the singlet field $S({\bf r}, {\bf R}, t)$ and the octet field $O^a({\bf r}, {\bf R}, t)$.
Here and throughout this paper we denote the quark-antiquark distance vector by ${\bf r}$, the center-of-mass position of the quark-antiquark system by ${\bf R}$, and the time by $t$.

Both, $h_s$ and the potential $V_s$ are operators acting on the Hilbert space of a heavy quark-antiquark system in the singlet configuration.\footnote{Therefore, in a more mathematical notation: $h \rightarrow \hat h$, $V_s({\bf r},{\bf p}) \rightarrow \hat V_s(\hat {\bf r},\hat {\bf p})$. We will however avoid this notation in order to facilitate the reading.}
According to the precision we are aiming for, the potentials have been displayed up to terms of order $1/m^2$.\footnote{Actually, 
we also have to include the leading correction to the nonrelativistic dispersion relation for our calculation of the spectrum:
\begin{align}
\delta V_s=-\left(\frac{1}{8m_1^3}+\frac{1}{8m_2^3}\right){\bf p}^4,
\end{align}
and use the fact there is no ${\cal O}(\als/m^3)$ potential.
}
The static and the $1/m$ potentials are real-valued functions of $r=|\bf r|$ only.
The $1/m^2$ potentials have an imaginary part proportional to
$\delta^{(3)}({\bf r})$, which we will drop in this analysis, 
and a real part that may be decomposed as:
\begin{align}
&
V^{(2,0)}=V^{(2,0)}_{SD}+V^{(2,0)}_{SI}, \qquad 
V^{(0,2)}=V^{(0,2)}_{SD}+V^{(0,2)}_{SI}, \qquad 
V^{(1,1)}=V^{(1,1)}_{SD}+V^{(1,1)}_{SI},
\label{decomSDSI}
\\[2 ex]
& 
V^{(2,0)}_{SI}=\frac{1}{ 2}\left\{{\bf p}_1^2,V_{{\bf p}^2}^{(2,0)}(r)\right\}
+V_{{\bf L}^2}^{(2,0)}(r)\frac{{\bf L}_1^2 }{ r^2} + V_r^{(2,0)}(r),
\label{v20sistrong}
\\
&
V^{(0,2)}_{SI}=\frac{1 }{ 2}\left\{{\bf p}_2^2,V_{{\bf p}^2}^{(0,2)}(r)\right\}
+V_{{\bf L}^2}^{(0,2)}(r)\frac{{\bf L}_2^2 }{ r^2}+ V_r^{(0,2)}(r),
\\
&
V^{(1,1)}_{SI}= -\frac{1 }{ 2}\left\{{\bf p}_1\cdot {\bf p}_2,V_{{\bf p}^2}^{(1,1)}(r)\right\}
-V_{{\bf L}^2}^{(1,1)}(r)\frac{({\bf L}_1\cdot{\bf L}_2+ {\bf L}_2\cdot{\bf L}_1) }{ 2r^2}+ V_r^{(1,1)}(r),
\\[2 ex]
&
V^{(2,0)}_{SD}=V^{(2,0)}_{LS}(r){\bf L}_1\cdot{\bf S}_1, 
\label{v20sdstrong}\\
&
V^{(0,2)}_{SD}=-V^{(0,2)}_{LS}(r){\bf L}_2\cdot{\bf S}_2,
\\
&
V^{(1,1)}_{SD}=
V_{L_1S_2}^{(1,1)}(r){\bf L}_1\cdot{\bf S}_2 - V_{L_2S_1}^{(1,1)}(r){\bf L}_2\cdot{\bf S}_1
+ V_{S^2}^{(1,1)}(r){\bf S}_1\cdot{\bf S}_2 + V_{{\bf S}_{12}}^{(1,1)}(r){\bf
  S}_{12}({\bf r}),
\label{v11sdstrong}
\end{align}
where, ${\bf S}_1=\bfsigma_1/2$, ${\bf S}_2=\bfsigma_2/2$, ${\bf L}_1 \equiv {\bf r} \times {\bf p}_1$, ${\bf L}_2 \equiv {\bf
  r} \times {\bf p}_2$ and ${\bf S}_{12}({\bf r}) \equiv \frac{
3 { {\bf r}}\cdot \bfsigma_1 \,{ {\bf r}}\cdot \bfsigma_2}{r^2} - \bfsigma_1\cdot \bfsigma_2$.
Note that neither ${\bf L}_1$ nor ${\bf L}_2$ correspond to the orbital angular momentum 
of the particle or the antiparticle.

Due to invariance under charge conjugation plus $m_1 \leftrightarrow m_2$ interchange we have 
\begin{align}
V^{(1,0)}(r) = V^{(0,1)}(r).
\end{align}
This allows us to write
\begin{align}
\frac{V^{(1,0)}}{ m_1}+\frac{V^{(0,1)}}{ m_2}=\frac{V^{(1,0)} }{ m_r}
\,.
\end{align}
 Invariance under charge conjugation plus $m_1\leftrightarrow m_2$ also implies
\begin{align}
&V_{{\bf p}^2}^{(2,0)}(r) =V_{{\bf p}^2}^{(0,2)}(r)\,, 
\qquad 
V_{{\bf L}^2}^{(2,0)}(r) =V_{{\bf L}^2}^{(0,2)}(r)\,, 
\qquad 
V_r^{(2,0)}(r)=V_r^{(0,2)}(r;m_2 \leftrightarrow m_1)\,, \nn\\
&V^{(2,0)}_{LS}(r)=V^{(0,2)}_{LS}(r; m_2 \leftrightarrow m_1)\,,
\qquad
V_{L_1S_2}^{(1,1)}(r)=V_{L_2S_1}^{(1,1)}(r; m_1 \leftrightarrow m_2)\,.  \label{potsym}
\end{align}

For the precision of the computation of the spectrum reached in this paper, we can neglect the center-of-mass momentum, i.e. we set ${\bf P}_{\bf R}=0$ in the following and thus ${\bf L}_1 \equiv {\bf r} \times {\bf p}_1=
{\bf r} \times {\bf p}\equiv{\bf L}$, ${\bf L}_2 \equiv {\bf r} \times {\bf p}_2=-
{\bf r} \times {\bf p}\equiv-{\bf L}$.

Expressions for the N$^3$LO potentials for the non-equal mass case can be found in \rcite{Peset:2015vvi} for different bases of potentials (on-shell, Wilson, Coulomb, Feynman matching schemes). For illustration, we will work with the on-shell basis of potentials where the potential proportional to ${\bf L}^2$ is set to zero (for ease of reference we list them in \app{app:potential}). Nevertheless, we emphasize that the results are independent on the chosen basis of potentials to N$^3$LL order.  In the following section we give the N$^3$LL potentials for the (un)equal mass case relevant for the P-wave spectrum (see also \cite{Pineda:2011aw}). The singlet potential $V_s$ depends on the factorization scales $\nuh$, $\nu$ and $\nuus$: $V_s(\nu;\nuh,\nuus)$.\footnote{Strictly speaking the $\nu$ dependence is traded off by a dependence in $1/r$ to the order we are working.} Throughout this paper we will use the notation $\als=\als(\nu)$, $\alus=\als(\nuus)$, $\alh=\als(\nuh)$. 
Large logarithms are resummed setting $\nuh \sim m$, $\nu \sim m\als$ and $\nuus \sim m\als^2$. We will generically split the RG improved potential to N$^i$LL into the fixed order result plus the correction generated by the resummation of logarithms:
\begin{align}
V^{\rm RG}_{s,\rm N^iLL} (\nu;\nuh,\nuus)= 
V_{s,\rm N^iLO} (\nu)+\delta V^{\rm RG}_{s,\rm N^iLL} (\nu;\nuh,\nuus)\,,
\end{align} 
such that $\delta V^{\rm RG}_{s,\rm N^iLL} (\nu;\nu,\nu)=0$,
and similarly for each individual potential: $V^{(1,0)}$, etc.

\section{Renormalization Group running}
\label{sec:RG}

We consider now the modifications of the N$^3$LO potentials needed to achieve the resummation of the large logarithms for the P-wave spectrum. 

\subsection{Ultrasoft Renormalization Group running}

The bare potential can be written in terms of the renormalized potential and its counterterm in the following way
\begin{align}
V_{s,B}=V_s+\delta V_s
\,.
\end{align}
If the counterterm is determined in terms of the Wilson coefficients of the EFT, it is possible to resum the large logarithms of the potentials associated to the ultrasoft scale by solving the associated renormalization group equation (RGE). 
 The counterterm $\delta V_s$ for the NLL ultrasoft running of $V_s$ was obtained in Eq.~(35) of \rcite{Pineda:2011aw}.\footnote{
Confirmation of the counterterm in the context of vNRQCD~\cite{Luke:1999kz} was 
obtained in \rcites{Hoang:2011gy,Hoang:2006ht} for the $\ord(1/m)$ and $\ord(1/m^2)$, 
potentials. Prior to this, the running of the static potential was computed at LL in \rcite{Pineda:2000gza} and at NLL in \rcite{Brambilla:2009bi} and confirmed in \rcite{Pineda:2011db}, whereas the complete  LL ultrasoft running of the $V_s$ was obtained in \rcite{Pineda:2001ra}. } 
The RGE then reads
\begin{align}
\label{RGE}
\nu\frac{d}{d\nu} V_{s,\MS}
=
B_{V_s}
\,,
\end{align}
where
\begin{align}
\nn
B_{V_s}&=
C_F
\Biggl(
{\bf r}^2(\Delta V)^3
-\frac{1}{2m_r^2}\left[{\bf p},\left[{\bf p},V^{(0)}_o\right]\right]
+\frac{1}{2m_r^2}\left\{{\bf p}^2,\Delta V\right\}
+\frac{2}{m_r}\Delta V  \left(r\frac{d}{dr}V^{(0)}\right)
\\
&
\nn
+\frac{1}{2m_r}\left[
4(\Delta V)^2+4\Delta V\left( \left(r\frac{d}{dr}\Delta V\right)+\Delta V \right)
+\left( \left(r\frac{d}{dr}\Delta V\right)+\Delta V \right)^2
\right]
\Biggr)
\\
&
\times
\left[
-\frac{2\als}{3\pi}
+
\frac{\als^2}{9\pi^2}
(
C_A(-\frac{47}{3}-2\pi^2)+\frac{10}{3}T_Fn_f
)
+
{\cal O}(\als^3)
\right]
\,,
\end{align}
and $\Delta V=V_o^\zero-V^\zero$.
Solving this RGE we obtain the RG improved (RGI) expressions for the static, the $1/m$ and the $1/m^2$ momentum-dependent spin-independent potentials (as they do not depend on the hard scale).\footnote{The contributions generated by \eq{RGE} to $V_r$ that contribute to the P-wave spectrum will be discussed in \Sec{Sec:delta}.} We obtain
\begin{align}
\label{VRG}
V^{(0)}_{\rm RG}(r;\nuus) & = V^{(0)}(r;\nu)+ \delta V^{(0)}_{\rm RG}(r;\nu,\nuus)\,,
\\
V^{(1,0)}_{\rm RG}(r;\nuus) & = V^{(1,0)}(r;\nu)+ \delta V^{(1,0)}_{\rm RG}(r;\nu,\nuus)\,,
\\
V^{(2,0)}_{{\bf p}^2,\rm RG}(r;\nuus) & = V^{(2,0)}_{{\bf p}^2}(r;\nu)+ \delta V^{(2,0)}_{{\bf p}^2,\rm RG}(r;\nu,\nuus)\,,
\\
V^{(1,1)}_{{\bf p}^2,\rm RG}(r;\nuus) & = V^{(1,1)}_{{\bf p}^2}(r;\nu)+ \delta V^{(1,1)}_{{\bf p}^2,\rm RG}(r;\nu,\nuus)\,,
\end{align}
where $V^{(0)}$, $V^{(1,0)}$, $V^{(2,0)}_{{\bf p}^2}$ and $V^{(1,1)}_{{\bf p}^2}$ are the fixed order potentials. We collect them in \eqss{VSDfo}{Vren5} for ease of reference. The symmetries in \eq{potsym} also apply to the N$^3$LL potentials.  

The functions $\delta V_{\rm RG}$ are the corrections generated by solving \eq{RGE}. They read:
\begin{align}
\label{deltaV0}
\delta V^{(0)}_{\rm RG}(r;\nu,\nuus)&=r^2\left(\frac{C_A \als }{2  r}\right)^3
\left(
1+3\frac{\als}{4\pi}\left(a_1+2\beta_0\ln(\nu e^{\gamma_E}r)\right)
\right)
F(\nu;\nuus),\\
\label{deltaV1}
\delta V^{(1,0)}_{\rm RG}(r;\nu,\nuus)&=
\left[2\left(\frac{C_A \als }{2  r}\right)^2
\left(
1+2\frac{\als}{4\pi}\left(a_1+2\beta_0\ln(\nu e^{\gamma_E+\frac{1}{2}}r)\right)
\right)
\right.
\\
&
\nn
\left.
+2\frac{C_AC_F \als^2 }{2  r^2}
\left(
1+2\frac{\als}{4\pi}\left(a_1+2\beta_0\ln(\nu e^{\gamma_E-\frac{1}{2}}r)\right)
\right)
\right]
F(\nu;\nuus),\\
\label{deltaVp11}
\delta V^{(1,1)}_{{\bf p}^2,\rm RG}(r;\nu,\nuus)&=\frac{C_A \als }{r}
\left(
1+\frac{\als}{4\pi}\left(a_1+2\beta_0\ln(\nu e^{\gamma_E}r)\right)
\right)
F(\nu;\nuus),\\
\label{deltaVp2}
\delta V^{(2,0)}_{{\bf p}^2,\rm RG}(r;\nu,\nuus)&=\frac{C_A \als }{2  r}
\left(
1+\frac{\als}{4\pi}\left(a_1+2\beta_0\ln(\nu e^{\gamma_E}r)\right)
\right)
F(\nu;\nuus),
\end{align}
where
\begin{align}
F(\nu;\nuus)
&=C_F
\frac{2\pi}{\beta_0}
\left\{
\frac{2}{3\pi}\ln\frac{\alus}{\als}
\right.
\\
\nn
&
\left.
-(\alus-\als)
\left(
\frac{8}{3}\frac{\beta_1}{\beta_0}\frac{1}{(4\pi)^2}-\frac{1}{27\pi^2}\left(C_A\left(47+6\pi^2\right)-10T_Fn_f\right)
\right)
\right\}
\,.
\end{align}
Note that these expressions should be truncated at the appropriate order in the expansion in $\als$ for a given accuracy. $\delta V^{(1,0)}_{\rm RG}(r;\nu,\nuus)$ corrects the NLL result in \rcite{Pineda:2011aw} because in $B_{V_s}$ some subleading terms in $\als(1/r)$ of the potentials were neglected, which are needed for a NLL precision. 
 
There are other operators in the pNRQCD Lagrangian that could potentially contribute to the P-wave spectrum to N$^3$LL. These are 
\begin{align}
\label{La}
\delta {\cal L}_a \sim \frac{c_F^{(1)}}{m_1} S^{\dagger} \bfsigma \cdot {\bf B}^a O^a+\cdots 
\end{align}
and 
\begin{align}
\label{Lb}
\delta {\cal L}_b \sim \frac{c_S^{(1)}}{m_1^2} S^{\dagger} \bfsigma \cdot ({\bf p} \times {\bf E}^a) O^a+\cdots 
\,,
\end{align}
where the dots stand for contributions needed to make the Lagrangian density hermitian as well as the contribution of the other heavy particle. Note that both operators are spin-dependent. Both operators can generate divergent contributions that are absorbed by $1/m^2$ delta-like potentials ($ \sim 1/r^3$). 

The contribution to the potential associated to the operator in \eq{La} is generated at 2nd order perturbation theory with ultrasoft gluons: 
 $\frac{{\bf \sigma}\cdot {\bf B}}{m} \cdots \frac{{\bf \sigma}\cdot {\bf B}}{m}$, and it produces the following divergence:
 \begin{align}
 \delta V_a \sim \frac{1}{m_1m_2}\frac{1}{\epsilon}c_F^{(1)}c_F^{(2)}\alus(\Delta V)^3 \bfsigma_1 \cdot \bfsigma_2.
 \end{align}
The contribution to the potential associated to the operator in \eq{Lb} is generated at 2nd order perturbation theory with ultrasoft gluons of the following type: 
 ${\bf r}\cdot {\bf E} \cdots \frac{c_S{\bf L}\cdot {\bf E}}{m^2}$, and produces the following divergence: 
 \begin{align}
 \delta V_b \sim \frac{1}{m_1^2}\frac{1}{\epsilon}c_S^{(1)}
 \alus(\Delta V)^3 \bfsigma_1 \cdot {\bf L}.
 \end{align}
For P-wave state energies, we know that the expectation value $\langle \frac{1}{r^3} \rangle_{l\not=0}$ is finite. This moves these contributions beyond the N$^3$LL accuracy we seek in this paper. Note however, that $\delta V_a$ actually contributes to the spectrum to N$^3$LL but only to S-wave energies (and in particular to the hyperfine splitting \cite{Kniehl:2003ap,Penin:2004xi}), since now $\langle \frac{1}{r^3} \rangle_{l=0}$ is divergent. $\delta V_b$ does not contribute to S-wave energies either; even though $\langle \frac{1}{r^3} \rangle_{l=0}$ is divergent, the overall contribution is multiplied by ${\bf L}$, which again moves the contribution beyond N$^3$LL.

Overall, we do not consider these contributions here as we are only interested in P-wave energies at N$^3$LL. Therefore one only needs to consider the ${\bf r}\cdot {\bf E} \cdots {\bf r}\cdot {\bf E}$ contributions up to two loops which we already discussed above. 

\subsection{Spin-dependent momentum-dependent potentials}
The spin-dependent potentials do not receive ultrasoft running, unlike the spin-independent ones. If we also restrict ourselves to the 
momentum-dependent potentials, they also do not receive potential running. Both statements hold true for N$^3$LL precision. 
On the other hand the spin-dependent momentum-dependent potentials receive non-trivial hard/soft running through the 
inherited NRQCD Wilson coefficients coming from spin-dependent operators. All boils down to a dependence on a single NRQCD Wilson coefficient: $c_F$ 
(the dependence on $c_S$ is transformed in a dependence on $c_F$ since $c_S=2c_F-1$, \cite{Manohar:1997qy}). For the precision we seek, we need $c_F$ with NLL precision, which is known at present  \cite{Amoros:1997rx,Czarnecki:1997dz}:
  \begin{align}
    c^{(i)}_{F,\rm NLL}(\nu,\nuh)&=
    z^{-\frac{\gamma_0}{2}}
    \left[ 1 + \frac{\alh}{4\pi}
      \left(c_1+\frac{\gamma_0}{2}\ln\frac{\nuh^2}{m_i^2}\right) 
      + \frac{\alh - \als}{4\pi}\left(
        \frac{\gamma_1}{2\beta_0} - \frac{\gamma_0\beta_1}{2\beta_0^2}
      \right)  \right] 
    \,,
  \end{align}
where  $c^{(i)}_{F,\rm LL}(\nu,\nuh)= z^{-\frac{\gamma_0}{2}}$, 
$z=\left(\als/\alh\right)^{1/\beta_0}$, 
$\nuh\sim m_i$ is the hard matching scale,  $c_1 = 2(C_A+C_F)$ 
and the one- and two-loop   anomalous dimensions read
\begin{align}
   \gamma_0 = 2 C_A \,, \qquad
   \gamma_1 = \frac{68}{9}\,C_A^2 - \frac{52}{9}\,C_A T_F\,n_f
   \,.
\end{align}

We will also need $c_F$ at fixed order in powers of $\als$, which can be obtained from the previous expression by fixing $\nuh=\nu$, i.e. $c_{F,\rm  NLO}^{(i)}(\nu)\equiv c^{(i)}_{F,\rm NLL}(\nu,\nu)$.
In this case $c^{(i)}_{F,\rm LO}(\nu)= 1$ is trivial. 

The spin-dependent potentials are unambiguous under the field redefinitions considered in \rcite{Peset:2015vvi} (at least to the order we are working at). They were originally computed in \rcite{Gupta:1981pd} at NNLO, in \rcite{Buchmuller:1981aj} for the N$^3$LO hyperfine splitting, 
and in \rcite{Pantaleone:1985uf} the complete expression for unequal masses was obtained. In principle, in order to obtain the RGI expressions of these potentials one should work in the EFT. We do not need to do that. 
The fact that we know the dependence of the potentials in terms of the NRQCD Wilson coefficients enables us to get them from old computations. The spin-dependent potentials have been defined in \eqss{v20sdstrong}{v11sdstrong}.
Their renormalized expressions read (renormalized NRQCD Wilson coefficients are understood)
\begin{align}
V_{LS,\rm RG}^{(2,0)}(r) = -\frac{c_F^{(1)} }{ r^2}i {\bf r}\cdot \lim_{T\rightarrow \infty}\int_0^{T}dt \, t \,  
\lla g{\bf B}_1(t) \times g{\bf E}_1 (0) \rra + \frac{c_S^{(1)}}{ 2 r^2}{\bf r}\cdot (\bfnabla_r V^{(0)}),
\label{vls20}
\end{align}
where
\begin{align}
 \frac{i {\bf r} }{ r^2}\cdot \lim_{T\rightarrow \infty}\int_0^{T}dt \, t \,  
\lla g{\bf B}_1(t) \times g{\bf E}_1 (0) \rra \Bigg|_{\MS}
&=
\frac{ C_FC_A  \als^2}{2\pi r^3  }\left(1+\ln\left(r\nu e^{\gamma_E-1}\right)\right)+{\cal O}(\als^3)
\,,
\end{align}
\begin{align}
\frac{{\bf r}}{  r^2}\cdot (\bfnabla_r V^{(0)})
=
\frac{ C_F  \als}{r^3}\left[1+\frac{\als}{4\pi}\left(a_1+2\beta_0\ln (r\nu  e^{\gamma_E-1}\right) \right]
+{\cal O}(\als^3);
\end{align}
\begin{align}
V_{L_2S_1,\rm RG}^{(1,1)}(r)= - \frac{c_F^{(1)} }{ r^2}i {\bf r}\cdot \lim_{T\rightarrow \infty}\int_0^{T}dt \, t \, 
\lla g{\bf B}_1(t) \times g{\bf E}_2 (0) \rra 
\label{vls11}
\,,
\end{align}
where
\begin{align}
&
\nn
 \frac{i {\bf r}}{r^2}\cdot \lim_{T\rightarrow \infty}\int_0^{T}dt \, t \, 
\lla g{\bf B}_1(t) \times g{\bf E}_2 (0) \rra  \Bigg|_{\MS}
\\
&
\hspace*{1cm}=
-C_F   \frac{\als(e^{1-\gamma_E}/r)}{r^3}\left\{ 1 +\frac{\als}{\pi }\left[\left(\frac{13 }{36}-\frac{1}{2} \ln\left(\frac{\nu r}{e^{1-\gamma_E}}\right)\right)C_A-\frac{5 }{9}n_f T_F\right]\right\},
\end{align}
and
\begin{align}
V_{{\bf S}_{12},\rm RG}^{(1,1)}(r)=
\frac{c_F^{(1)} c_F^{(2)}}{ 4}i {\hat {\bf r}}^i{\hat {\bf r}}^j
\lim_{T\rightarrow \infty}\int_0^{T} dt \, 
\left[
\lla g {\bf B}^i_1(t) g {\bf B}^j_2 (0) \rra  - \frac{\delta^{ij}}{ 3}\lla g{\bf B}_1(t)
\cdot g{\bf B}_2 (0) \rra
\right],
\label{vs12}
\end{align}
where
\begin{align}
&
i {\hat {\bf r}}^i{\hat {\bf r}}^j
\lim_{T\rightarrow \infty}\int_0^{T} dt \, 
\left[
\lla g {\bf B}^i_1(t) g {\bf B}^j_2 (0) \rra  - \frac{\delta^{ij}}{3}\lla g{\bf B}_1(t)
\cdot g{\bf B}_2 (0) \rra
\right]
\Bigg|_{\MS}
\\
&
\hspace*{1cm}=
\frac{C_F\als(e^{4/3-\gamma_E}/r)}{r^3}\left\{1+\frac{\als }{\pi } \left[ \left(\frac{13}{36}- \ln\left(\frac{\nu r}{e^{4/3-\gamma_E}}\right)\right)C_A -\frac{5}{9} n_f T_F \right] \right\}.
\end{align}
The other potentials follow from the symmetry relations in \eq{potsym}.

Note that the above potentials have N$^3$LL accuracy. This is a new result. Additionally, we give expressions with N$^3$LO accuracy, ${\cal O}(\als^2)$, for the Wilson loops with chromomagnetic (and/or chromoelectric) insertions in the $\MS$. One can easily change to other schemes by changing e.g. $c_F^{(i)}$ from the $\MS$ to the lattice scheme (since the whole potential is scheme independent). This enables a more detailed comparison with lattice simulations at short distances. This research will be carried out elsewhere. 

Overall, with very few new computations we have been able to obtain the spin-dependent momentum-dependent $1/m^2$ potentials with N$^3$LL accuracy. The NNLL result was originally obtained in \rcite{Chen:1994dg}.

\subsection{$V_r$ and $V^{(1,1)}_{S^2}$ potentials}
\label{Sec:delta}

The remaining potentials we need to consider are $V_r$ 
and $V^{(1,1)}_{S^2}$. 
At ${\cal O}(\als)$ they are proportional to $\delta({\bf r})$, which does not contribute to the spectrum of $l \not=0$ 
states to the order we work (the delta-like potential contribution vanishes at first and 2nd order in perturbation theory). At ${\cal O}(\als^2)$,  potentials proportional to
$\ln k$ (or reg $1/r^3$ in position space) are generated in the NRQCD-pNRQCD matching. Such potentials generate 
non-zero contributions to the spectrum of $l \not=0$ states. We know them at leading nonvanishing order, which is all we need. We need them both for the spin-dependent and the spin-independent potentials. 

The spin-dependent potential has been computed with N$^3$LL accuracy in \rcite{Kniehl:2003ap,Penin:2004xi}. We are only interested in the term proportional to ${ \rm reg} \frac{1}{r^3}$, which reads 
\begin{align}
 V^{(1,1)}_{S^2,\rm RG}(r)&\dot=
\frac{8\pi C_F}{3} \left[- \frac{1}{4\pi} { \rm reg} \frac{1}{r^3}\right] c_F^\one c_F^\two \frac{\als^2}{\pi}
\left(-\frac{\beta_0}{2}+\frac{7}{4}C_A\right)
  ,
\label{V11LS}
\end{align}
where
\begin{align}
- \frac{1}{4\pi} { \rm reg} \frac{1}{r^3} 
\equiv  \int \frac{d^3k}{(2\pi)^3} e^{-i{\bf k} \cdot {\bf r}}\ln k 
\, .
\end{align}
The correction  to the fixed order potential  comes from considering the difference between $c_F^\one c_F^\two$ evaluated at $\nuh$ and at $\nuh=\nu$.

The spin-independent $V_r$ is at present unknown with N$^3$LL accuracy (indeed, it is the missing link to obtain the complete N$^3$LL spectrum for a general S-wave energy), since the ${\cal O}(\als^2)$ of the delta potential is not known. This does not affect our analysis, since the term proportional to $\delta^{(3)}({\bf r})$ does not contribute to the energy of P-wave states. On the other hand, we know the term 
proportional to 
 ${ \rm reg} \frac{1}{r^3}$  with enough accuracy, as it can be deduced from the $k$ dependence of the NNLL result.
It reads
\begin{align}
\frac{ V^{(2,0)}_{r,\rm RG}(r)}{m_1^2}+\frac{ V^{(0,2)}_{r,\rm RG}(r)}{m_2^2}
+\frac{ V^{(1,1)}_{r,\rm RG}(r)}{m_1 m_2}
\dot=&
\frac{\pi C_F}{m_1 m_2}
\left[- \frac{1}{4\pi} { \rm reg} \frac{1}{r^3}\right] 
\left( k \frac{d}{d k}\tilde D_{d}^{(2)}\right)\Bigg|_{k=\nu}^{LL}
\,,
\label{deltaVreg}
\end{align}
where
\begin{align}
\label{Ddreg}
  k \frac{d}{ d k}\tilde D_{d}^{(2)}\Bigg|_{k=\nu}^{LL}
  &=
  -\beta_0\frac{\als^2}{2\pi}+
  \frac{\als^2}{\pi}\left(2C_F-\frac{ C_A}{
2}\right)c_k^{(1)}c_k^{(2)} 
\\
\nn
&
+
\frac{\als^2}{\pi}\left[
\frac{m_1}{ m_2}\left(\frac{1}{3}T_fn_f \bar c_1^{hl(2)}- \frac{ 4 }{3}(C_A+C_F)[c_k^{(2)}]^2-\frac{ 5 }{ 12}C_A[c_F^{(2)}]^2\right)
\right.
\\
\nn
&
\left.
\qquad
+\frac{m_2 }{m_1}\left(\frac{1}{3}T_fn_f \bar c_1^{hl(1)}- \frac{ 4 }{ 3}(C_A+C_F)[c_k^{(1)}]^2-\frac{ 5 }{ 12}C_A[c_F^{(1)}]^2\right)
 \right]
  \\
  &
  \nn
  -\frac{(m_1+m_2)^2}{m_1m_2}\frac{4}{3}\left(\frac{C_A}{2}-C_F\right)\frac{\als^2}{\pi}\left[\ln\left(\frac{\als}{\alus}\right)+1\right]
\,,
\end{align}
($c_k^{(i)}=1$ because of reparameterization invariance \cite{Luke:1992cs}) and the gauge independent combination of NRQCD Wilson coefficients
\begin{align}
\bar c_1^{hl(i)}(\nu)&\equiv c_1^{hl(i)}(\nu)+c_D^{(i)}(\nu)=z^{-2C_A}+\left(\frac{20 }{ 13}+\frac{32 }{ 13}\frac{C_F }{C_A}\right)\left[1-z^\frac{-13C_A }{ 6}\right]
\end{align}
was computed in \rcites{Bauer:1997gs,Blok:1996iz}.

Finally, note that the ultrasoft contribution to $V_r$ in \eq{Ddreg} is $1/N_c^2$ suppressed and that \eq{Ddreg} is the expression in the on-shell scheme. 

\section{Total shift on the energy levels}

The P-wave spectrum at N$^3$LO was obtained in \rcites{Kiyo:2013aea,Kiyo:2014uca} for the equal mass case and in \rcite{Peset:2015vvi} for the unequal mass case. The resulting expression for $E_{\rm N^3LO}$ can be found in \app{app:spectrum}. 
From the RGI potentials discussed in \Sec{sec:RG} we obtain the N$^i$LL shift in the energy levels
\begin{align}
E_{\rm N^iLL}(\nu,\nuh,\nuus) & = E_{\rm N^iLO}+ \delta E_{\RG}(\nu,\nuh,\nuus)\Big|_{\rm N^iLL}.
\end{align}
where $E_{\rm N^iLO}=E_{\rm N^iLL}(\nu,\nu,\nu)$. The explicit expressions of the fixed order and resummed energies can be found in \apps{app:spectrum}{sec:NNNLL}.

The LO and NLO energy levels are unaffected by the RG improvement, i.e.
\begin{align}
\delta E_{\RG}\Big|_{\rm LL}=\delta E_{\RG}\Big|_{\rm NLL}=0.
\end{align}
We now determine the variations with respect to the NNLO and N$^3$LO results.
 We are here interested in the corrections associated to the resummation of logarithms.
In order to obtain the spectrum of a P-wave at NNLL and N$^3$LL we need to add the following energy shift to the NNLO and N$^3$LO spectrum (strictly speaking we only compute the piece that contributes to the P-wave spectrum):
\begin{align}
&\delta E_{\RG}\Big|_{\rm NNLL}=
\langle nl|
\delta V^{\RG}_{s,\rm NNLL}
|nl\rangle=E_n^C\left(\frac{\als}{\pi}\right)^2\delta c_2
\,,
\end{align}
which was computed in \rcite{Pineda:2001ra} for equal masses, and 
\begin{align}
&\delta E_{\RG}\Big|_{\rm N^3LL}=\label{deltaEN3LLdef}
\langle nl|
\delta V^{\RG}_{s,\rm NNLL}
|nl\rangle+\langle nl|
\delta V^{\RG}_{s,\rm N^3LL}|nl\rangle\\
&\hspace*{0.5cm}+
2\langle nl|[V_1^{(0)}-V_0^{(0)}] \frac{1}{\left(E_n^C-h\right)'} \delta V^{\RG}_{s,\rm NNLL}|nl\rangle
+\left[\delta E_{\US}(\nu,\nuus)-\delta E_{\US}(\nu,\nu)\right]
\nn
\\
&\hspace*{2cm}=E_n^C\left[\left(\frac{\als}{\pi}\right)^2\delta c_2+\left(\frac{\als}{\pi}\right)^3\left(2 \beta _0 \delta c_2 L_{\nu } +\delta c_3\right)\right]
\,.
\label{deltaEN3LL}
\end{align}
Let us note that the second term in \eq{deltaEN3LLdef}, besides the N$^3$LO ultrasoft corrections to the momentum-dependent potentials, also contains the $\ln k$ contributions associated to $V_r$ and $V_{S^2}$ discussed in \Sec{Sec:delta}. In the 3rd term of \eq{deltaEN3LLdef} we only have to consider the LL ultrasoft running of the momentum-dependent potentials and the LL (hard) running of the spin-dependent potentials. 
The 2nd and 3rd terms in \eq{deltaEN3LLdef} are computed in the same way we did in \rcite{Peset:2015vvi}. 
We add the last term in \eq{deltaEN3LLdef} in order to account for the evaluation at the ultrasoft scale of the ultrasoft energy:
\begin{align}
\label{EnlUS}
\delta E_{nl}^{\US}(\nu,\nuus)&=-E_n^C\frac{\als\alus}{\pi}\left[\frac{2}{3} C_F^3 L^E_{nl}+\frac{1}{3} C_A\left(L_{\nuus}-L_{\US}+\frac{5}{6}\right) \left(\frac{C_A^2}{2}+\frac{4 C_A C_F}{(2 l+1) n}\right.\right.\nn\\
&\quad +\left.\left.2C_F^2 \left(\frac{8}{(2 l+1) n}-\frac{1}{n^2}\right)\right)+\frac{8\delta_{l0} }{3 n}C_F^2  \left(C_F-\frac{C_A}{2}\right) \left(L_{\nuus}-L_{\US}+\frac{5}{6}\right)\right],
\end{align}
where $L_{\US} = \ln\frac{C_F\als\, n}{2}+S_1(n+l)$, and $L^E_n$
are the non-Abelian Bethe logarithms. 
Numerical determinations of these non-Abelian Bethe logarithms for $l\not=0$ can be found in \rcite{Kiyo:2014uca}.

In \eq{deltaEN3LL}, $E_n^C=-\frac{m_rC_F^2\als^2}{2n^2}$, $L_{\nu} =\ln\frac{n \nu}{2 C_F m_r \als}+S_1(n+l)$. We split the $\delta c_i$ coefficients into a Coulomb-like and a non-Coulomb like contributions
\begin{align}
\delta c_i=\delta c_i^{\rm c}+\delta c_i^{\rm nc}.\label{deltaci}
\end{align} 
for $i=2,3$. The corrections $\delta c_i^{\rm c}$ are given in \eqs{deltac2c}{deltac3c}. They are generated by the ultrasoft corrections to the static potential. The relativistic ultrasoft contribution to $\delta c_{2/3}^{\rm nc}$ is produced by \eqss{deltaV1}{deltaVp2}, plus the ultrasoft part of \eq{Ddreg} and \eq{EnlUS}, to the appropriate order.  Explicit expressions for these quantities can be found in \eqs{c2ncus}{c3ncus}. The hard contribution to $\delta c_{2/3}^{\rm nc}$ is generated by the (non-trivial) hard/soft running of the relativistic potentials encoded in the NRQCD Wilson coefficients. The explicit expressions can be found in \eq{c2ncSDh} and \eq{deltac3nch}. The former only receives contributions from the spin-dependent potentials, whereas the latter receives contributions from both the spin-dependent and the spin-independent potentials. The spin-(in)dependent contributions from the running of the Wilson coefficients can be found in \eq{c3ncSDh} (\eq{c3ncSIh}).

Note that, throughout this paper, we have introduced a change of the basis of spin operators with respect to the basis used in \rcite{Peset:2015vvi} to compute the spectrum for different masses: $\{{\bf S},{\bf S_1}\}\longrightarrow \{{\bf S},{\bf S}^-\}$ where the symmetric and antisymmetric spin operators are ${\bf S}={\bf S_1}+{\bf S_2}$ and ${\bf S}^-={\bf S_1}-{\bf S_2}$. We find that the latter basis suits better the description of the heavy quarkonium spectrum since $\langle {\bf S^-}\rangle=0$. We give the expressions of the N$^3$LO energy in the new spin basis in \app{app:spectrum}.

From the N$^3$LL computation we can obtain the large logs of $\mathcal{O}(\als^6)$ for the expansion of the mass in powers of $\als$ at the scale $\nu=mC_F\als$. For 
the $n=2$, $l=1$ state and equal masses, it reads (with $n_f=3$)
\begin{align}
\delta E_{21}&=E_2^{C}\left(\frac{\als}{\pi}\right)^4\left[\ln \frac{1}{C_F\als } (81.4171 D_s-2.19325 \mathcal{S}_{12}+160.084 X_{LS}-7160.10)\right.\nn\\
&\left.\hspace*{2.5cm}+\ln ^2\frac{1}{C_F\als }(-8.22467  D_s-13.1595 X_{LS}-244.684)\right].
\end{align}

\section{Fine and hyperfine splitting}

The results of the previous section apply to a general state with $l\not=0$. Now we would like to study in more detail the fine and hyperfine splittings of P-wave states. Note that these splittings do not depend on the ultrasoft scale at the order at which we are working. In principle, this means that we do not have to rely on the assumption that the ultrasoft scale can still be handled in the weak-coupling approximation (otherwise the power counting of the nonperturbative corrections changes). If one assumes that $mv^2 \gg \lQ$, the complete expression for the leading nonperturbative expression was computed in \cite{Pineda:1996nw}\footnote{We profit to correct Eq.~(3.6) of that reference that should read
$$
\Delta _{\rm HF}({\rm new})
=-
\frac{\pi \langle \als G^2 \rangle}{m^3(C_F\tilde \als)^2}\frac{\als}{\tilde \als}
\frac{79139056}{1437897825}
\,.
$$
The change is produced by an algebraic mistake in $V_8^{\rm HF}(\rm annihilation)$ (the ``-3'' should be zero). This makes the $1/N_c^2$ correction vanish in Eqs.~(1.7) and (3.1), changes 29 $\rightarrow$ 32 in Eq.~(3.2) and Eq.~(3.6) to the expression above.} (earlier partial results can be found in \cite{Dosch}). In any case, we will not try to incorporate nonperturbative effects in this paper,  lacking a more clear understanding of the behavior of the perturbative series.

\subsection{Fine splitting}

In general,  we find for $s=1$ and $l\neq 0$ (following standard heavy quarkonium spectroscopy we define $n_r=n-1$ for P-wave states): 
\begin{align}
E(n_r^3L_j)-E(n_r^3L_{j'})&=E_n^C \left(\frac{\als}{\pi}\right)^2\left[\delta c_2^{\rm SD,h}\Big|_j-\delta c_2^{\rm SD,h}\Big|_{j'}\right]\nn\\
&+E_n^C \left(\frac{\als}{\pi}\right)^3\left[\left(\delta c_3^{\rm SD,h}+2\beta_0 \delta c_2^{\rm SD,h}L_\nu+\mathcal{E}_{\rm h} \left(\frac{\alh}{\als}L_{\nuh}-L_\nu\right)\right)\Big|_j\right.\nn\\
&\left.-\left(\delta c_3^{\rm SD,h}+2\beta_0 \delta c_2^{\rm SD,h}L_\nu+\mathcal{E}_{\rm h} \left(\frac{\alh}{\als}L_{\nuh}-L_\nu\right)\right)\Big|_{j'}\right]
\,,
\end{align}
where $j$ is the quantum number associated to the combination of operators ${\bf J}={\bf L}+{\bf S}$.

For different masses and $n=2$ we find:
\begin{align}
&E(1^3P_j)-E(1^3P_{j'})=\frac{\als ^4 C_F^4 m_r^3 }{192 m_1 m_2}\left\{4(D_s\big|_{j}-D_s\big|_{j'}) z^{-\gamma_0} \right.\nn\\
&\times\left.\left[1+\frac{\als  }{2 \pi }\left(2 \left(\frac{209}{36}-\frac{\pi ^2}{3}\right) \beta_0+\frac{\alh-\als }{\als }\left(\frac{\gamma_1}{2 \beta_0}-\frac{\beta_1 C_A}{\beta_0^2}+C_A \ln \left(\frac{\nuh^2}{m_1 m_2}\right)+2 C_A+2 C_F\right)\right.\right.\right.\nn\\
&\left.\left.\left.+2 (2 \beta_0-C_A) \ln \frac{\nu }{ m_rC_F\als  }-C_A \ln \frac{m_1 m_2}{\nuh^2}-\frac{11 C_A}{3}+2 C_F\right)\right]\right.\nn\\
&\left.-(j (j+1)-j'(j'+1)) \frac{m_1^2+m_2^2 }{m_1 m_2}\left[1+\frac{\als  }{2 \pi }\left(\beta_0 \left(4 \ln \frac{\nu }{ m_r C_F\als  }-\frac{2 \pi ^2}{3}+\frac{215}{18}\right)-\frac{16 C_A}{3}\right)\right]\right.\nn\\
&\left.+(j (j+1)-j'(j'+1))\frac{2 m_1 m_2}{m_r^2} z^{-\frac{\gamma_0}{2}} \left[1+\frac{\als }{4 \pi } \left(2 \beta_0 \left(4 \ln \frac{\nu }{ \als C_F m_r}-\frac{2 \pi ^2}{3}+\frac{215}{18}\right)+2 C_F\right.\right.\right.\nn\\
&\left.\left.\left.-2C_A \left(\ln\frac{\nu }{   m_rC_F\als}+\frac{m_r}{m_1m_2}\left(m_2\ln\frac{m_1}{\nuh}+m_1\ln\frac{m_2}{\nuh}\right)+\frac{16}{3}\right)\right.\right.\right.\nn\\
&\left.\left.\left.+\frac{\alh-\als }{\als }\left(\frac{\gamma_1}{2 \beta_0}-\frac{\beta_1 C_A}{\beta_0^2}-\frac{2m_r}{m_1m_2} C_A \left(m_2\ln \frac{m_1}{\nuh}+m_1\ln \frac{m_2}{\nuh}\right)+2 C_A+2 C_F\right)\right)\right]\right\}
\,,
\end{align}
where $D_s=1/2\langle S_{12}({\bf r})\rangle$.
Note that the equal mass case is obtained just by taking $m_1=m_2=m$.

We have checked that in the limit $\nuh=\nu$ we recover the result at N$^3$LO obtained in Eq.~(26) of \rcite{Brambilla:2004wu}.

Finally we can obtain the leading large logarithms for the fixed order contribution by expanding $\alh$ in terms of $\als$. The leading logarithmic resummation contains all terms of order $\als^{4+n}\ln^n \als$, while the NLO resummation contains all terms of order $\als^{5}\als^{n}\ln^n\als$. Setting $\nuh=\sqrt{m_1m_2}$ and $\nu=2C_Fm_r\als$ we obtain the higher order logarithms:
\begin{align}
&E(1^3P_j)-E(1^3P_{j'})\Big|_{>\mathcal{O}(\als^5)\times \rm logs}=\nn\\
&\hspace*{1cm}E_2^C\left(\frac{\als }{\pi }\right)^4
\ln\left(\frac{1}{C_F\als}\right)\frac{ \pi ^2 C_F^2}{96 }   \left\{\frac{8 m_r^2 }{ m_1 m_2}(D_s\big|_{j}-D_s\big|_{j'})\left[-\frac{23 C_A^2}{4}-\frac{2}{3} \pi ^2 \beta_0 C_A+\frac{493 \beta_0 C_A}{36}\right.\right.\nn\\
&\hspace*{1.5cm}\left.\left.-2 C_A (C_A-2 \beta_0) \ln2+2 C_A C_F+\beta_0 C_F-\frac{C_A}{2} ( \beta_0+2 C_A) \ln \frac{m_1m_2}{4 m_r^2}\right]\right.\nn\\
&\hspace*{1.5cm}\left.+(j (j+1)-j' (j'+1))\left[-\frac{89 C_A^2}{6}-\frac{4}{3} \pi ^2 \beta_0 C_A+\frac{505 \beta_0 C_A}{18}+2 C_A (4 \beta_0-C_A) \ln 2\right.\right.\nn\\
&\hspace*{1.5cm}\left.\left.+2 C_A C_F+2 \beta_0 C_F+C_A(\beta_0+C_A)\left(\frac{m_1-m_2}{m_1+m_2}\ln\frac{m_1}{m_2}- \ln \frac{m_1m_2}{4 m_r^2}\right)\right]\right\}\nn\\
&\hspace*{1cm}-E_2^C\left(\frac{\als}{\pi}\right)^4\ln ^2\left(\frac{1}{C_F\als}\right)\frac{ \pi ^2 C_A C_F^2 }{96}\bigg\{(j (j+1)-j' (j'+1))(\beta_0+C_A)\nn\\
&\hspace*{1.5cm}\left.+\frac{4 m_r^2  }{ m_1m_2}( \beta_0+2 C_A)(D_s\big|_{j}-D_s\big|_{j'})\right\}
\,.
\end{align}

\subsection{Hyperfine splitting}
The hyperfine splitting of P-wave states is defined in the following way:
\begin{align}
\Delta_{n,l} \equiv E(n^1l_{j=l})
-E(n^3l)_{c.o.g.}
\,,
\end{align}
where the ''center of gravity'' average reads
\begin{align}
E(n^3l)_{c.o.g.}
=
\frac{2l-1}{3(2l+1)}E(n^3l_{j=l-1})
+
\frac{2l+1}{3(2l+1)}E(n^3l_{j=l})
+
\frac{2l+3}{3(2l+1)}E(n^3l_{j=l+1})
\,.
\end{align}
In practice we will use this expression only 
for the case $n=2$ and $l=1$:
\begin{align}
\label{eq:Delta}
\Delta \equiv \Delta_{2,1}=
E(1^1P_1)-\frac{1}{9}\left( 5 E(1^3P_2)+3E(1^3P_1)+E(1^3P_0)\right)
\,.
\end{align}

For general radial and $l\not=0$ angular quantum numbers, and different masses, we find at fixed order
\begin{align}
\Delta_{n,l}&=-\frac{m_r^3 C_F^4\als^5 (1-\delta_{l0}) }{9 m_1m_2 \pi  l (l+1) (2 l+1) n^3}(C_A-8n_fT_F)
\,.
\label{DeltaHFnl}
\end{align}
We have checked that \eq{DeltaHFnl} for $m_1=m_2=m$ recovers the result obtained in \rcite{Titard:1994id}. Note that the hyperfine splitting for P-wave states is ${\cal O}(\als)$ suppressed compared with the hyperfine splitting of S-wave states.  This suppression is indeed seen experimentally (actually, experimentally, the suppression is stronger than expected. For a 
discussion on this issue see \cite{Lebed:2017yme}). 

The resummation of logarithms can be easily obtained by incorporating the nontrivial running of the NRQCD Wilson coefficients. The general N$^3$LL result for a P-wave state reads 
\begin{align}
\Delta_{n,l}^{\rm RG}&=-\frac{m_r^3 C_F^4\als ^5 (1-\delta_{l0}) }{9 m_1m_2 \pi  l (l+1) (2 l+1) n^3}(C_A-8n_fT_F)z^{-\gamma_0}
\,.
\end{align}
Note that this quantity is positive, because it is one of the few places where light-fermion effects are more important than nonabelian effects.

Since we only have the first oder in the logarithmic expansion, we can only compute terms that are $\als^{5+n}\ln^n\als$. Setting $\nuh=\sqrt{m_1m_2}$ and $\nu=2m_rC_F\als$, we obtain for a P-wave
\begin{align}
\Delta_{n,l}\Big|_{\ln\als}&=\nn\\
&\hspace*{-0.3cm}
=\frac{m_r^3C_F^4\als^6 C_A (C_A-8 T_Fn_f )}{9\pi ^2m_1m_2 n^3}\frac{(1-\delta_{l0})}{l (l+1) (2 l+1)}  \ln \frac{1}{C_F\als}\left\{1-\frac{\als}{4\pi}\left(2C_A+3\beta_0\right)\ln\frac{1}{C_F \als }\right\}
\,.
\end{align}

\section{Phenomenology of $n=2$, $l=1$ states at strict weak coupling}
\label{Sec:Phen}

We now confront our results with the experimental values of the spectrum \cite{Patrignani:2016xqp} for $n=2$, $l=1$ states, which we list in \tab{tab:DeltaVals}. We use the bottom and charm quark masses determined in \rcite{Peset:2018ria}. For the strong coupling we take $\als(M_z)=0.1184(12)$ from \rcite{Patrignani:2016xqp}.

\begin{table*}
\begin{center}
  \caption{Experimental values of the heavy quarkonium masses, $\Delta$ and the fine splittings in MeV.}
\label{tab:DeltaVals}
\begin{tabular}{c|c ||c||c}
  \hline\hline
  System & $b\bar b(1P)$ (exp) &$c\bar b(1P)$ (exp) &$c\bar c(1P)$  (exp)\\
 \hline\hline 
   $h ({}^1 \! P_1)$ & 9899.3(8) & --- & 3525.38(11)  \\
  
   $\chi_{0} ({}^3 \! P_0)$ & 9859.44(42)(31) &  --- & 3414.71(30) \\
  
  $\chi_{1} ({}^3 \! P_1)$ &  9892.78(26)(31) &  --- & 3510.67(5) \\
  
  $\chi_{2} ({}^3 \! P_2)$ &  9912.21(26)(31) &  --- &  3556.17(7)\\
  $\Delta$  & $-0.57(84)$ &  --- & +0.08(13)\\
  $\chi_{1} ({}^3 \! P_1)-\chi_{0} ({}^3 \! P_0)$ &32.49(93) & --- &95.96 (30) \\
  $\chi_{2} ({}^3 \! P_2)-\chi_{1} ({}^3 \! P_1)$ & 19.10(25) & --- & 45.5 (1)\\
  \hline \hline
\end{tabular}
\end{center}
\end{table*}


\subsection{Spin-independent energy combination}
\begin{figure}[!htb]
	\begin{center}
	\includegraphics[width=0.49\textwidth]{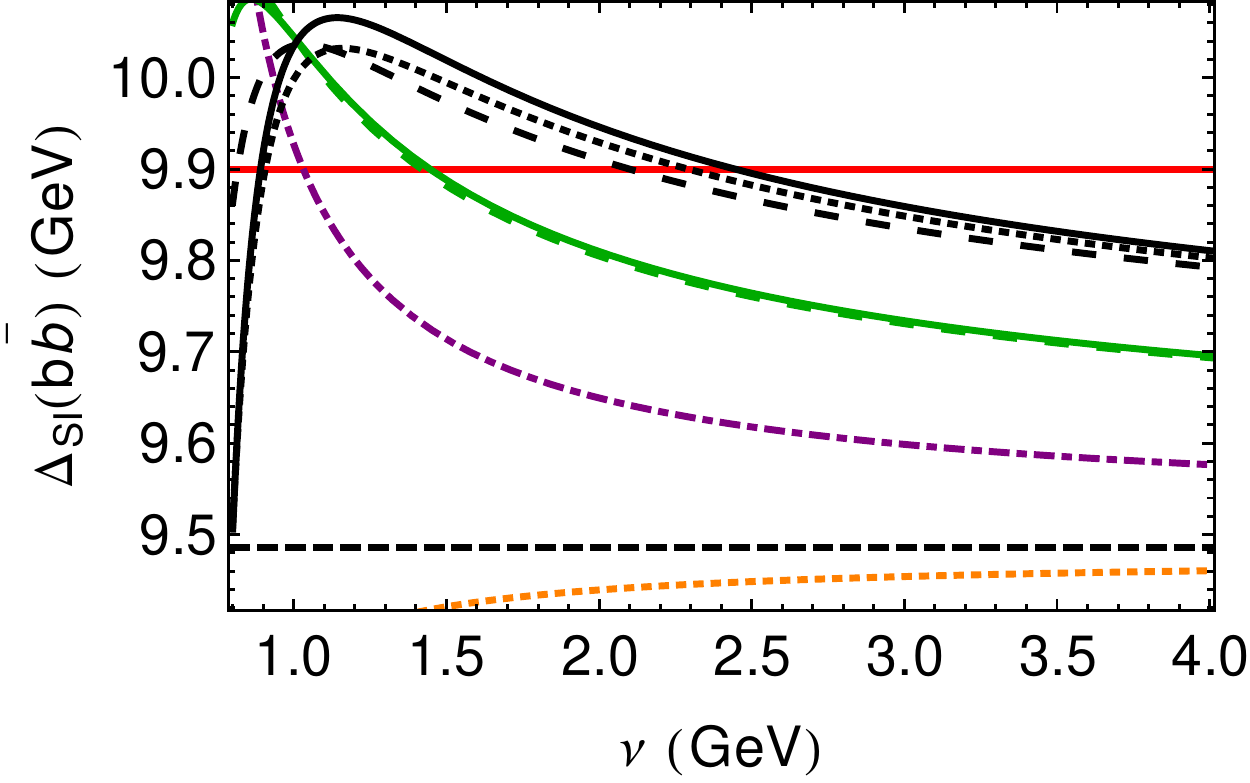}      
	\includegraphics[width=0.49\textwidth]{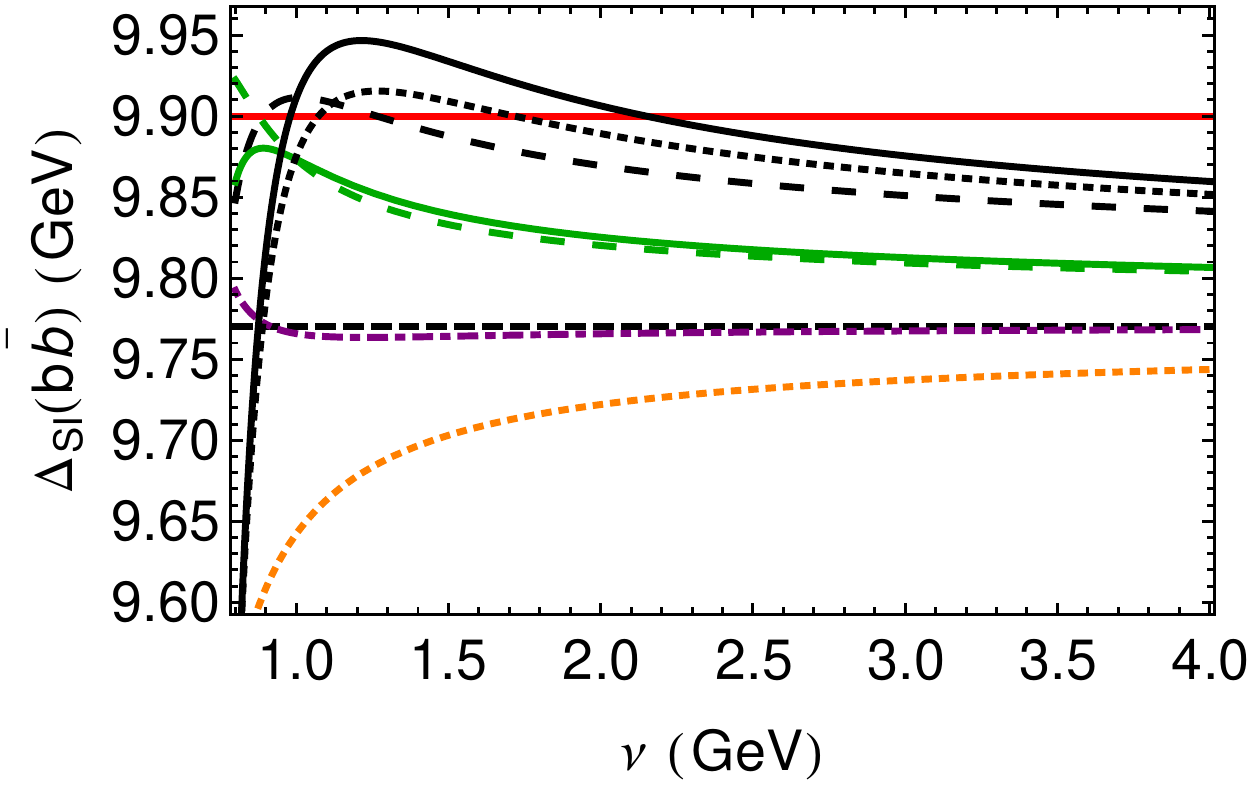}\\
	\includegraphics[width=0.49\textwidth]{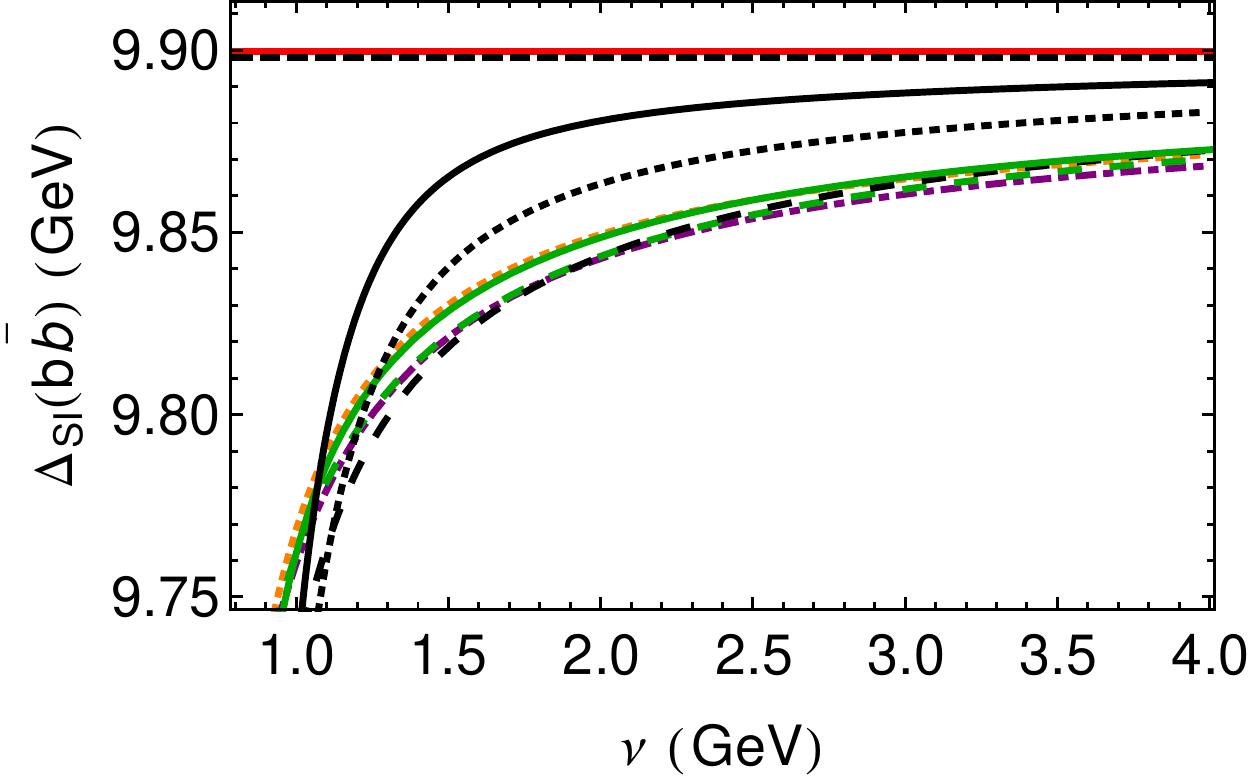}
	\includegraphics[width=0.49\textwidth]{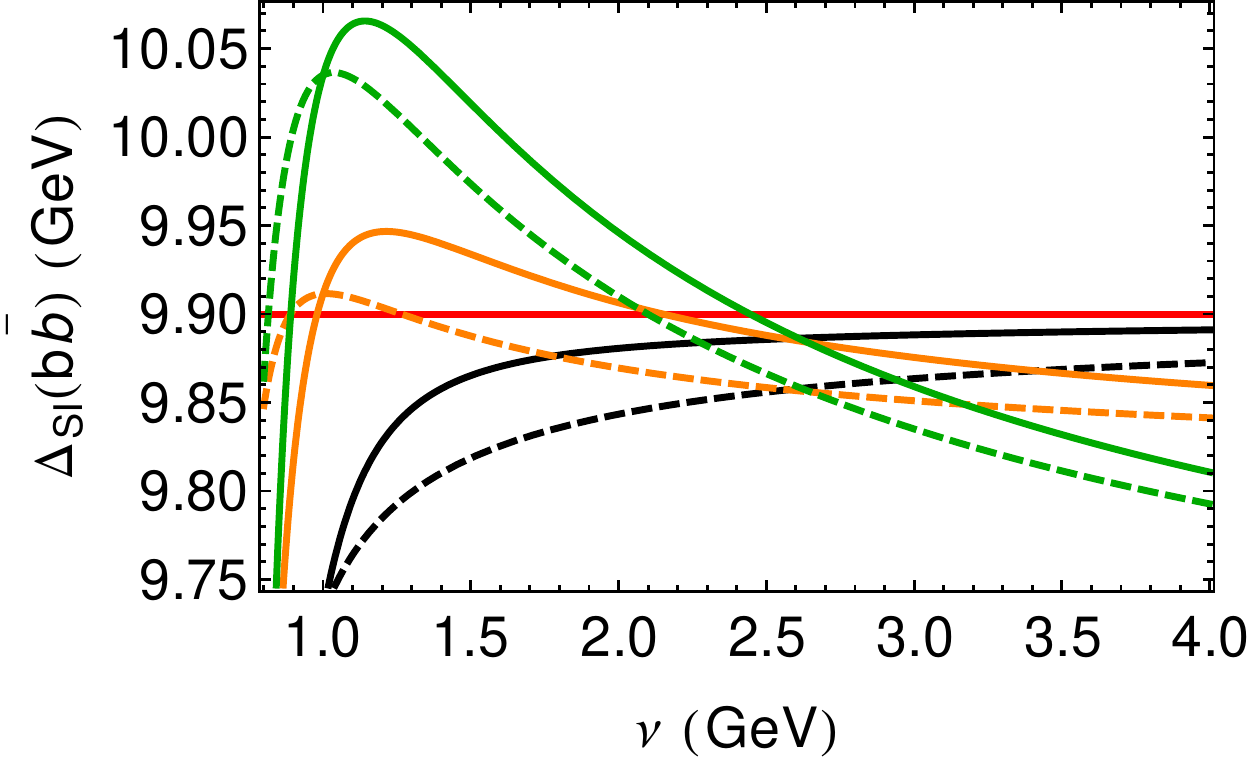}
\caption{Plots for $\Delta_{SI}$ in the RS' scheme with $\nuus=1$ GeV for bottomonium.  {\bf Upper left, upper right and lower left panels:} Plots for $\nu_f=2$, 1 and 0.7 GeV respectively. The red line is the experimental value, the black-dashed line is $2m_{b,\RS'}$. The orange-dotted, purple dot-dashed, green-dashed and black-dashed lines are $\Delta_{SI}$ evaluated at LO-N$^3$LO, respectively. The solid-green and solid-black lines are the NNLL and N$^3$LL result respectively, and the dotted-black line is the N$^3$LL result without $\delta E_{21}^{\US}$. 
{\bf Lower right panel:} Comparison of the $\nu_f=2$ GeV (green),  $\nu_f=1$ GeV (orange) and $\nu_f=0.7$ GeV (black) lines. For each case, the dashed line is the  N$^3$LO result and the solid line the N$^3$LL one.
\label{Fig:hb}}   
\end{center}
\end{figure}
\begin{figure}[!htb]
	\begin{center}
	\includegraphics[width=0.49\textwidth]{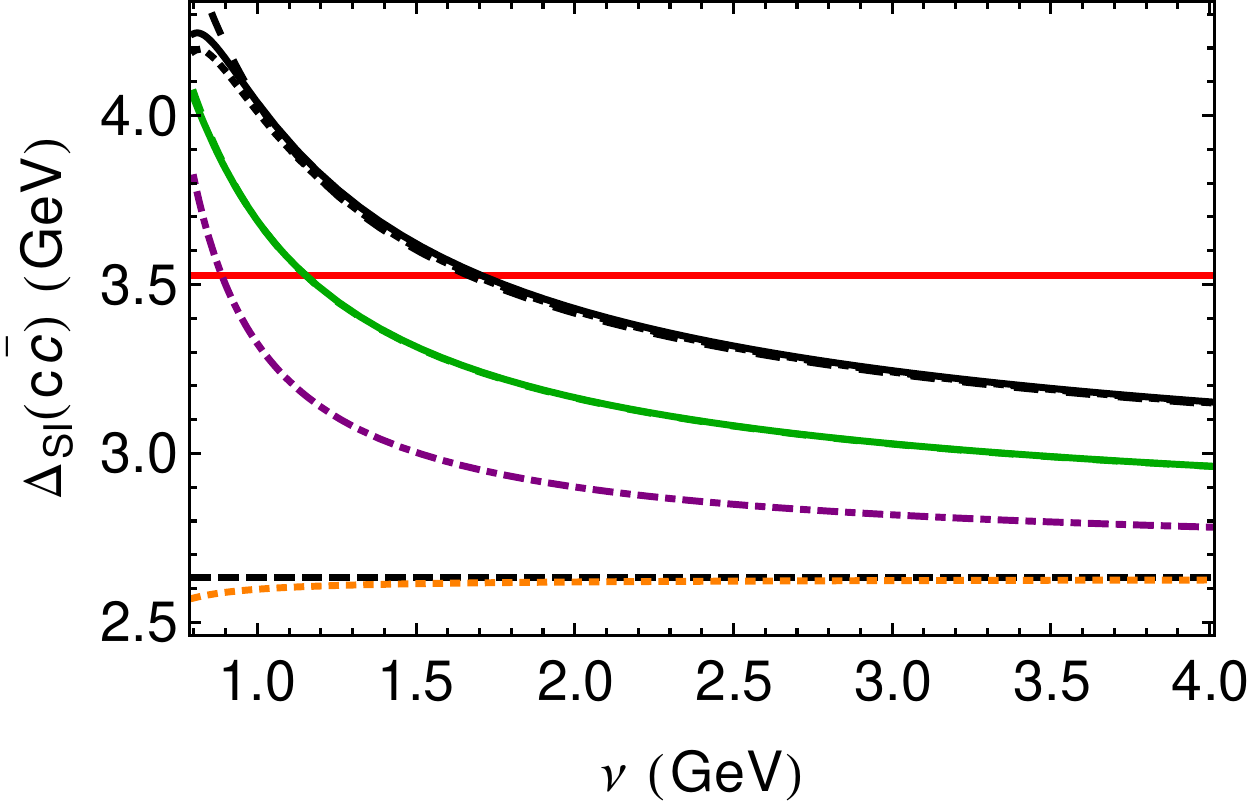}
	\includegraphics[width=0.49\textwidth]{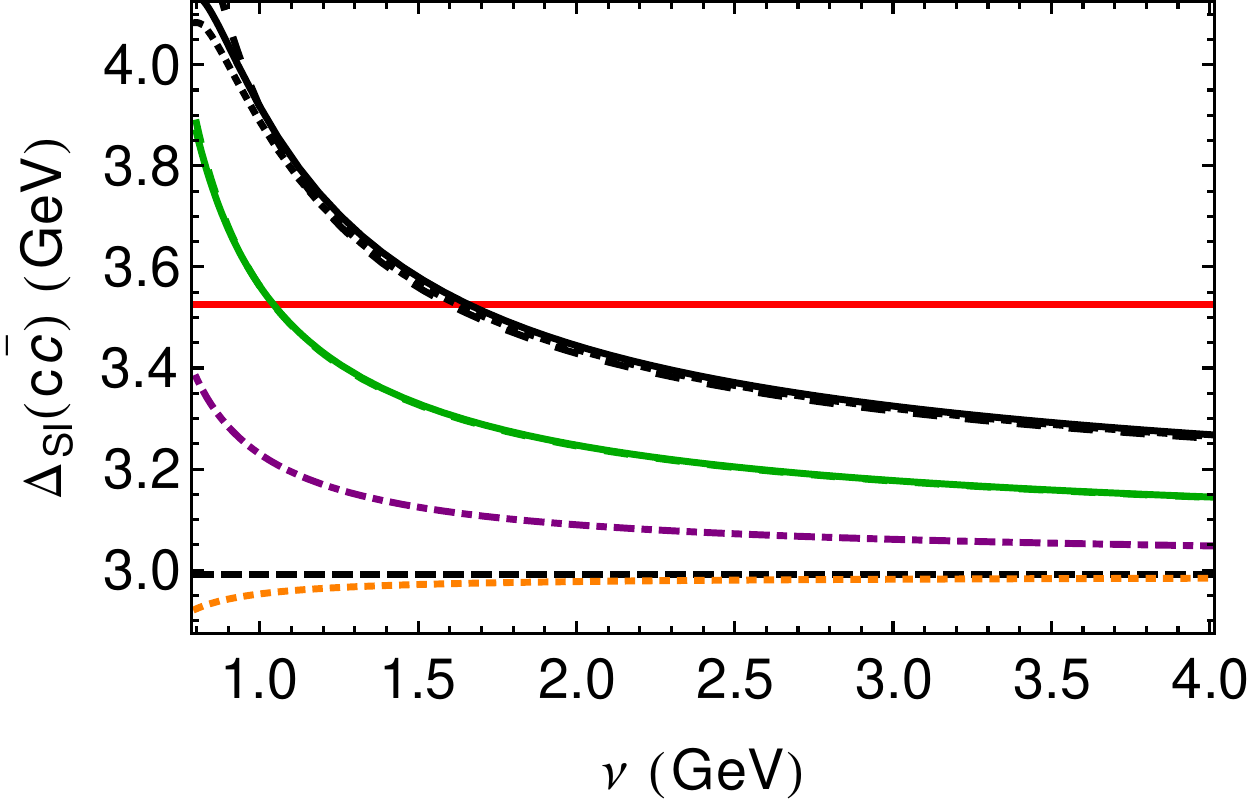}\\
	\includegraphics[width=0.49\textwidth]{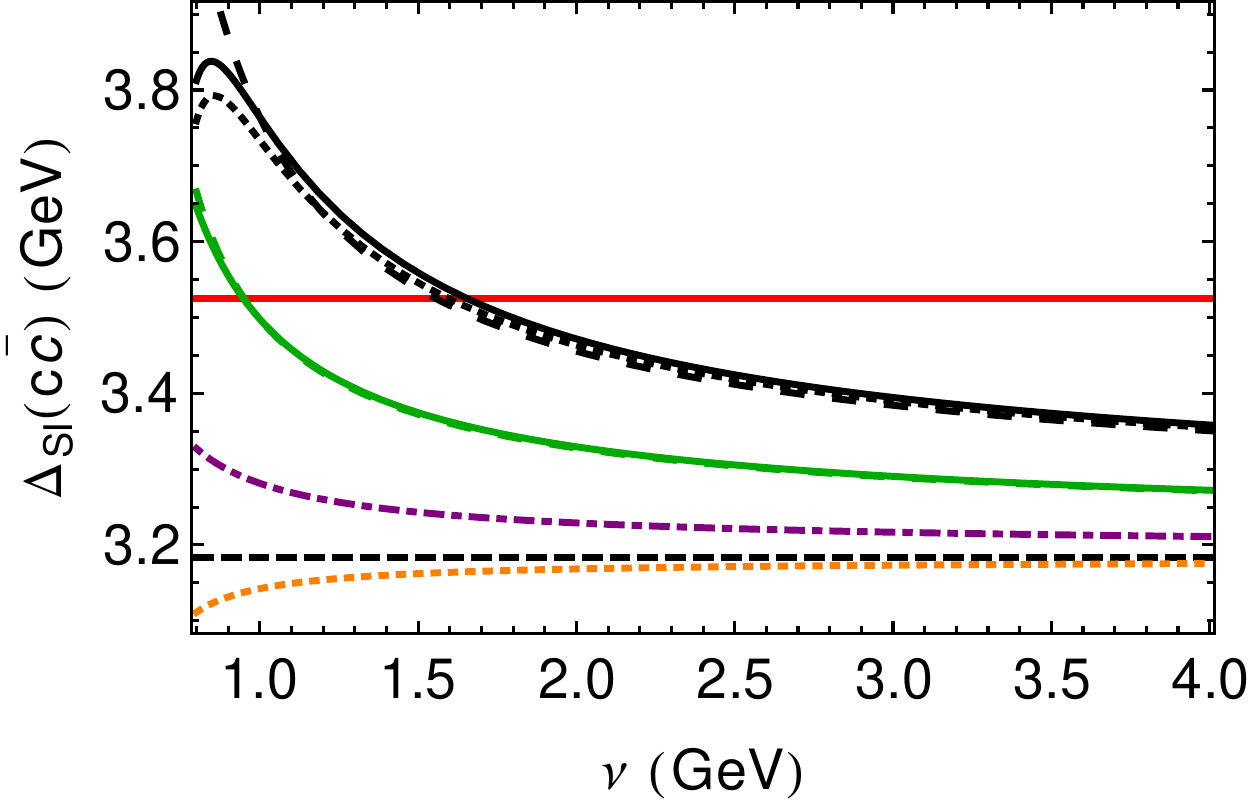}
	\includegraphics[width=0.49\textwidth]{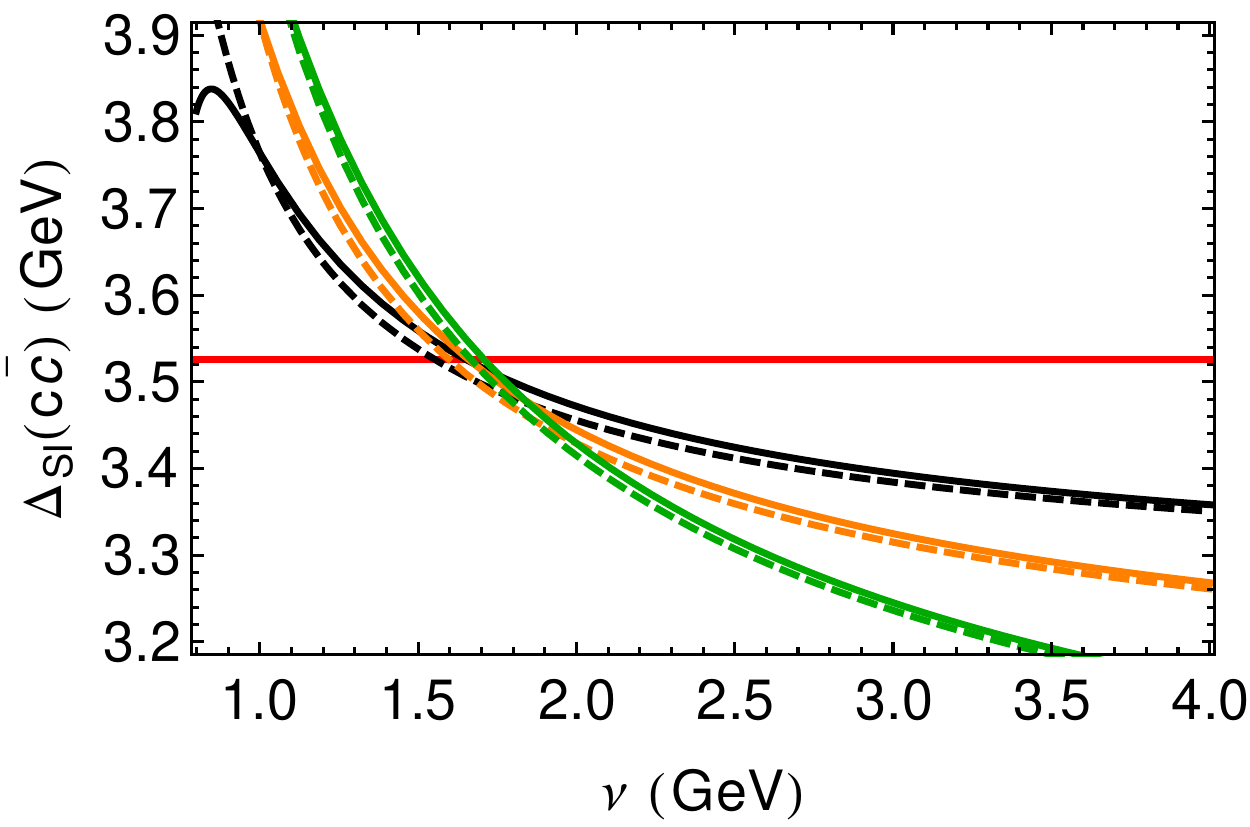}
	%
\caption{Plots for $\Delta_{SI}$ in the RS' scheme with $\nuus=1$ GeV for charmonium.  {\bf Upper left, upper right and lower left panels:} Plots for $\nu_f=2$, 1 and 0.7 GeV respectively. The red line is the experimental value, the black-dashed line is $2m_{b,\RS'}$. The orange-dotted, purple dot-dashed, green-dashed and black-dashed lines are $\Delta_{SI}$ evaluated at LO-N$^3$LO, respectively. The solid-green and solid-black lines are the NNLL and N$^3$LL result respectively, and the dotted-black line is the N$^3$LL result without $\delta E_{21}^{\US}$. 
{\bf Lower right panel:} Comparison of the $\nu_f=2$ GeV (green),  $\nu_f=1$ GeV (orange) and $\nu_f=0.7$ GeV (black) lines. For each case, the dashed line is the  N$^3$LO result and the solid line the N$^3$LL one.
\label{Fig:hc}}   
\end{center}
\end{figure}
\begin{figure}[!htb]
	\begin{center}
	\includegraphics[width=0.49\textwidth]{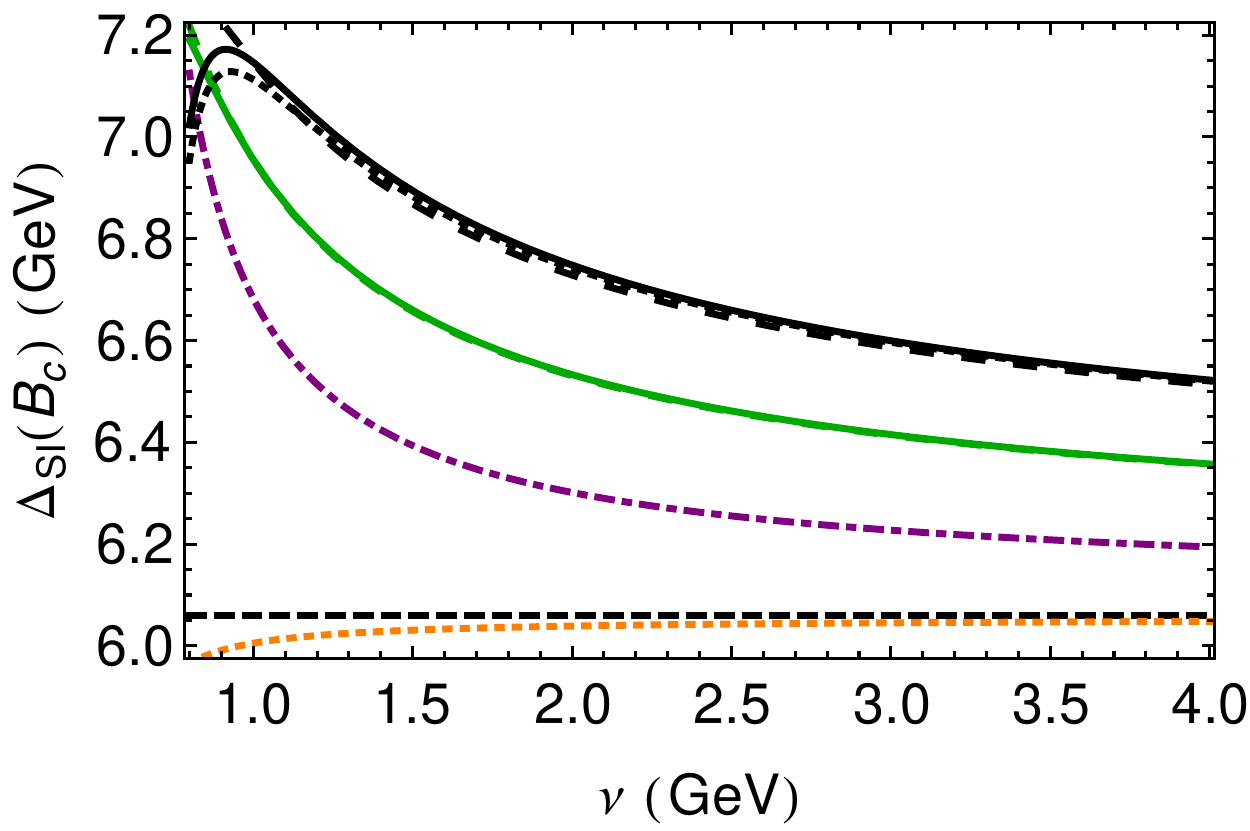}	
    \includegraphics[width=0.49\textwidth]{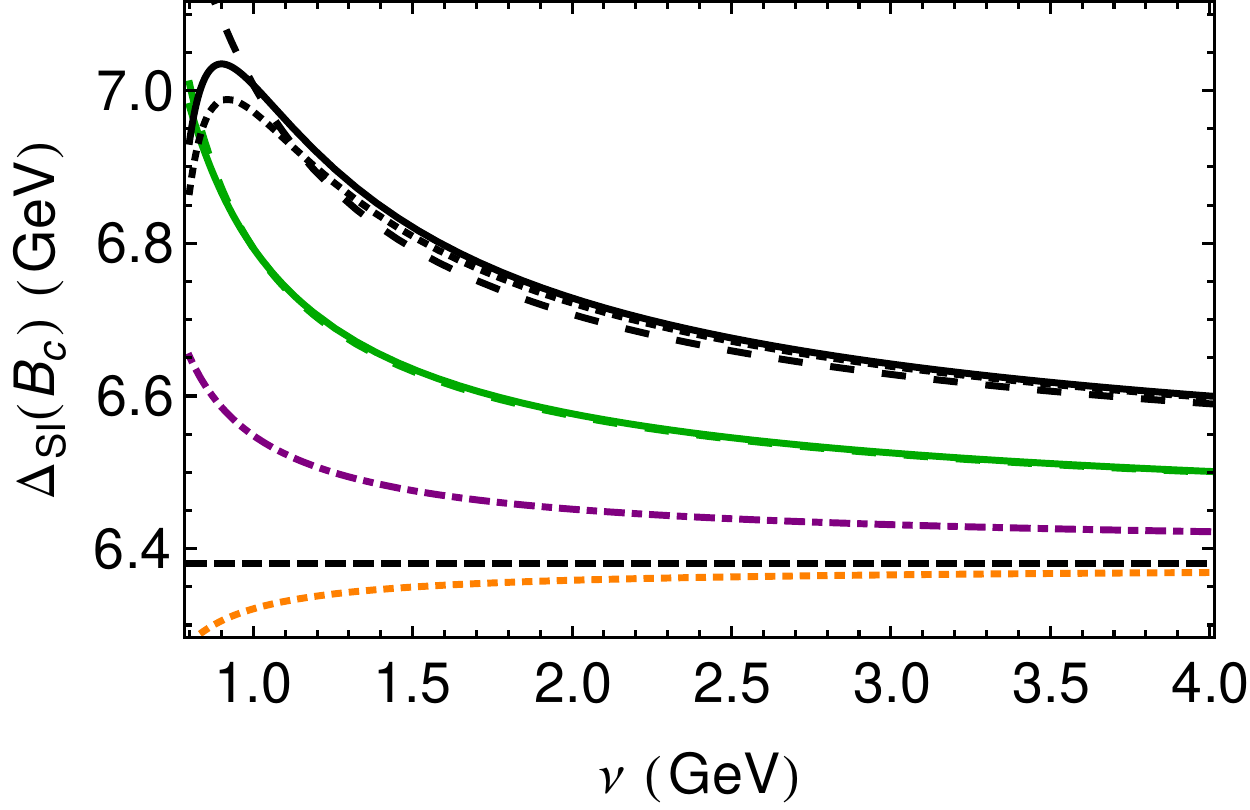}\\
	\includegraphics[width=0.49\textwidth]{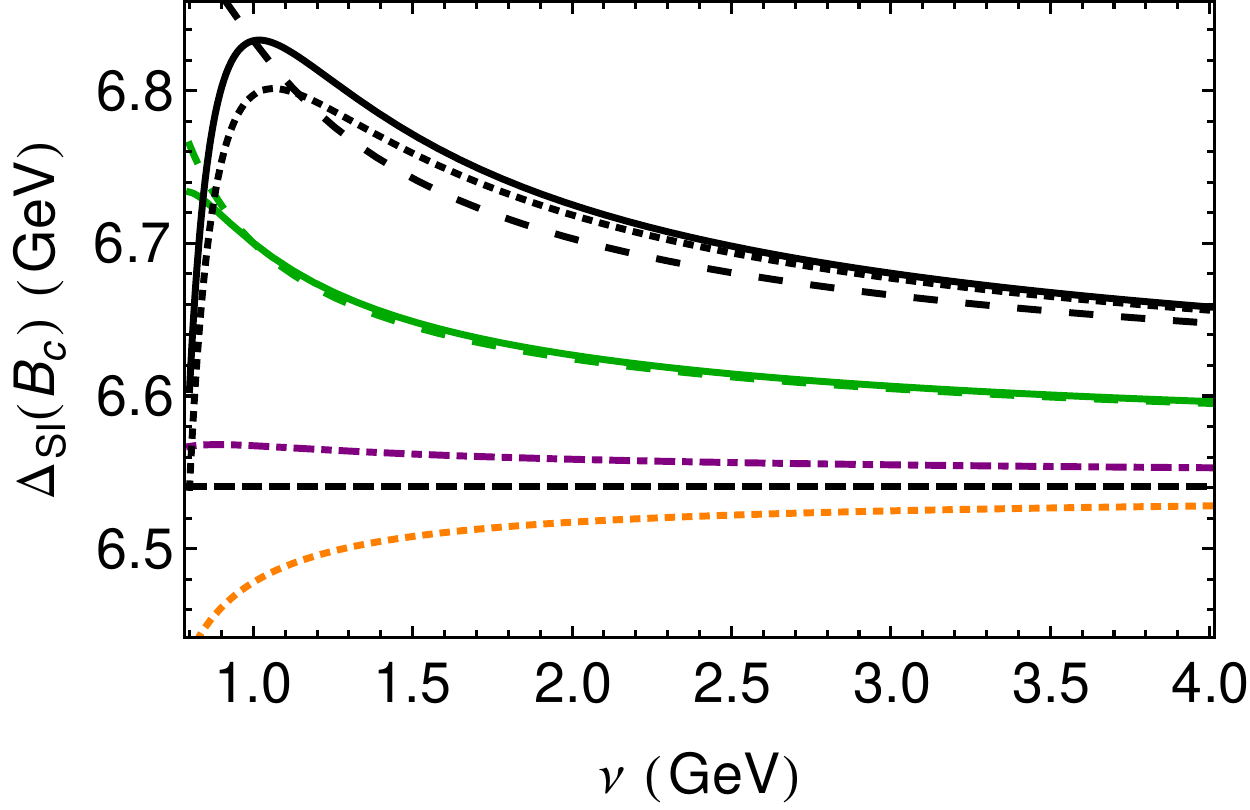}
	\includegraphics[width=0.49\textwidth]{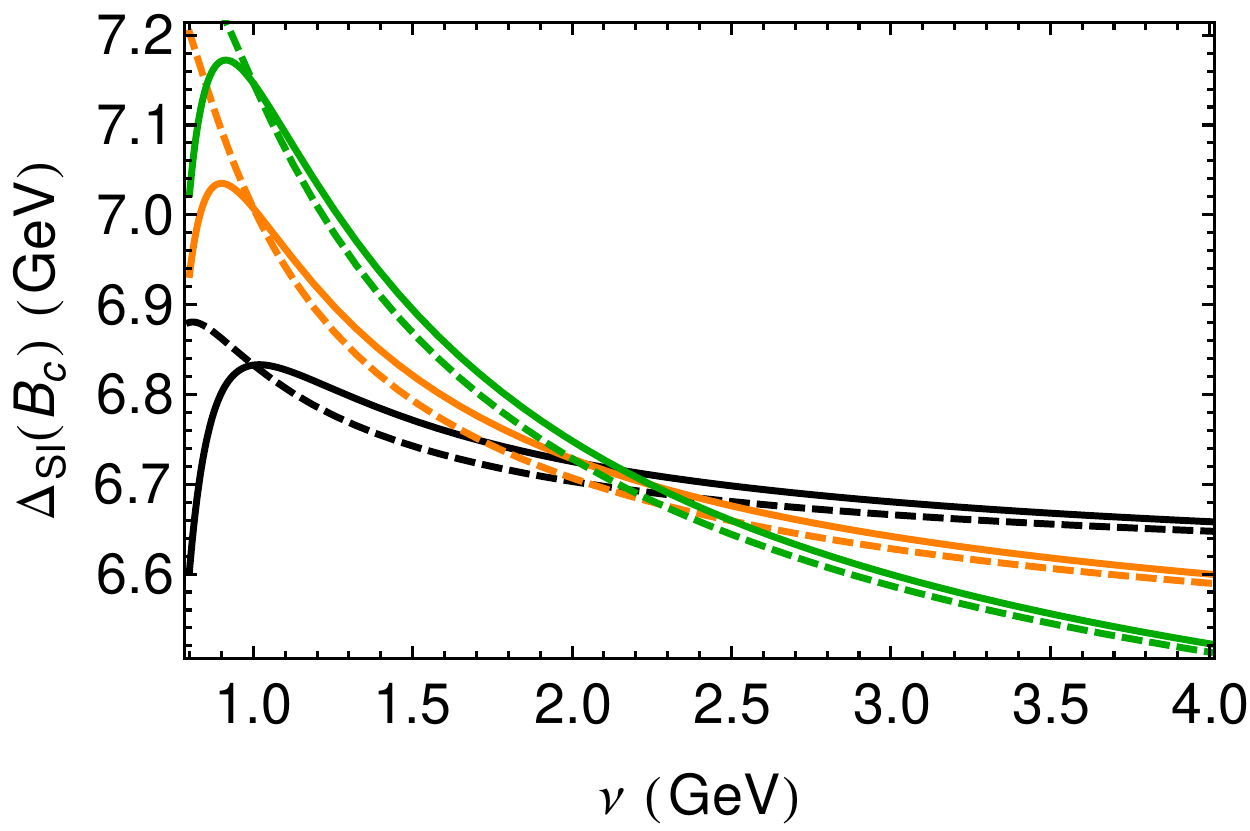}
\caption{Plots for $\Delta_{SI}$ in the RS' scheme with $\nuus=1$ GeV for $B_c$.  {\bf Upper left, upper right and lower left panels:} Plots for $\nu_f=2$, 1 and 0.7 GeV respectively. The black-dashed line is $m_{b,\RS'}+m_{c,\RS'}$. The orange-dotted, purple dot-dashed, green-dashed and black-dashed lines are $\Delta_{SI}$ evaluated at LO-N$^3$LO, respectively. The 
solid-green and solid-black lines are the NNLL and N$^3$LL result respectively, and the dotted-black line is the N$^3$LL result without $\delta E_{21}^{\US}$. 
{\bf Lower right panel:} Comparison of the $\nu_f=2$ GeV (green),  $\nu_f=1$ GeV (orange) and $\nu_f=0.7$ GeV (black) lines. For each case, the dashed line is the  N$^3$LO result and the solid line the N$^3$LL one.
\label{Fig:Bc}}   
\end{center}
\end{figure}
We first consider the following energy combination, which is free of spin-dependent effects:
\begin{align}
\label{MSIPwave}
 \Delta_{SI} \equiv \frac{1}{12} \left(5M_{\chi_{b2}}+3M_{\chi_{b1}}+M_{\chi_{b0}}\right)+\frac{1}{4 }M_{h_b}
 \,,
\end{align}
and similarly for charmonium and $B_c$. 

This quantity allows us to visualize the gross features of the spectrum of any P-wave state. We consider first bottomonium. In \fig{Fig:hb} we compare the strict weak-coupling prediction with experiment. We show both the fixed order and RGI expressions. The former can be found in \eq{EnN$^3$LO} and the latter in \eq{EnN$^3$LL}. We have explored the dependence of the result with $\nu_f$, $\nu$ and the order of truncation of the computation. We produce plots with $\nu_f=2$ GeV, $\nu_f=1$ GeV and $\nu_f=0.7$ GeV. For reference we take the $\nu_f=1$ GeV case. In this case, the fixed-order result approaches the experimental number as we increase the order of truncation of the computation (albeit the size of the consecutive terms is almost equal, i.e. the convergence is marginal). Indeed the N$^3$LO result agrees with experiment at $\nu \sim 1.2$ GeV and shows a relatively mild scale dependence. The resummation of logarithms produces nontrivial results at NNLL and N$^3$LL. We observe that most of the effect of the RGI is due to the ultrasoft gluons. At NNLL the effect of resummation of logarithms is marginal. At N$^3$LL the effect is important. At this order there is relatively good agreement with experiment. At $\nu \sim 2.2$ GeV there is agreement with experiment and the scale variation is of order $\sim \pm 50$ MeV in the range $\nu=$1-4 GeV. In this respect, the resummation of logarithms (in particular ultrasoft logarithms) does not spoil the agreement with data, though it makes the shift between the NNLL and N$^3$LL bigger putting into question the convergence of the perturbative expansion. Finally, the biggest point of concern is the applicability of the weak-coupling computation at the ultrasoft scale. We roughly asses the importance of this effect by subtracting $\delta E_{us}$ to the N$^3$LL result. The effect is small (this happens both for the RGI and the fixed order computation). Overall, the uncertainties of the computation do not allow to see if the resummation of the large logarithms improves the result or not. We have also explored the dependence of the result on $\nu_f$. Choosing a larger value, $\nu_f=2$ GeV, does not change the qualitative picture. It makes it slightly more convergent but at the prize of making the corrections and scale dependence bigger (note though that $\nu_f=2$ GeV is an unnatural value for $\nu_f$, as the power counting demands $\nu_f <$ soft scale, which we do not expect to happen for 
$\nu_f=2$ GeV). Remarkably, for the smaller value $\nu_f=0.7$ GeV, the size of the higher order corrections is very small, except for the N$^3$LL result, where the incorporation of the large (ultrasoft) logarithms and of the ultrasoft correction brings the result quite close to experiment. In the last plot in \fig{Fig:hb}, we combine the N$^3$LL and 
N$^3$LO results for different values of $\nu_f$. We observe that smaller values of $\nu_f$ produce smaller $\nu$ scale dependence (we remark again the warning of choosing a too high value of $\nu_f$). They are all consistent among them and with experiment. Indeed the three N$^3$LL lines cross at 
\begin{align}
\Delta_{SI} \sim 9.885 \; {\rm GeV}
\,,
\end{align} 
quite close the experimental value $\Delta_{SI} \sim 9.900 \text{ GeV}\sim M_{h_b}$. The 
 three N$^3$LO lines cross at $\Delta_{SI} \sim 9.850$ GeV, also quite close the experimental value. As a final remark, in all cases, at $\nu\lesssim 1$ GeV, there is a strong scale dependence. 

\begin{figure}[!htb]
	\begin{center}      
	\includegraphics[width=0.49\textwidth]{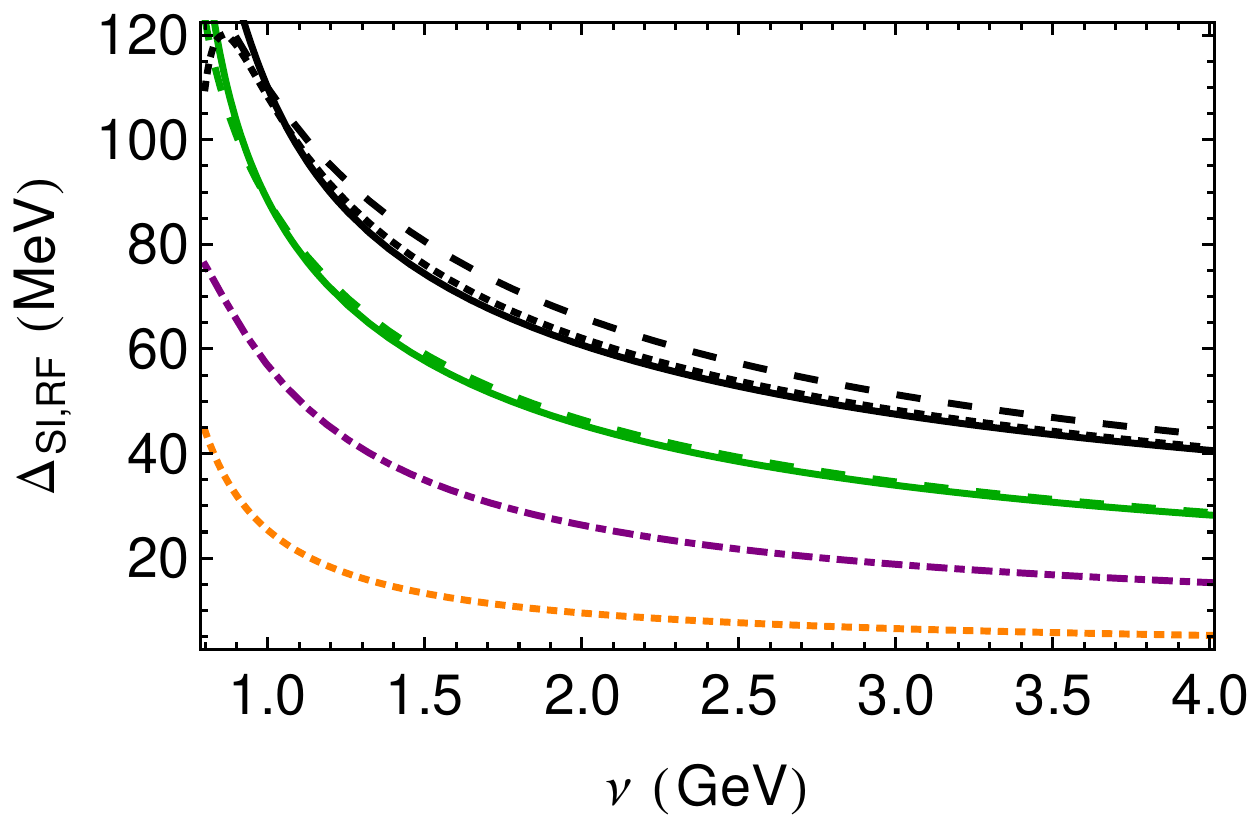}	
   	\includegraphics[width=0.49\textwidth]{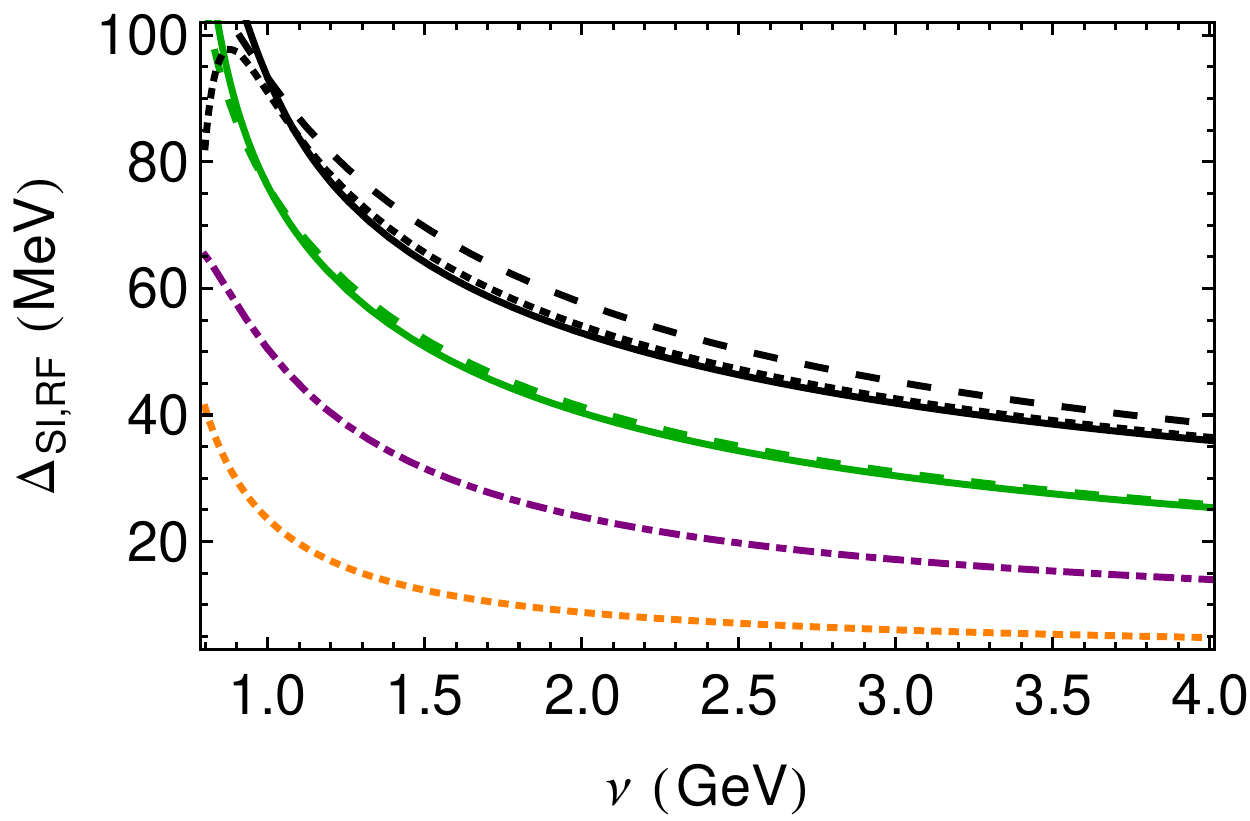}\\
	\includegraphics[width=0.49\textwidth]{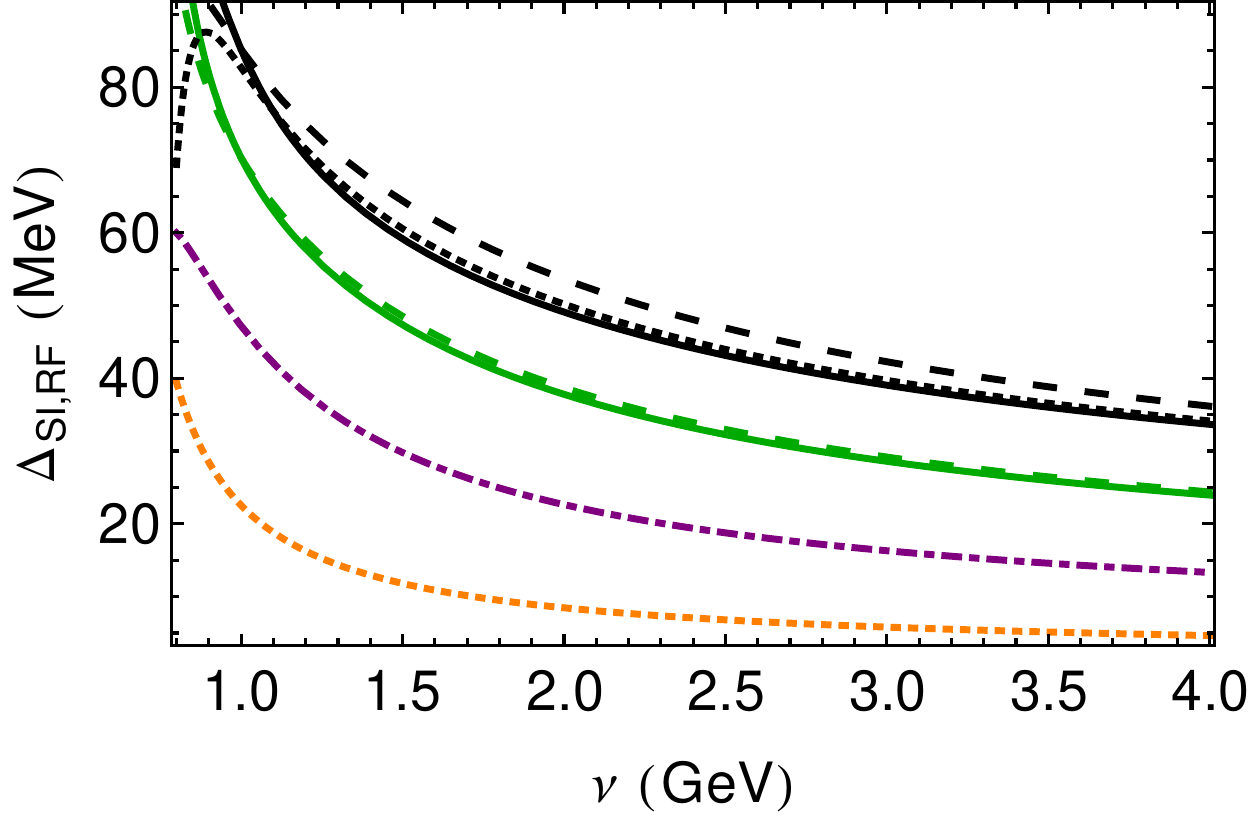}
	\includegraphics[width=0.49\textwidth]{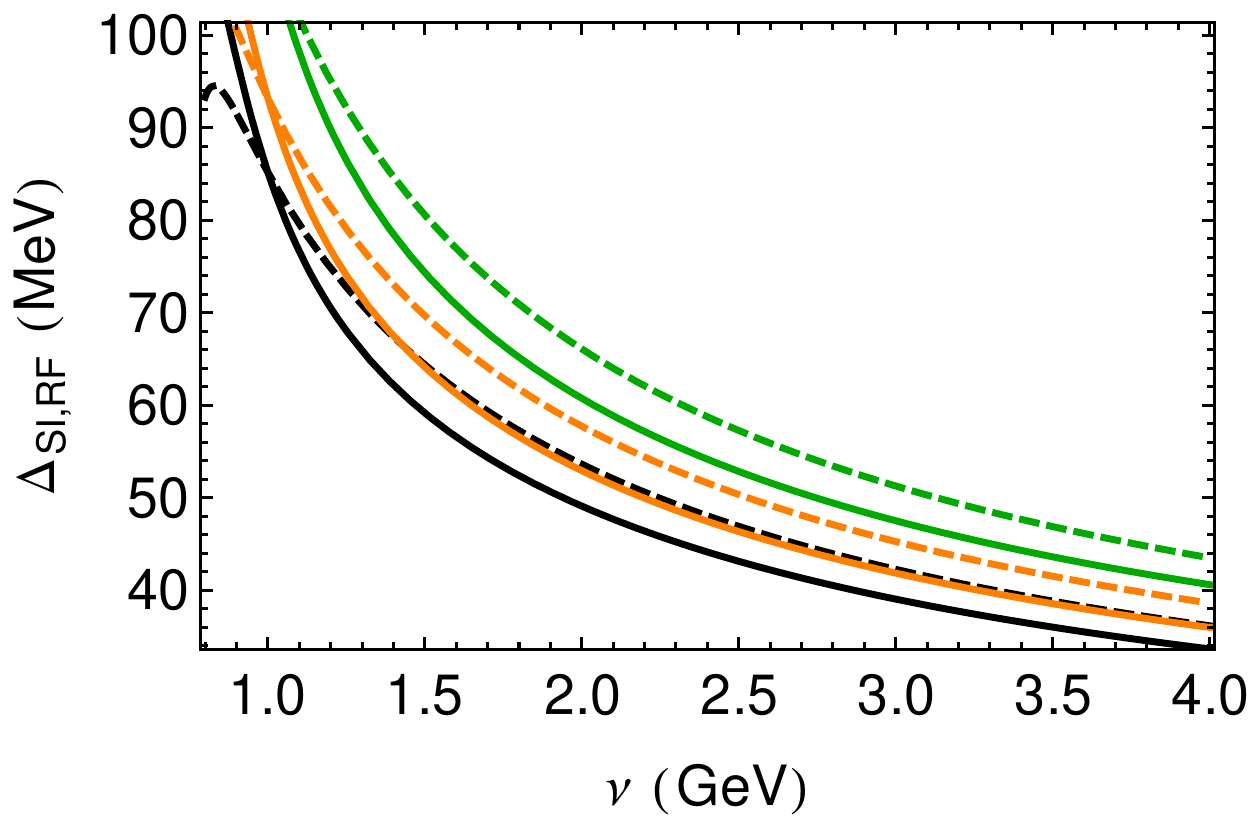}
	%
\caption{Plots for $\Delta_{SI}^{(B_c)}-\Delta_{SI}^{(bb)}/2-\Delta_{SI}^{(cc)}/2$ in the RS' scheme with $\nuus=1$ GeV.  {\bf Upper left, upper right and bottom left panels:} Plots for $\nu_f=2$, 1 and 0.7 GeV respectively. The orange-dotted, purple dot-dashed, green-dashed and black-dashed lines are the LO-N$^3$LO results respectively. The solid-green and solid-black lines are the NNLL and N$^3$LL result respectively, and the dotted-black line is the N$^3$LL result without $\delta E_{21}^{\US}$. 
{\bf Bottom right panel:} Comparison of the $\nu_f=2$ GeV (green),  $\nu_f=1$ GeV (orange) and $\nu_f=0.7$ GeV (black) lines. For each case, the dashed line is the N$^3$LO result and the solid line the N$^3$LL one. 
\label{Fig:SPwavebb}}   
	\end{center}
\end{figure}

For completeness, we also show the results for charmonium and $B_c$ in \figs{Fig:hc}{Fig:Bc} (and for the renormalon-free combination $\Delta_{SI}^{(B_c)}-\Delta_{SI}^{(bb)}/2-\Delta_{SI}^{(cc)}/2$ in \fig{Fig:SPwavebb}) but in those cases the errors are so large that we do not aim to any serious phenomenological analysis. At most we can give an estimate of 
\begin{align}
\Delta_{SI}^{(B_c)} \sim 6.75 \text{ GeV}
\,.
\end{align}
This number is obtained from the approximate crossing of the three different curves in the lower-right panel in \fig{Fig:Bc}. For the case of bottomonium and charmonium this gave a reasonable estimate. Such value and the experimental masses of bottomonium and charmonium yields $\Delta_{SI}^{(B_c)}-\Delta_{SI}^{(bb)}/2-\Delta_{SI}^{(cc)}/2 \sim 60$ MeV.

\subsection{Fine splitting}

The $E(1^3P_j)-E(1^3P_{j'})$ energy differences are interesting objects for study, they are free of renormalon effects (we take $\nu_f=1$ GeV for reference but the result is quite insensitive to this) and 
also of ultrasoft effects. In this paper, we have obtained, for the first time, theoretical expressions with relative NLL precision (i.e. we have two terms of the weak-coupling expansion and we also know the RGI expression for them). We would like to see how well our theoretical predictions compare with experiment. 

\begin{figure}[!htb]
	\begin{center}      
	\includegraphics[width=0.49\textwidth]{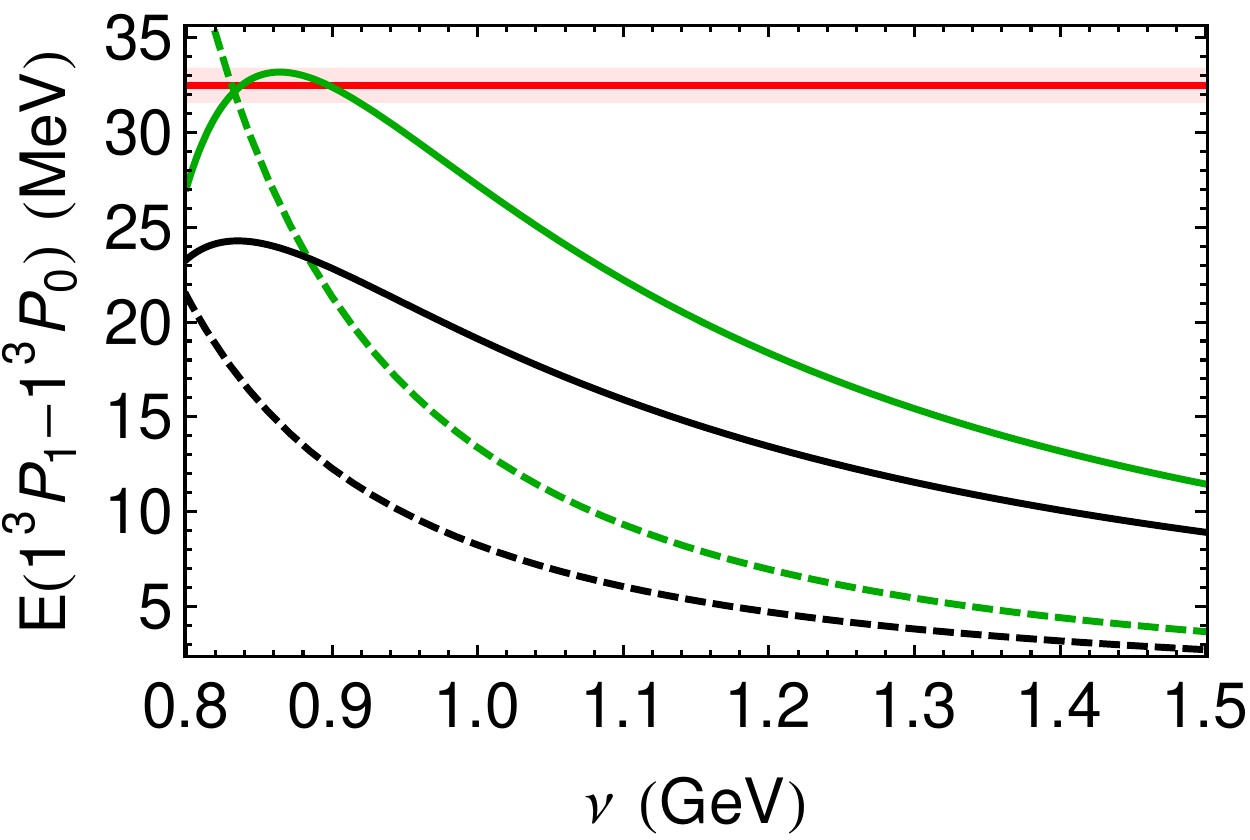}
	\includegraphics[width=0.49\textwidth]{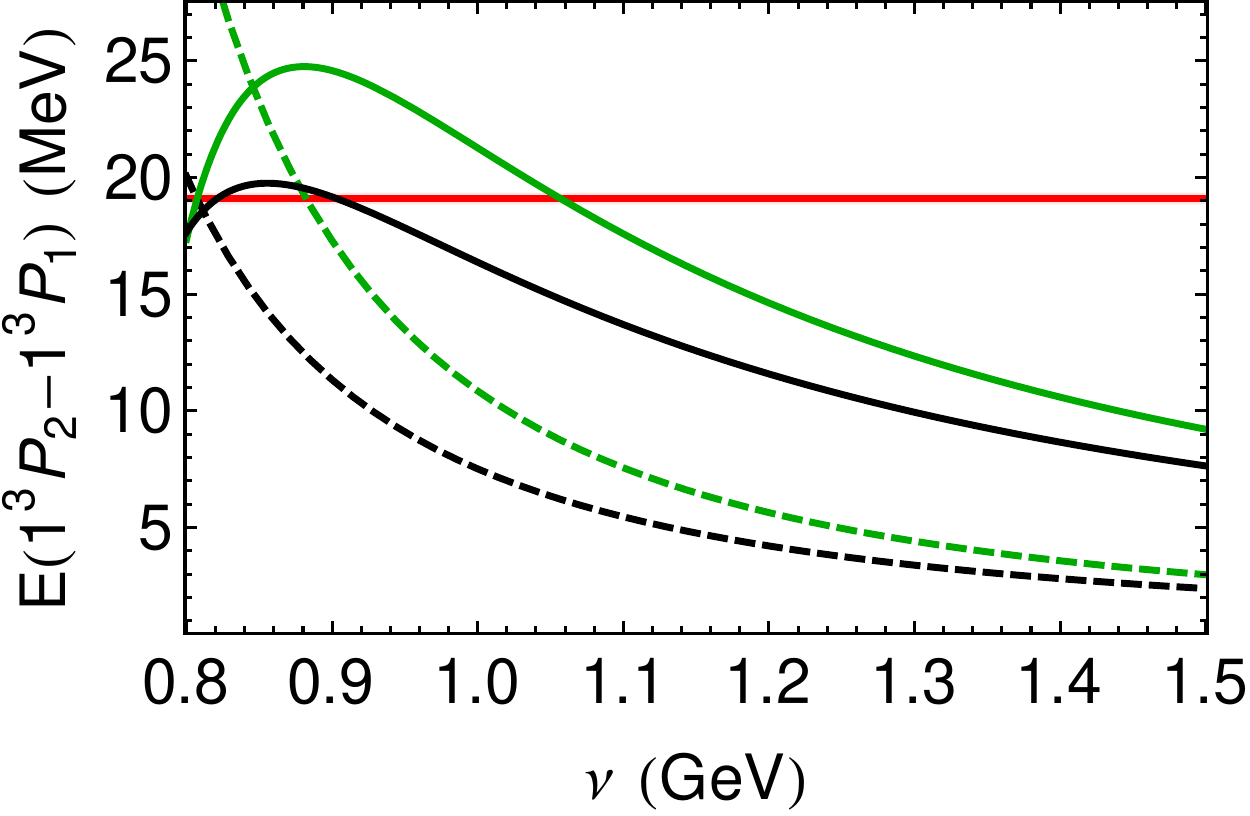}
	%
\caption{Plots for the P-wave fine splittings in bottomonium in the RS' scheme with $\nu_f=1$ GeV and $\nuh=m_{b,{\rm RS}'}$. The red band is the experimental value. The dashed-green, dashed-black, solid-green, and the solid-black lines are the NNLO, NNLL, N$^3$LO and N$^3$LL results respectively. {\bf Left panel:} Plot of $M_{\chi_{b_1}}-M_{\chi_{b_0}}$. {\bf Right panel:} Plot of $M_{\chi_{b_2}}-M_{\chi_{b_1}}$.
\label{Fig:PwaveFine}}   
	\end{center}
\end{figure}
\begin{figure}[!htb]
	\begin{center}      
	\includegraphics[width=0.49\textwidth]{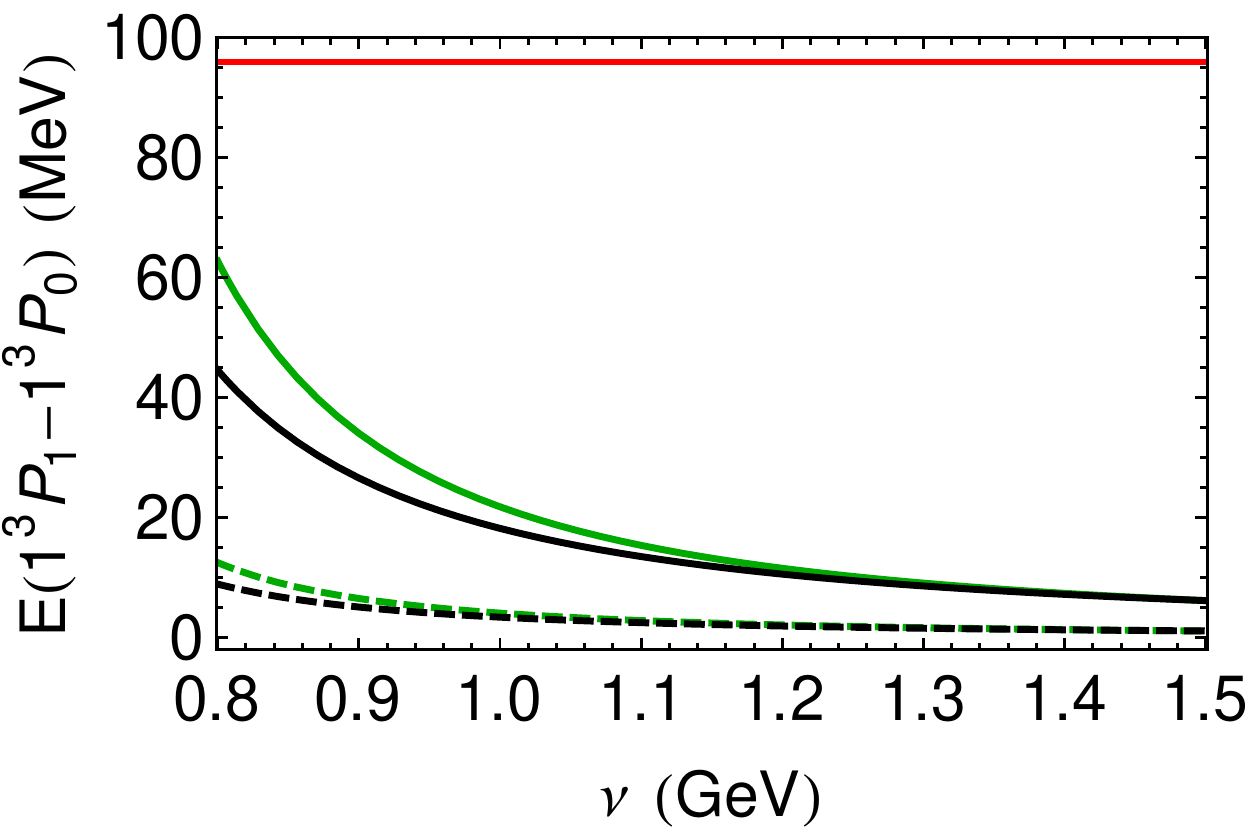}
	\includegraphics[width=0.49\textwidth]{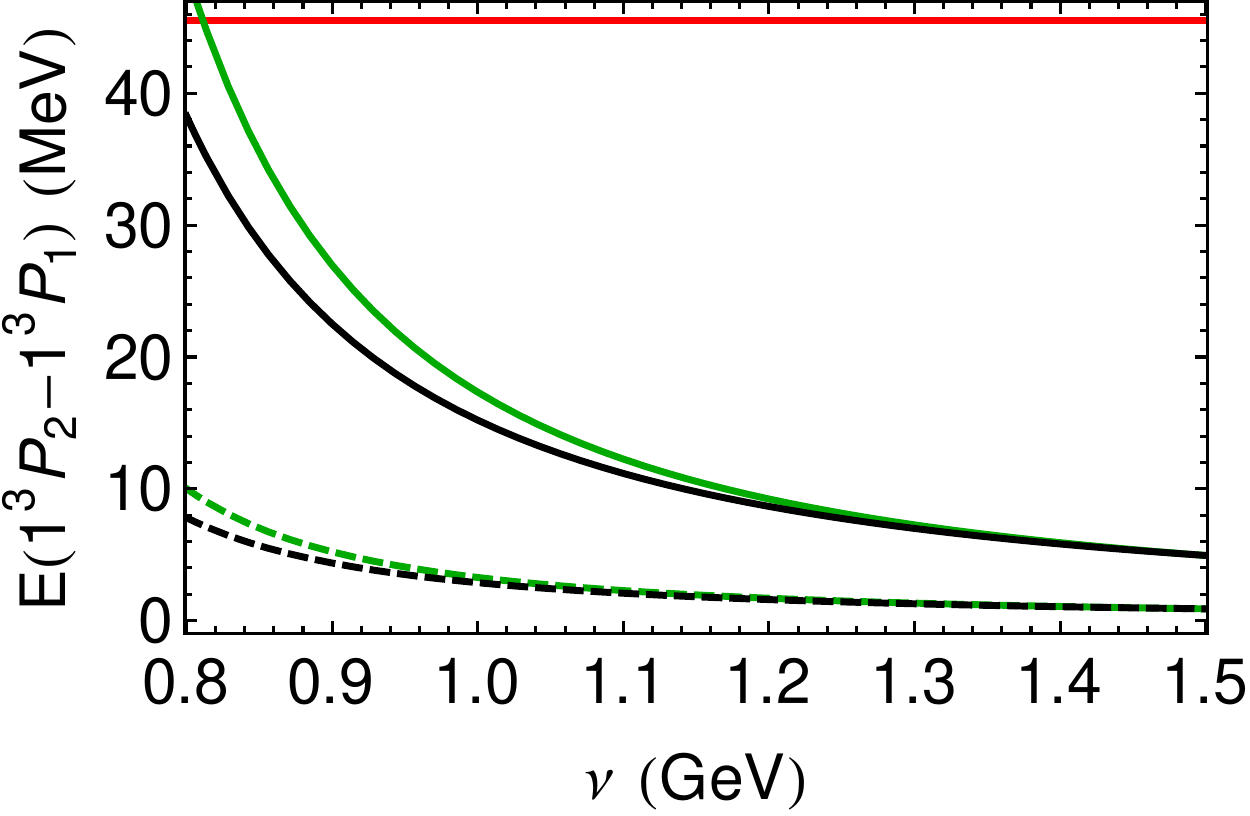}
	%
\caption{Plots for the P-wave fine splittings in charmonium in the RS' scheme with $\nu_f=1$ GeV and $\nuh=m_{c,{\rm RS}'}$. The red band is the experimental value.  The dashed-green, dashed-black, solid-green, and the solid-black lines are the NNLO, NNLL, N$^3$LO and N$^3$LL results respectively. {\bf Left panel:} Plot of $M_{\chi_{c_1}}-M_{\chi_{c_0}}$. {\bf Right panel:} Plot of $M_{\chi_{c_2}}-M_{\chi_{c_1}}$.
\label{Fig:PwaveFinecharm}}   
	\end{center}
\end{figure}
\begin{figure}[!htb]
	\begin{center}      
	\includegraphics[width=0.49\textwidth]{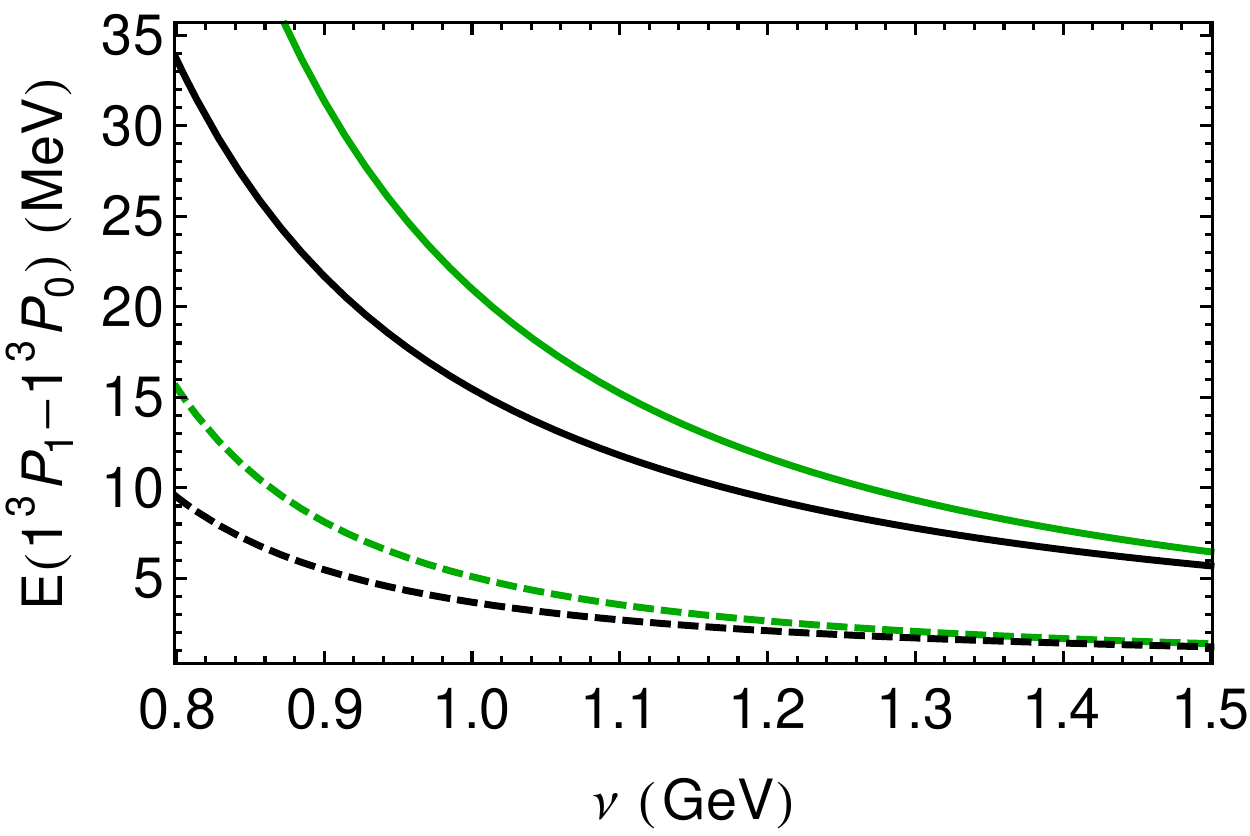}
	\includegraphics[width=0.49\textwidth]{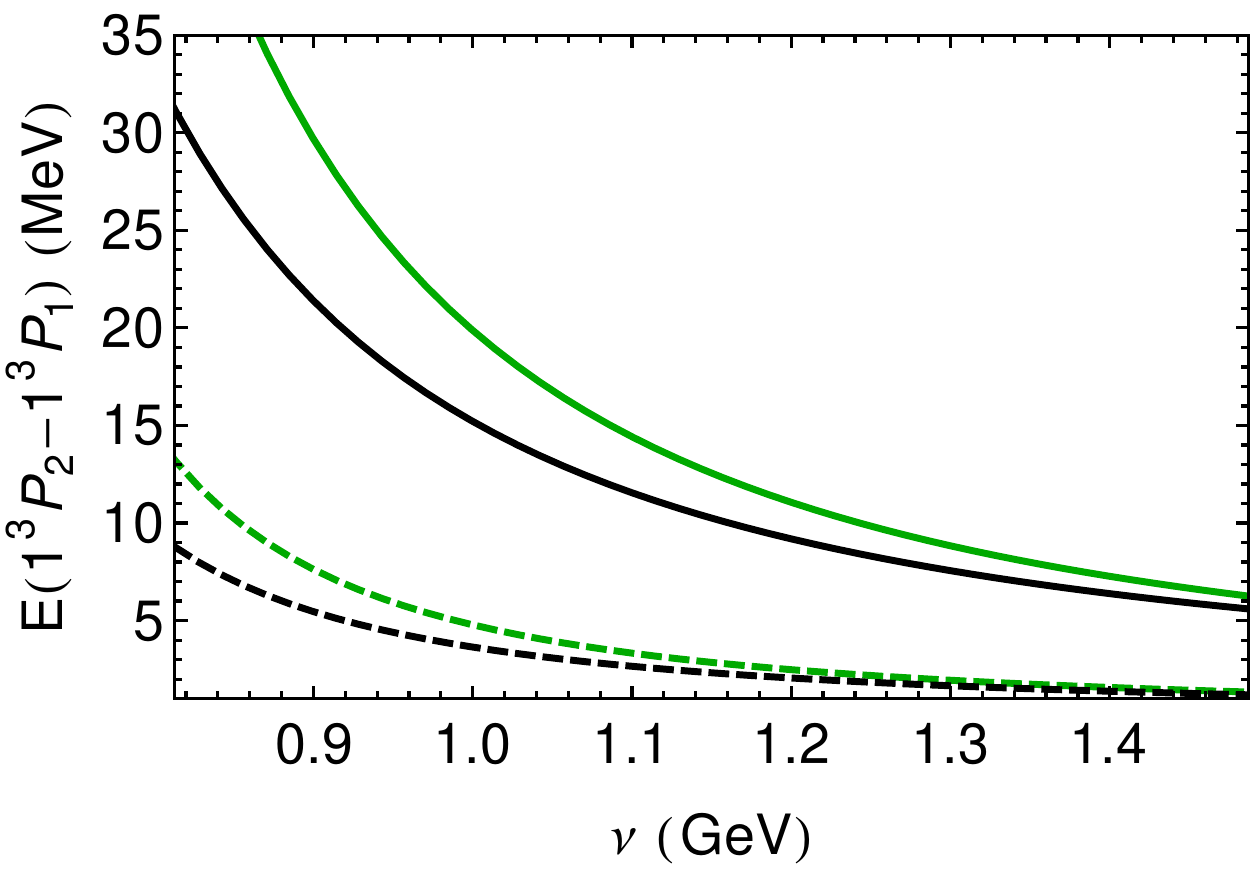}
	%
\caption{Plots for the P-wave fine splitting in $B_c$ in the RS' scheme with $\nu_f=1$ GeV and $\nuh=2m_{b,{\rm RS}'}m_{c,{\rm RS}'}/(m_{b,{\rm RS}'}+m_{c,{\rm RS}'})$. The dashed-green, dashed-black, solid-green, and the solid-black lines are the NNLO, NNLL, N$^3$LO and N$^3$LL results respectively. {\bf Left panel:} Plot of $M_{B_{c}(1^3P_1)}-M_{B_{c}(1^3P_0)}$. {\bf Right panel:} Plot of $M_{B_{c}(1^3P_2)}-M_{B_{c}(1^3P_1)}$.
\label{Fig:PwaveFineBc}}   
	\end{center}
\end{figure}
We first start with bottomonium, which, in principle, is the system where the weak-coupling approach should work better. We plot the strict weak-coupling predictions in \fig{Fig:PwaveFine}. We expect the large logarithms to be resummed around $\nu \sim$ soft scale, of order 1 GeV. Indeed we observe a much better agreement at those scales, and results compatible with experiment within the expected uncertainties. We also observe that the resummation of (hard) logarithms produces a sizable effect but of the order of uncertainties. At those scales we also observe convergence of the expansion (the N$^3$LL correction is smaller than the NNLL correction). If we repeat the analysis for charmonium the numbers we get are quite low when compared with experiment. We show them in \fig{Fig:PwaveFinecharm}. In principle, this confirms the expectation that charmonium P-wave states can not be described by a weak-coupling analysis. For completeness, we also show the prediction of the strict weak-coupling analysis for the fine splitting of the P-wave $B_c$ states in \fig{Fig:PwaveFineBc}. 

 \begin{figure}[!htb]
	\begin{center}      
\includegraphics[width=0.49\textwidth]{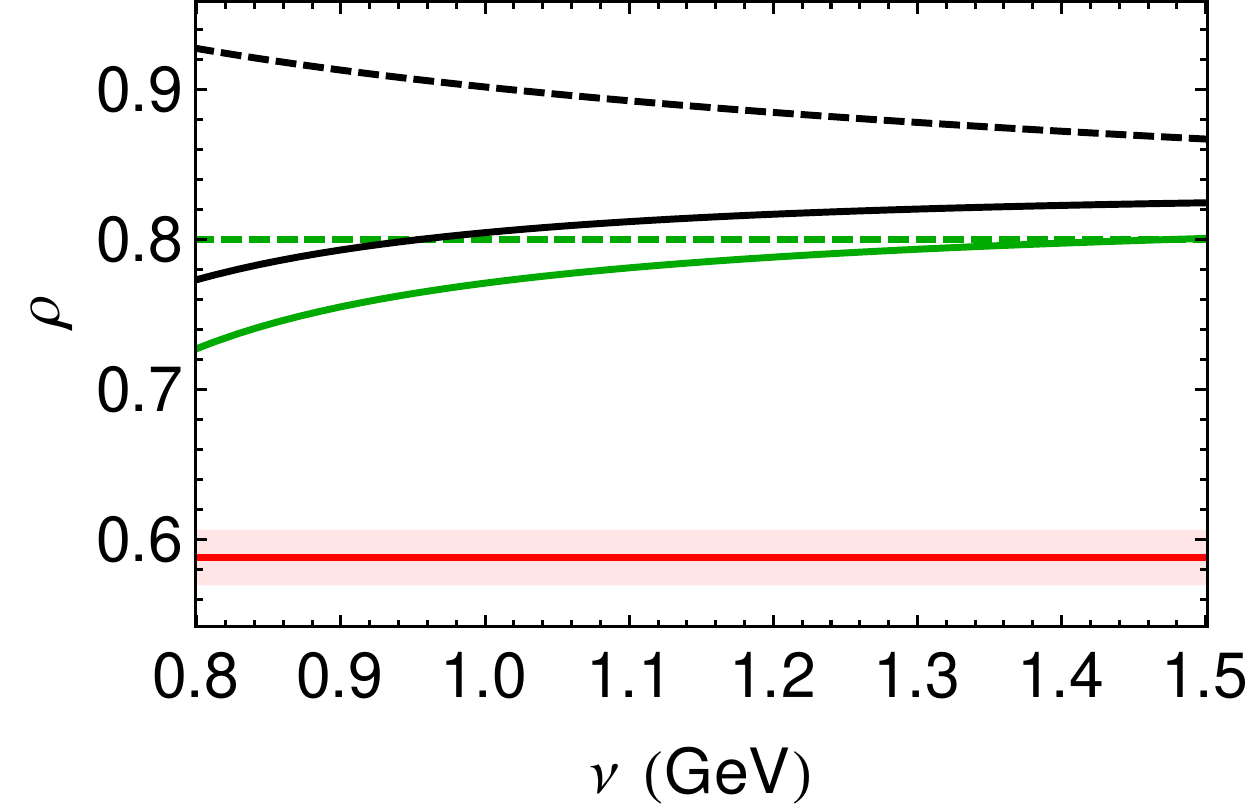}
	\includegraphics[width=0.49\textwidth]{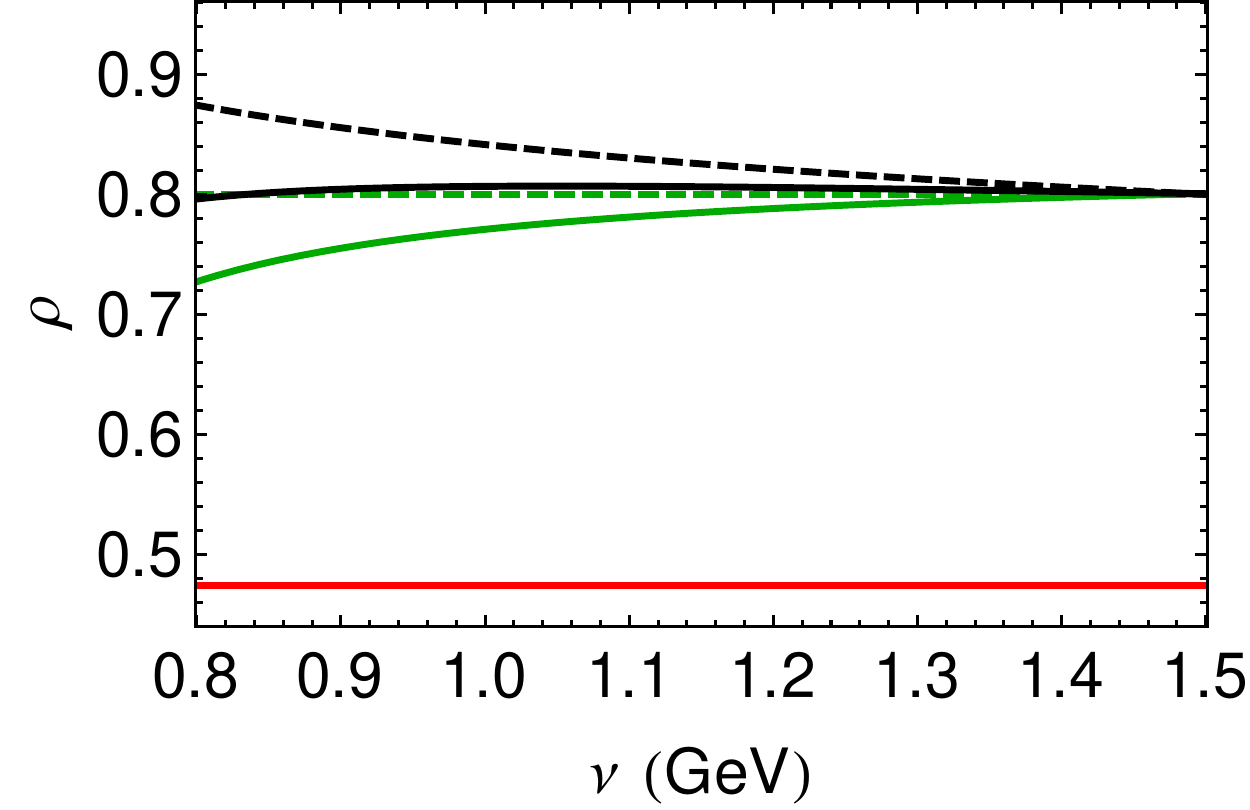}\\
\includegraphics[width=0.49\textwidth]{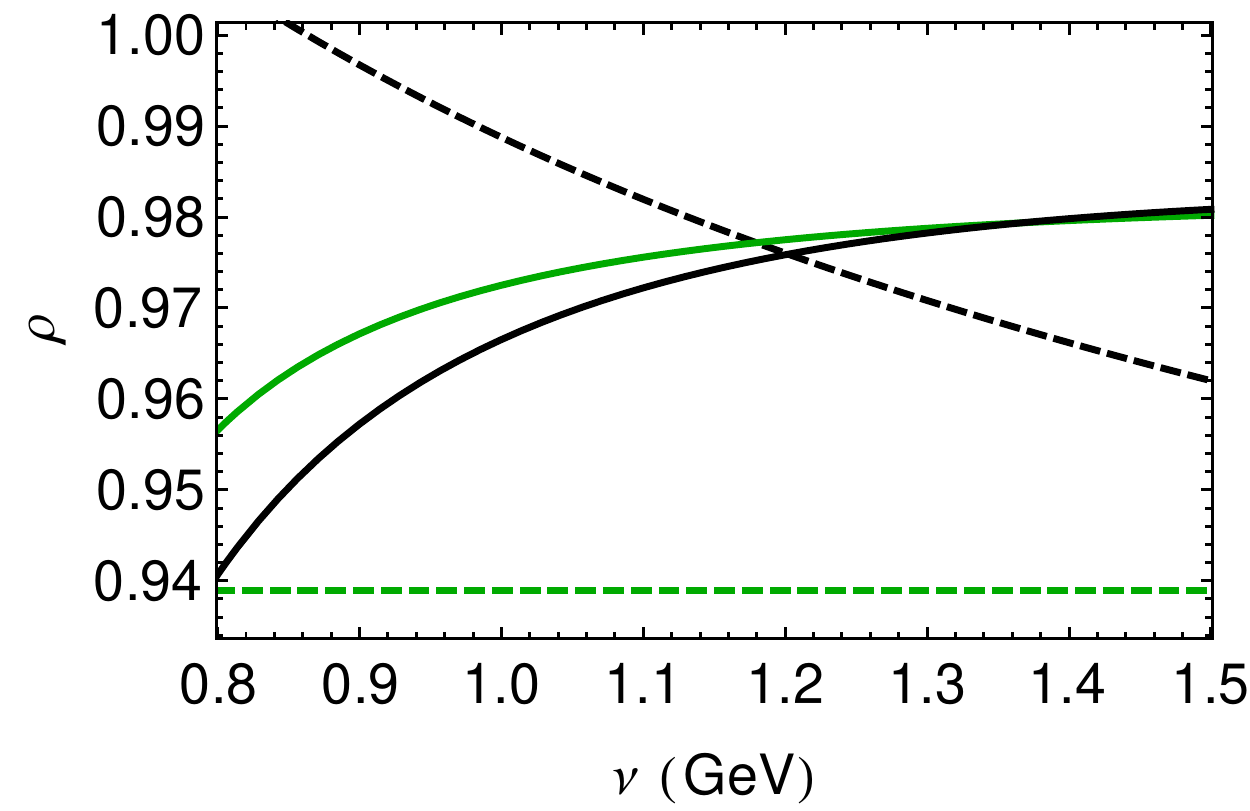}
	%
\caption{Plots of $\rho$ expanded in powers of $\als$ in the RS' scheme with $\nu_f=1$ GeV. The red line is the experimental value. We start the counting at the leading nonvanishing order. Then, the dashed-green, dashed-black, solid-green, and solid-black lines are the LO, LL, NLO and NLL results respectively. {\bf Upper left panel:} Plot for bottomonium with $\nuh=m_{b,{\rm RS}'}$. {\bf Upper right panel:} Plot for charmonium with $\nuh=m_{c,{\rm RS}'}$. {\bf Bottom panel:} Plot for $B_c$ with $\nuh=2m_{b,{\rm RS}'}m_{c,{\rm RS}'}/(m_{b,{\rm RS}'}+m_{c,{\rm RS}'})$.
\label{Fig:PwaveFineratiocharm}}   
\end{center}
\end{figure}
We also study the ratio
\begin{align}
\rho=\frac{E(1^3P_2)-E(1^3P_1)}{E(1^3P_1)-E(1^3P_0)}=\frac{4}{5}+\frac{6 (m_1-m_2)^2}{5 \left(m_1^2+10 m_1m_2+m_2^2\right)}+{\cal O}(\als)\label{def:rho}
\,.
\end{align}
One can speculate that this observable is cleaner in the sense that the NR matrix element cancels in the ratio at the leading nonvanishing order. Nevertheless, this observable is also sensitive to the wave function at the next order. 
We show the result in \fig{Fig:PwaveFineratiocharm}. There is a difference with experiment of order 25\%. The resummation of hard logarithms does not improve the agreement with data (it actually makes it slightly worse, specially for the LL result). The difference between theory and experiment in the case of charmonium is larger, since the theoretical prediction is more or less equal as for bottomonium but the experimental value of $\rho$ is significantly different for bottomonium and charmonium. For completeness, we also show the prediction of the strict weak-coupling analysis of $\rho$ for the $B_c$ states. Note that, in this case, the leading order theoretical prediction is different to the equal mass case (c.f. \eq{def:rho}). This provides an extra motivation to measure this ratio.

\subsection{Hyperfine splitting}
\begin{figure}[!htb]
	\begin{center}      
	\includegraphics[width=0.49\textwidth]{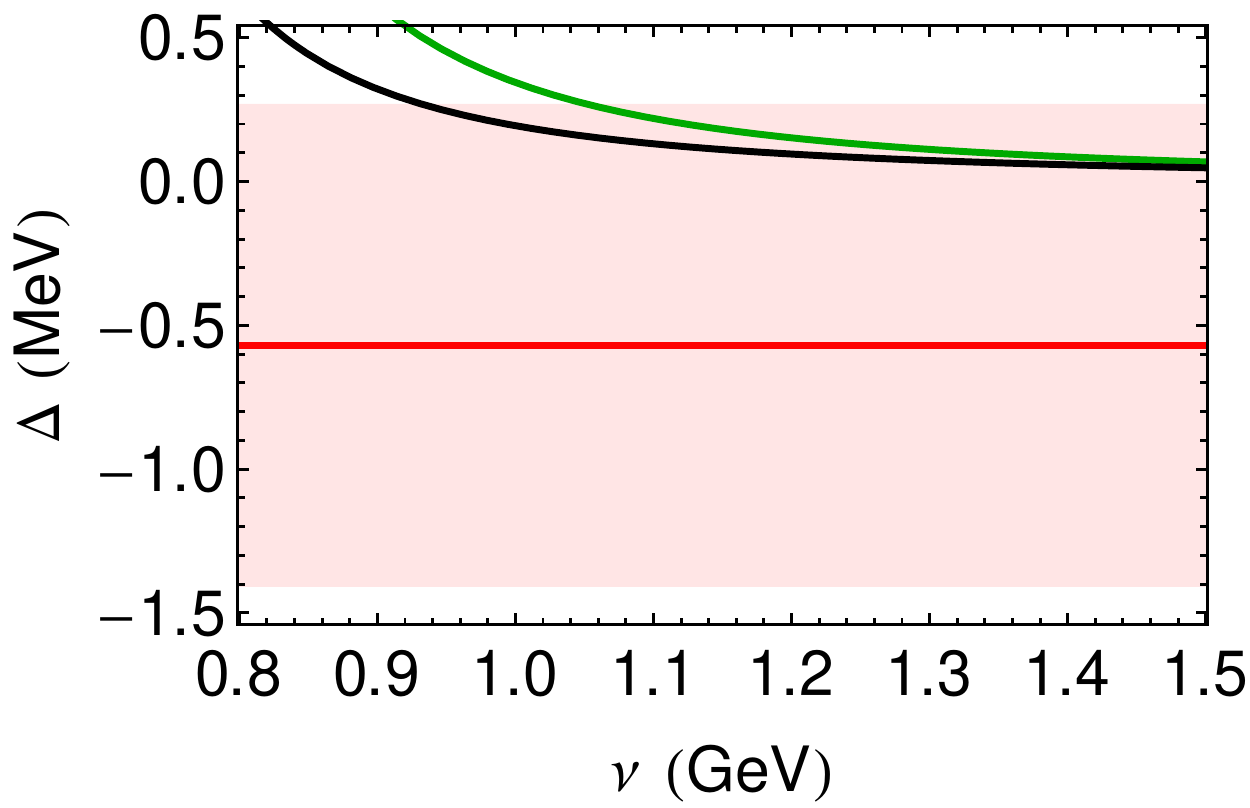}
	\includegraphics[width=0.49\textwidth]{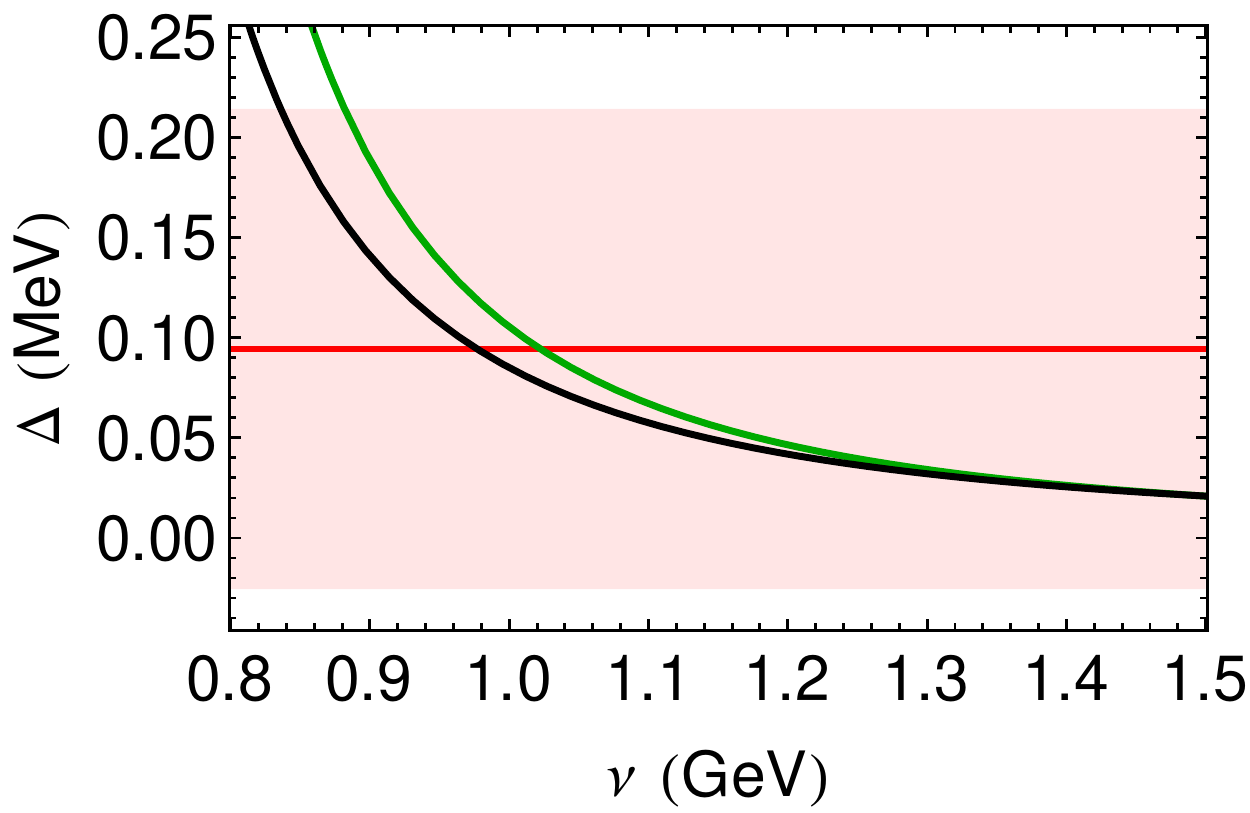}\\
	\includegraphics[width=0.49\textwidth]{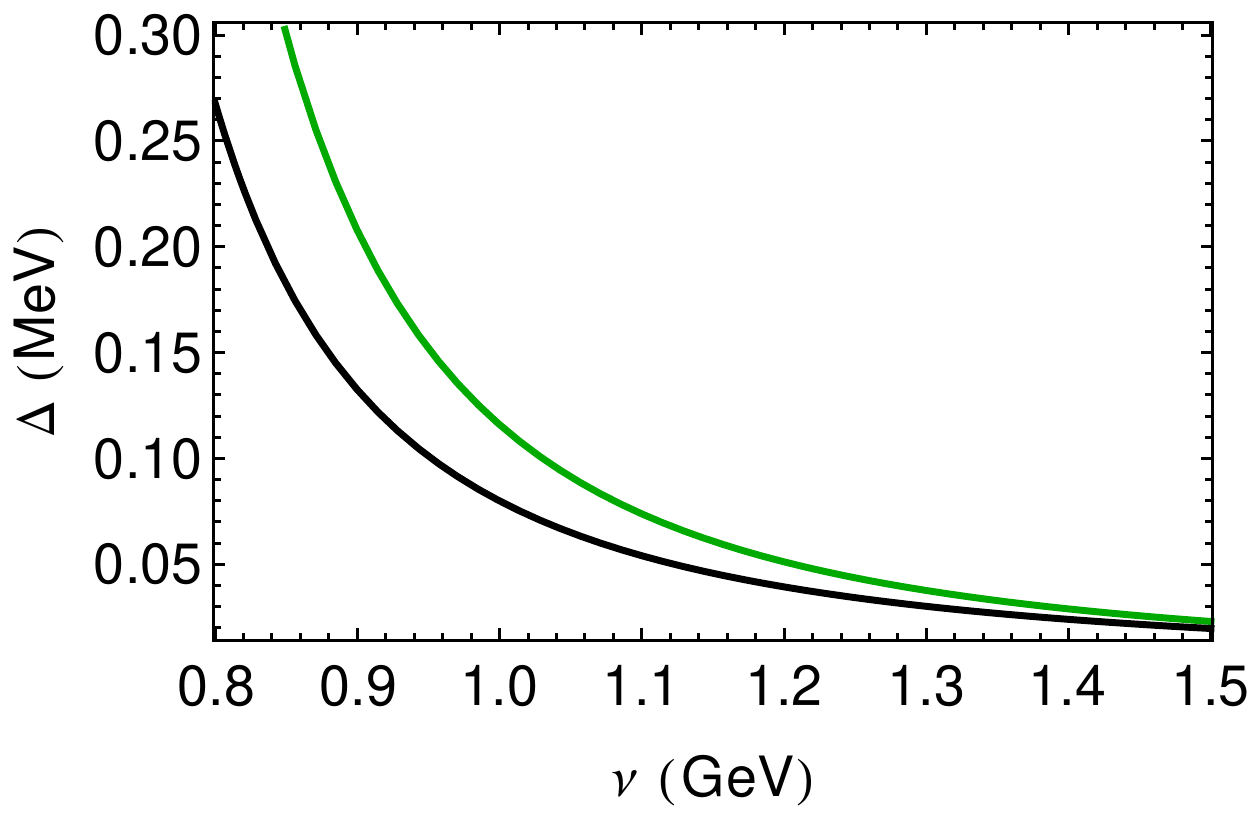}
\caption{Plots for the P-wave hyperfine splitting $\Delta$ in the RS' scheme with $\nu_f=1$ GeV. The red line is the experimental value, the solid-green line the N$^3$LO result and the 
solid-black line is the N$^3$LL result. {\bf Upper left panel:} Plot for bottomonium with $\nuh=m_{b,{\rm RS}'}$. {\bf Upper right panel:} Plot for charmonium with $\nuh=m_{c,{\rm RS}'}$. {\bf Lower panel:} Plot for $B_c$ with $\nuh=2m_{b,{\rm RS}'}m_{c,{\rm RS}'}/(m_{b,{\rm RS}'}+m_{c,{\rm RS}'})$.
\label{Fig:PwaveHF}}   
\end{center}
\end{figure}
Finally, we consider the hyperfine splitting of the P-wave states. We show our results in \fig{Fig:PwaveHF}. The strict weak-coupling prediction of the hyperfine splitting is perfectly compatible with experiment. The resummation of (hard) logarithms is a tiny effect and does not affect this conclusion. Surprisingly enough, this is also true for charmonium (then we conjecture that the prediction we give for the P-wave $B_c$, compatible with zero, is also robust). This could be accidental. The key issue for the agreement is that the expectation value of the relativistic potential is small. 
We ellaborate on this issue in Sec. \ref{Sec:PhenImpr}.  

\section{Alternative counting approach}

In the previous section we have confronted the strict weak-coupling theoretical predictions with the experimental values of the masses of the $n=2$, $l=1$ excitations for bottomonium, charmonium and $B_c$. For bottomonium, the convergence was somewhat marginal. On the other hand the predictions were consistent with experiment (for the $\rho$ ratio the situation was somewhat worse but still consistent with the expected size of higher order relativistic corrections). For charmonium and $B_c$ the situation was significantly worse. Only for the hyperfine case there was agreement with experiment.

We now study a computational scheme that reorganizes the perturbative expansion such that it performs a selective sum of higher order corrections (such scheme was already applied in  \cite{Pineda:2013lta,Kiyo:2010jm,Peset:2018ria}). We want to test if such scheme could improve/accelerate the convergence. In this method we incorporate the static potential exactly (to a given order) in the leading order Hamiltonian (the explicit $\nu$ dependence of the static potential appears at N$^3$LO and partially cancels with the explicit $\nu$ dependence of \eq{EnlUS}, the ultrasoft correction): 
\begin{align}
\label{eq:Schroedinger}
\left[\frac{{\bf p}^2}{2m_r}+V_{N,\RS'}^{(0)}(r;\nu)
\right]\phi^{(0)}_{nl}({\bf r})=E_{nl}^{(0)}\phi^{(0)}_{nl}({\bf r})
\,,
\end{align}
where the static potential will be approximated by a polynomial of 
order $N+1$, 
\begin{align}
\label{VRS}
V^{(0)}_{N,\RS'}(r;\nu)=
\,\left\{
\begin{array}{ll}
&
\displaystyle{
(V^{(0)}_{N}+2\delta m^{(N)}_{\RS'})|_{\nu=\nu}\equiv
 \sum_{n=0}^{N}V_{\RS',n}\als^{n+1}(\nu)
\qquad {\rm if} \quad r>\nu_r^{-1} }
\\
&
\displaystyle{
(V^{(0)}_{N}+2\delta m^{(N)}_{\RS'})|_{\nu=1/r}\equiv
 \sum_{n=0}^{N}V_{\RS',n}\als^{n+1}(1/r)
\qquad {\rm if} \quad r<\nu_r^{-1}. }
\end{array} \right.
\end{align}
$V^{(0)}_{N}$ is the static potential defined in \eq{VSDfo}. We implement the renormalon cancellation working in 
 the RS' scheme. Expressions for $\delta m_{\RS'}$ can be found in \rcite{Pineda:2013lta}. In principle we would like to take $N$ as large as possible (though we also want to explore the dependence on $N$). 
In practice we take the static potential at most up to N=3, i.e. up to
${\cal O}(\als^4)$. This is the order to which the coefficients $V_{\RS',n}$ are completely known.

Taking different values for $\nu_r$ and $\nu_f$ in \eq{VRS} we obtain the most relevant limits:

{\bf (a)}.
 The case $\nu_r=\infty$, $\nu_f=0$ is nothing but the on-shell static potential at fixed order, i.e. \eq{VSDfo}. 
 Note that the $N=0$ case reduces to a standard computation with a Coulomb potential, for which we can compare with analytic results for the matrix elements. We use this fact to check our numerical solutions of the Schr\"odinger equation. 
 
 {\bf (b)}.
The case $\nu_r=\infty$ (with finite non-zero $\nu_f$) is nothing but adding an $r$-independent constant to the static potential. 

{\bf (c)}. The case $\nu_r=$ finite (and, for consistency, $\nu_r \geq \nu_f$). We expect this case to improve over the previous results, as it incorporates the correct (logarithmically modulated) short distance behavior of the potential. This has to be done with care in order not to spoil the 
renormalon cancellation. For this purpose it is compulsory to keep a finite, non-vanishing, $\nu_f$,  otherwise the renormalon cancellation is not achieved order by order in $N$, as it was discussed in detail in 
\rcite{Pineda:2002se}. 

We have explored the effect of different values of $\nu_f$ in our analysis. Large values of $\nu_f$ imply a large infrared cutoff. In this way, our scheme becomes closer to an $\MS$-like scheme. Such schemes still achieve renormalon cancellation, yet they jeopardize the power counting, as the residual mass $\delta m_{\RS'}$ does not count as $mv^2$. As a consequence, consecutive terms of the perturbative series become bigger. Therefore, we prefer values of $\nu_f$ as low as possible, with the constraints that one should still obtain the renormalon cancellation, and that it is still possible to perform the expansion in powers of $\als$.

\begin{figure}[!htb]
	\begin{center}      
	\includegraphics[width=0.49\textwidth]{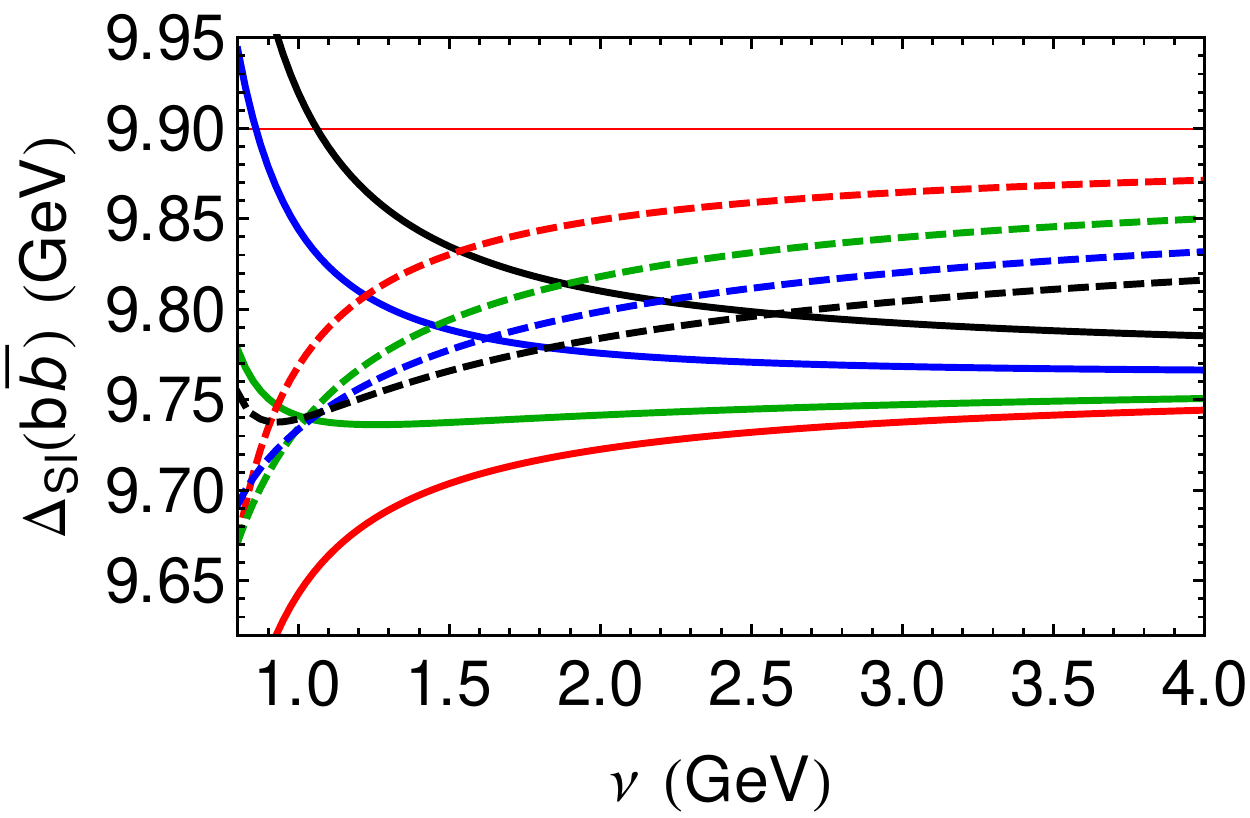}
	\includegraphics[width=0.49\textwidth]{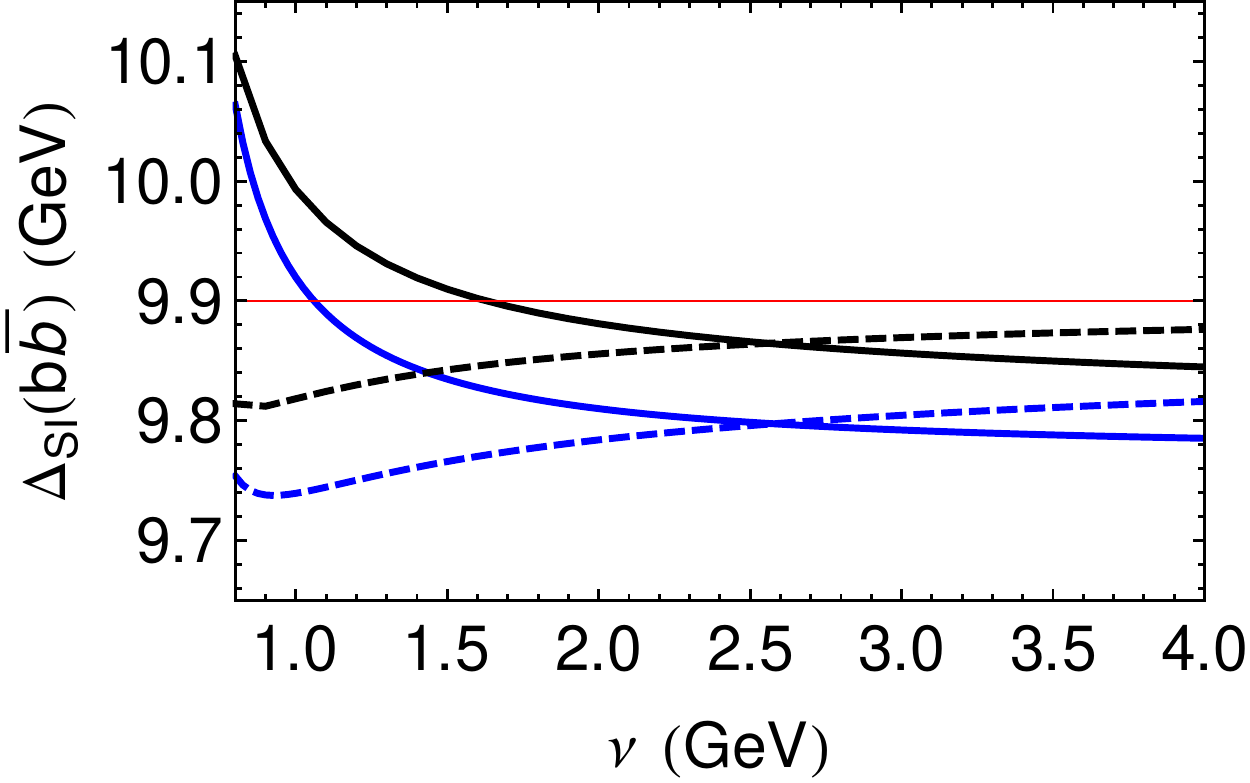}
\caption{
{\bf Left panel:} Plot of $2m_b+E^{(0)}_{21}$ for bottomonium using $V_{N,\rm RS'}^{(0)}$ for N=0 (red), 1 (green), 2 (blue), 3 (black). Dashed lines are computed with $\nu_f=0.7$ GeV and continuous lines with $\nu_f=1$ GeV.  
{\bf Right panel:} Plot of $2m_b+E^{(0)}_{21}$ for bottomonium using $V_{N,\rm RS'}^{(0)}$ for N=3. Dashed lines are computed with $\nu_f=0.7$ GeV and continuous lines with $\nu_f=1$ GeV.  Blue lines correspond to the black lines of the left panel. Black lines are computed at strict weak coupling. 
\label{Fig:hbstaticImproved}}   
\end{center}
\end{figure}

The energy $E_{nl}^{(0)}$ in \eq{eq:Schroedinger} correctly incorporates the N$^N$LO corrections to the spectrum associated to the static potential. It also includes higher order corrections (those generated by the iteration of the static potential). In order for this computational scheme to make sense, it first requires that the $N \rightarrow \infty$ converges, or at least that the error is small compared with the relativistic correction. We show the result of the computation of $E_{21}^{(0)}$ for bottomonium in \fig{Fig:hbstaticImproved} setting $\nu_r=\infty$ (setting $\nu_r=1$ GeV does not change the qualitative picture) and $\nu_f=1$ or 0.7 GeV. We do not see convergence for $\nu_f=1$ GeV but we get it for $\nu_f=0.7$ GeV. Either way, it is worth emphasizing that, for $N=3$, the $\nu_f=1$ and 0.7 GeV are consistent with each other, as we can see in \fig{Fig:hbstaticImproved} (left). This shows a mild dependence on $\nu_f$. On the other hand the dependence on $\nu$ is still large. We can also compare with the strict weak-coupling expansion result. We do so in \fig{Fig:hbstaticImproved} (right). We find a difference of order 60 MeV. This difference appears to be very stable under $\nu$ or $\nu_f$ variations. The origin of this constant shift is not clear to us at present. Setting $\nu_r=1$ GeV does not qualitatively change the picture.  Overall, we take $E_{nl}^{(0)}\sim 9.8$ GeV as the leading ${\cal O}(v^2)$ solution. Note that this number still suffers from sizable uncertainties ($\sim \pm 60$ MeV if looking to the scale variation or the difference with the strict weak-coupling evaluation). 

Once we have the leading ${\cal O}(v^2)$ solution, we can consider the incorporation of the relativistic and ultrasoft corrections, which will scale, at most, as ${\cal O}(v^4)$. With the accuracy of this work, we only have to take the expectation value of $\delta V$ where 
\begin{align}
\label{deltaV}
\delta V=V_s-V^{(0)}
\end{align}
stands for the relativistic potential ($V_s$ is the total singlet potential) that contributes up to N$^3$LL
and also add the ultrasoft correction from \eq{EnlUS}. Overall the mass of the bound states reads
\begin{align}
\label{Mnlfull}
M(n,l,j)=m_1+m_2+E_{nl}^{(0)}+{}^{(0)}\langle n,l|\delta V|n,l \rangle^{(0)}+\delta E_{nl}^{\US}
\,,
\end{align}
where $E_{nl}^{(0)}$ counts as $v^2$, ${}^{(0)}\langle n,l|\delta V|n,l \rangle^{(0)}$ counts as $v^4$ (including also $v^4 \als$ corrections) and $\delta E_{nl}^{\US}$ as $v^5$. 
\eq{Mnlfull} is numerically correct with N$^3$LL precision and incorporates extra subleading terms (albeit in an incomplete way). If one sets 
$\nu_h=\nuus=\nu$ one also recovers the N$^3$LO result incorporating some extra subleading terms. For notation purposes we will label the results obtained using \eq{Mnlfull} as N$^i$LL(N) where N stands for the order at which the static potential (we introduce exactly in the Schr\"odinger equation) is truncated and $i$ will be 2 or 3 depending on the order at which the relativistic and ultrasoft corrections are included. A similar counting will apply to the 
N$^i$LO(N) result, where we do not perform the logarithmic resummation by setting $\nu_h=\nuus=\nu$.

Note that the correction to the static potential generated by the resummation of ultrasoft logarithms obtained in \eq{deltaV0} is not incorporated in 
\eq{VRS} but rather added to \eq{deltaV} as part of the correction\footnote{Adding \eq{deltaV0} to 
\eq{VRS} would redefine the leading ${\cal O}(v^2)$ solution. We do not explore this line of research in this paper.}. 

Overall, this computational scheme resums a subset of subleading corrections in the hope that they would account for the bulk of such subleading terms. This could be so if the higher order corrections that we infer from our knowledge of the static potential are indeed responsible for the leading corrections.
 
The expressions we use for the relativistic potential  (valid also in the unequal mass case) are taken from \rcite{Peset:2015vvi}, which uses the computation of the $1/m$ potential obtained in \rcite{Kniehl:2001ju}. For ease of reference we quote them in \app{app:potential}. We can use any of the bases for the potentials presented in that paper, which were referred as: Wilson, onshell, Coulomb or Feynman. At strict N$^3$LO they all yield the same result. Since the computational scheme we implement in this section partially resums higher orders some dependence on the basis of potentials shows up. We have checked that, for the set of bases we consider, the dependence is quite small.

\begin{figure}[!htb]
	\begin{center}      
	\includegraphics[width=0.49\textwidth]{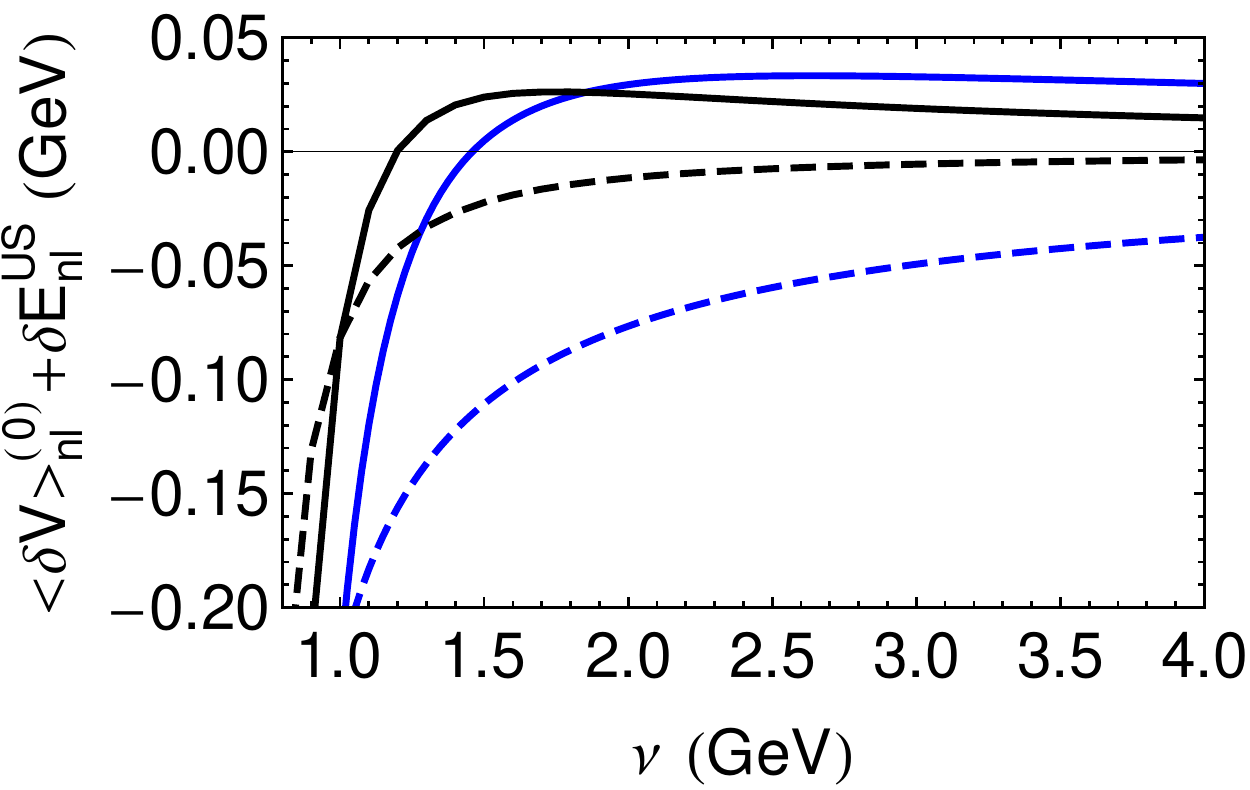}
	%
\caption{Plot of the ${}^{(0)}\langle n,l|\delta V|n,l \rangle^{(0)}+\delta E_{nl}^{\US}$ contribution to $\Delta_{SI}$ for bottomonium with N$^3$LL (continuous black line), N$^3$LL(3) (continous blue line), N$^3$LO (dashed black line), N$^3$LO(3) (dashed blue line) precision, evaluated with 
$\nu_r=\infty$ GeV, $\nu_f=1$ GeV, $\nuus=1$ GeV. Alternative plots with $\nu_r=1$ GeV or $\nu_f=0.7$ GeV change little.
\label{Fig:hbRelativistic}}   
\end{center}
\end{figure}

The computation of the relativistic corrections opens new issues compared with the static potential. In the case of the static potential the natural scale is $\nu \sim 1/r$, except in the ${\cal O}(\als^4)$ term where also the ultrasoft scale $\nu_{us}$ appears\footnote{If one considers the RGI expression ultrasoft logs already appear at ${\cal O}(\als^3)$.}. The case of the relativistic potentials is quite different. They are much more dependent on the hard, and above all, the ultrasoft scale (on the other hand they are formally insensitive to the pole mass renormalon). Moreover, in order for the computation with the static potential to be a more or less reasonable approximation we need to have at least three or more terms (also important is the resummation of soft logarithms). For the case of the relativistic potentials, we have at most two terms. This, together with a much stronger scale dependence, can trigger that inefficiencies of the description of the relativistic potentials get amplified when computing the expectation values. In this respect, for the first time, we have two terms of the perturbative expansion of the (relativistic) potentials, for which the complete resummation of large logarithms is known. This allows us to compare fixed order with RGI results. We do this comparison in \fig{Fig:hbRelativistic}. We observe that the resummation of logarithms happens to be crucial to get consistent results between the strict and the alternative counting scheme. It makes the correction much smaller too (which is good news for the validity of the velocity expansion), and, as we will see next, it helps in getting reasonable agreement with experiment.

\section{Phenomenology: $n=2$, $l=1$. Alternative counting approach}
\label{Sec:PhenImpr}

\begin{figure}[!htb]
	\includegraphics[width=0.49\textwidth]{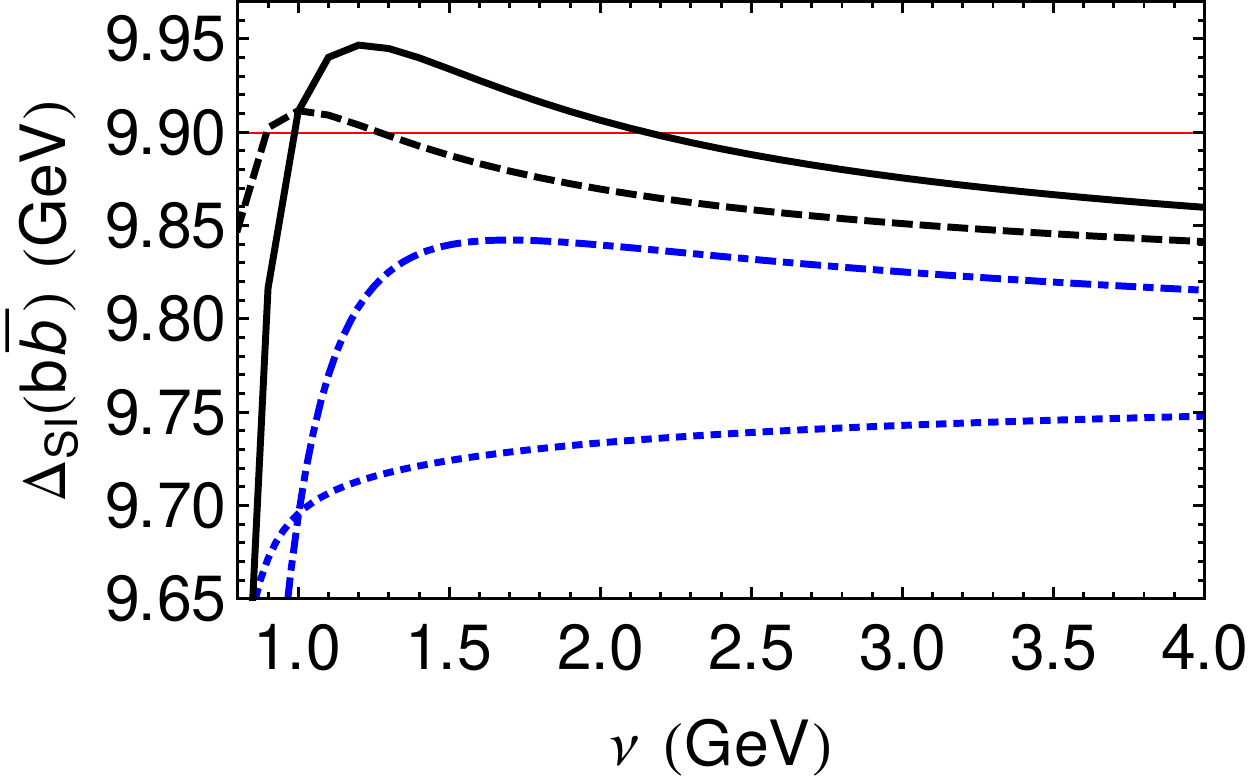}
	\includegraphics[width=0.49\textwidth]{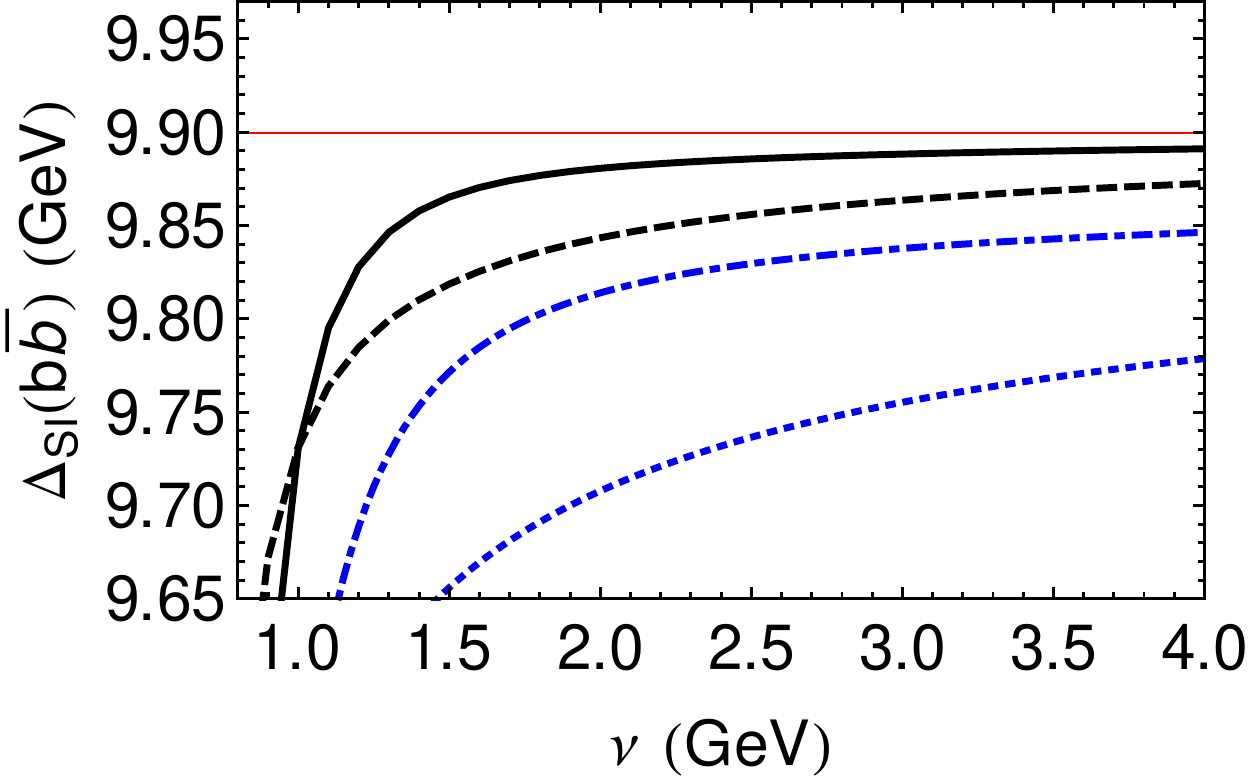}
	%
\caption{Plot of $\Delta_{SI}$ evaluated with N$^3$LL (continuous black line), N$^3$LL(3) (dash-dotted blue line), N$^3$LO (dashed black line) and N$^3$LO(3) (dotted blue line) precision. All lines are computed with $\nu_r=\infty$ GeV and $\nuus=1$ GeV and 
{\bf Left panel} with $\nu_f=1$ GeV and  
{\bf Right panel} with $\nu_f=0.7$ GeV.
\label{Fig:hbImproved}}   
\end{figure}
We now repeat the analysis of \Sec{Sec:Phen} but using the predictions obtained in the previous section. We first plot our prediction of $\Delta_{SI}$ in \fig{Fig:hbImproved}. The bulk of the difference with the strict weak-coupling computation comes from the different results of the static solution. On the other hand, the relativistic corrections are similar. We emphasize again that for this to be the case, the resummation of large logarithms is crucial. The final result is compatible with experiment within uncertainties. 

\begin{figure}[!htb]
	\begin{center}      
	\includegraphics[width=0.49\textwidth]{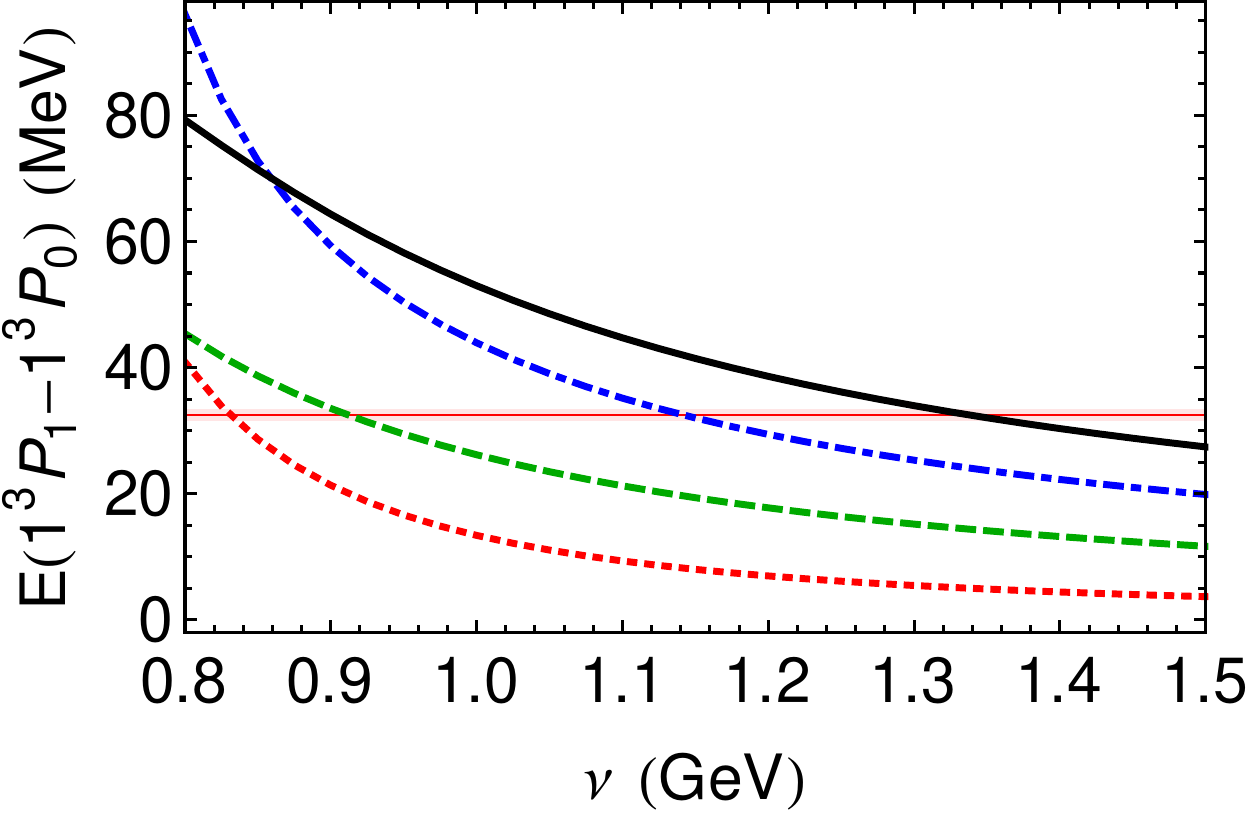}
	\includegraphics[width=0.49\textwidth]{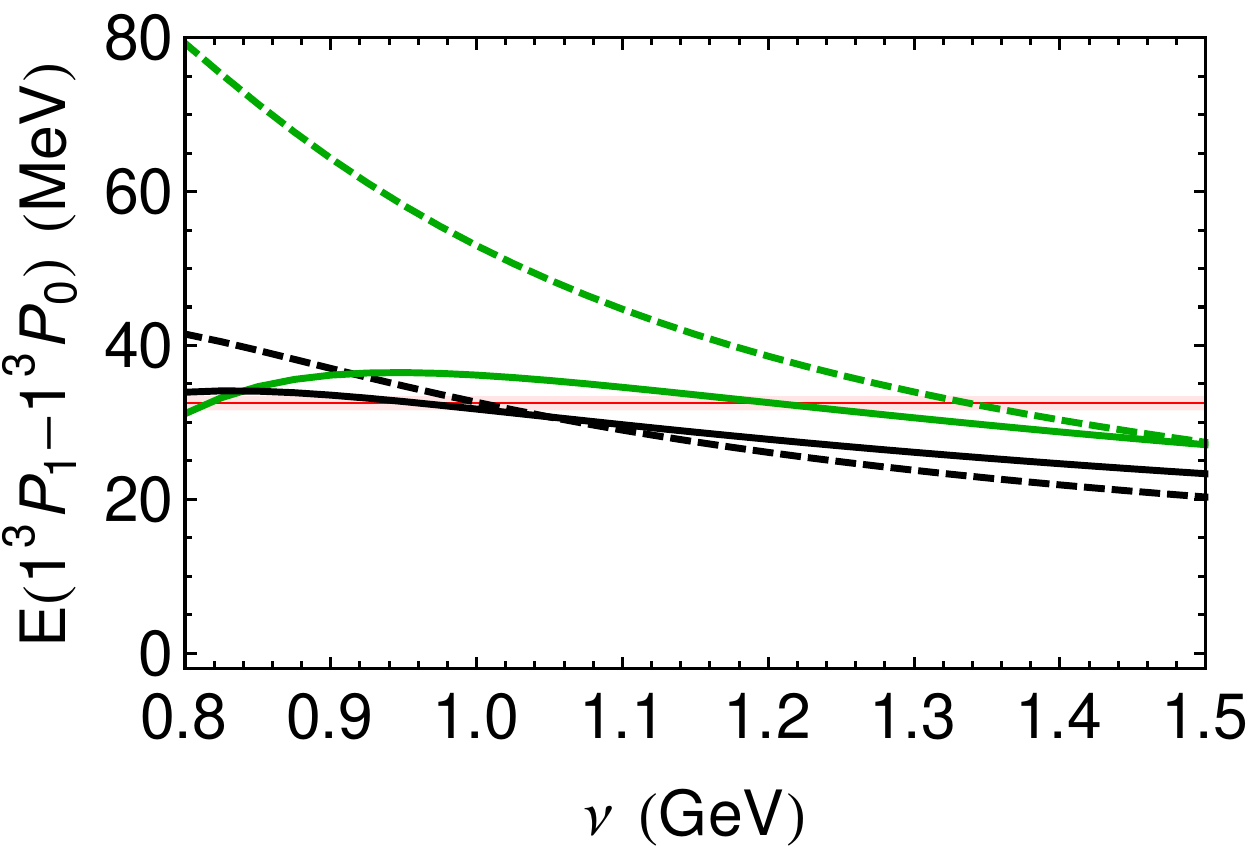}\\
	\includegraphics[width=0.49\textwidth]{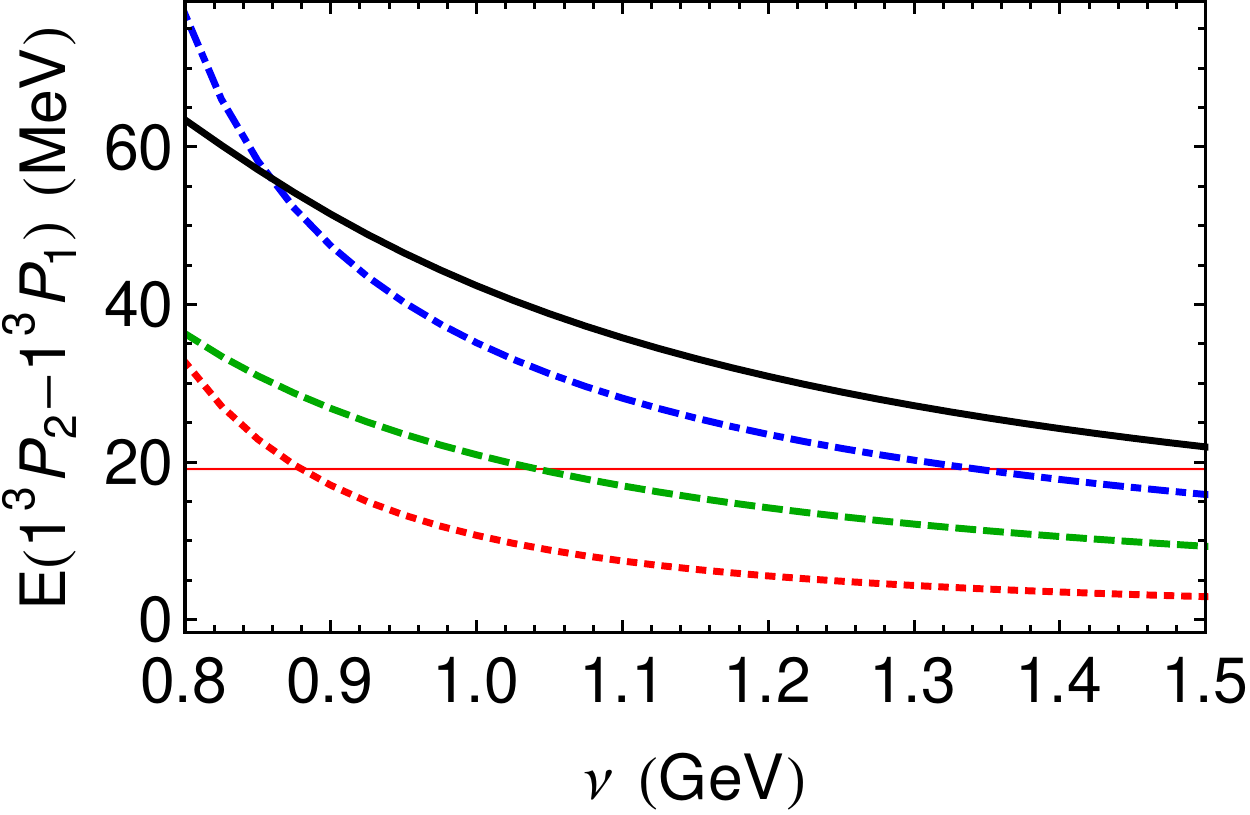}
	\includegraphics[width=0.49\textwidth]{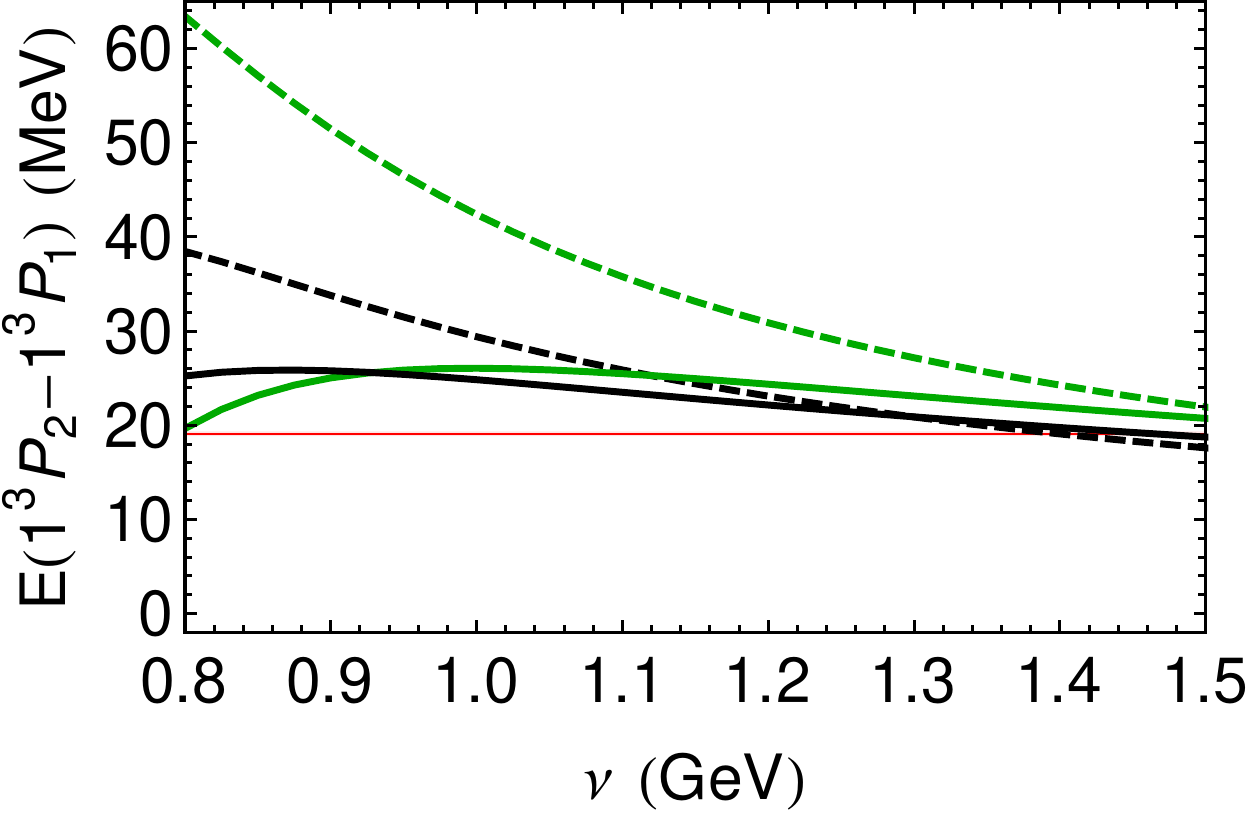}
	%
\caption{
Plots for the P-wave fine splittings in bottomonium in the RS' scheme with $\nu_f=1$ GeV and 
$\nuh=m_{b,{\rm RS}'}$. 
Red band is the experimental value. 
{\bf Left-up panel:} Plot of $E(1^3P_1)-E(1^3P_0)$ with NNLO(N) accuracy with N=0 (dotted red line), 1 (dashed green line), 2 (dash-dotted blue line), 3 (continuous black line). {\bf Left-bottom panel:} Plot of $E(1^3P_2)-E(1^3P_1)$ with NNLO(N) accuracy with N=0  (dotted red line), 1 (dashed green line), 2 (dash-dotted blue line), 3 (continuous black line). {\bf Right-up panel:} Plot of $E(1^3P_1)-E(1^3P_0)$ with NNLO(3) accuracy (dashed green line), NNLL(3) accuracy (dashed black line), 
N$^3$LO(3) accuracy (continuous green line) and N$^3$LL(3) accuracy (continuous black line). {\bf Right-bottom panel:} Plot of $E(1^3P_2)-E(1^3P_1)$ with NNLO(3) accuracy (dashed green line), NNLL(3) accuracy (dashed black line), 
N$^3$LO(3) accuracy (continuous green line) and N$^3$LL(3) accuracy (continuous black line).
\label{Fig:PwaveFineRGI}}   
\end{center}
\end{figure}
We now turn to the fine splittings. We remark that they are renormalon-free observables. Indeed the results are virtually insensitive to $\nu_f$, so by default we will use $\nu_f=1$ GeV. Therefore, they are a cleaner place than $E_{21}^{(0)}$ to test the convergence of truncating at $N$ the static potential. We show such plot in \fig{Fig:PwaveFineRGI} (left). In the left figures we plot the fine splitting with NNLO(N) accuracy. This figure effectively draws (up to a constant) $\langle 1/r^3 \rangle_{21}$ for bottomonium using different $N$'s, which allows us to check the convergence associated to the static potential. The convergence is somewhat marginal. Things improve considerably when we include higher order corrections to the NNLO(3) result. We show the results in \fig{Fig:PwaveFineRGI} (right). Moving from NNLO(3) to NNLL(3) (incorporating the resummation of large hard logarithms) makes the result more scale independent and closer to experiment. Going to N$^3$LL(3) or N$^3$LO(3) improves the result. They are quite scale independent, quite close among them, and in quite good agreement with experiment. For $E(1^3P_1)-E(1^3P_0)$ the N$^3$LL(3) theoretical result hits the experimental value at the scale of minimal sensitivity. For $E(1^3P_2)-E(1^3P_1)$, the N$^3$LL(3) theoretical result is around 5 MeV above the experimental value at the scale of minimal sensitivity.  Overall, the agreement with experiment is quite remarkable. If we compare with the strict weak-coupling results, they typically yield smaller values than the alternative computational approach, and show a larger factorization scale dependence. Nevertheless, the N$^3$LO and N$^3$LL results in the strict weak-coupling approximation are in reasonable agreement with the  N$^3$LO(3) and N$^3$LL(3) result obtained in this alternative computational approach at the scale of minimal sensitive of both.  If we look to the difference between the NNLL(3) and N$^3$LL(3) the difference is small. On the other hand the difference between NNLO(3) and N$^3$LO(3) is bigger.  In both cases they converge to experimental value. 

Note that the N$^3$LO/LL(N) result is the sum of a N$^3$LO/LL(N) contribution coming from the potential and a N$^3$LO/LL(N) contribution coming from the NRQCD Wilson coefficients, neglecting crossed terms, which are subleading. 

\begin{figure}[!htb]
	\begin{center}      
	\includegraphics[width=0.49\textwidth]{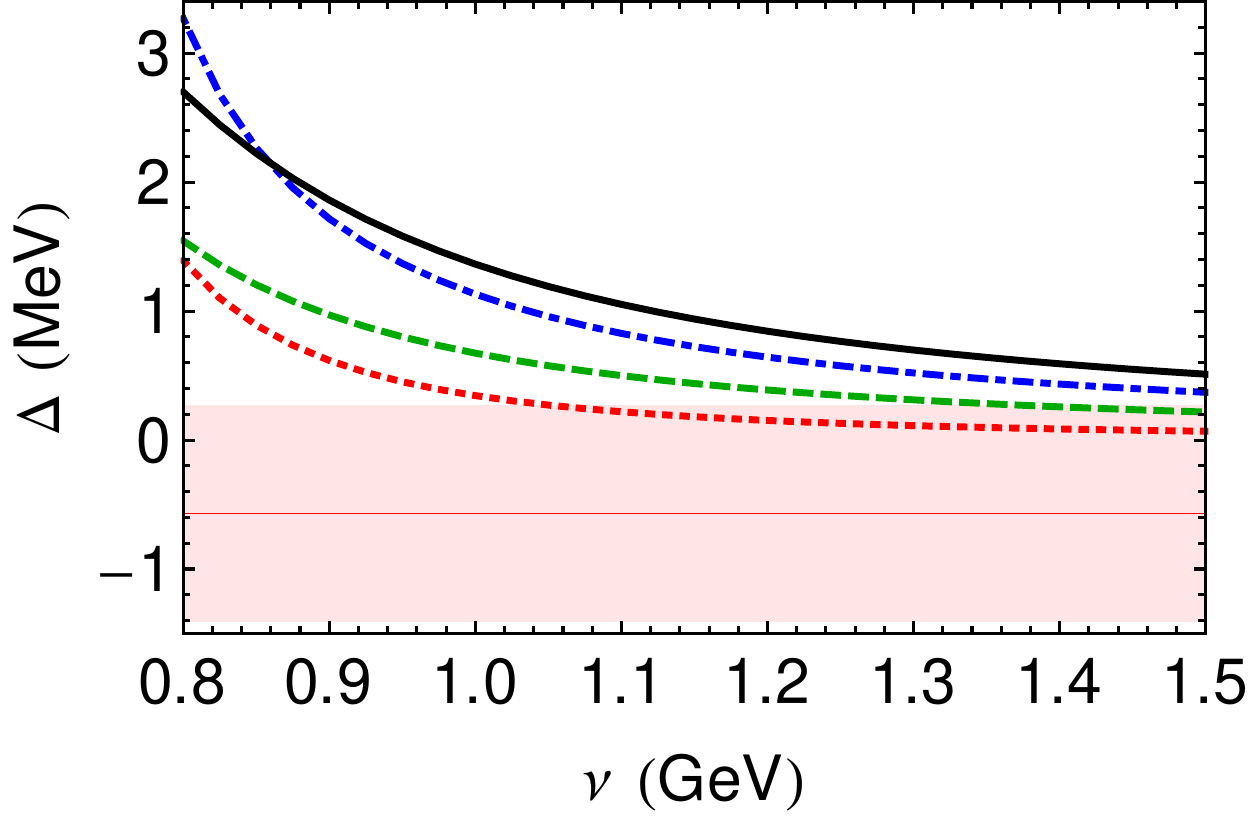}
	\includegraphics[width=0.49\textwidth]{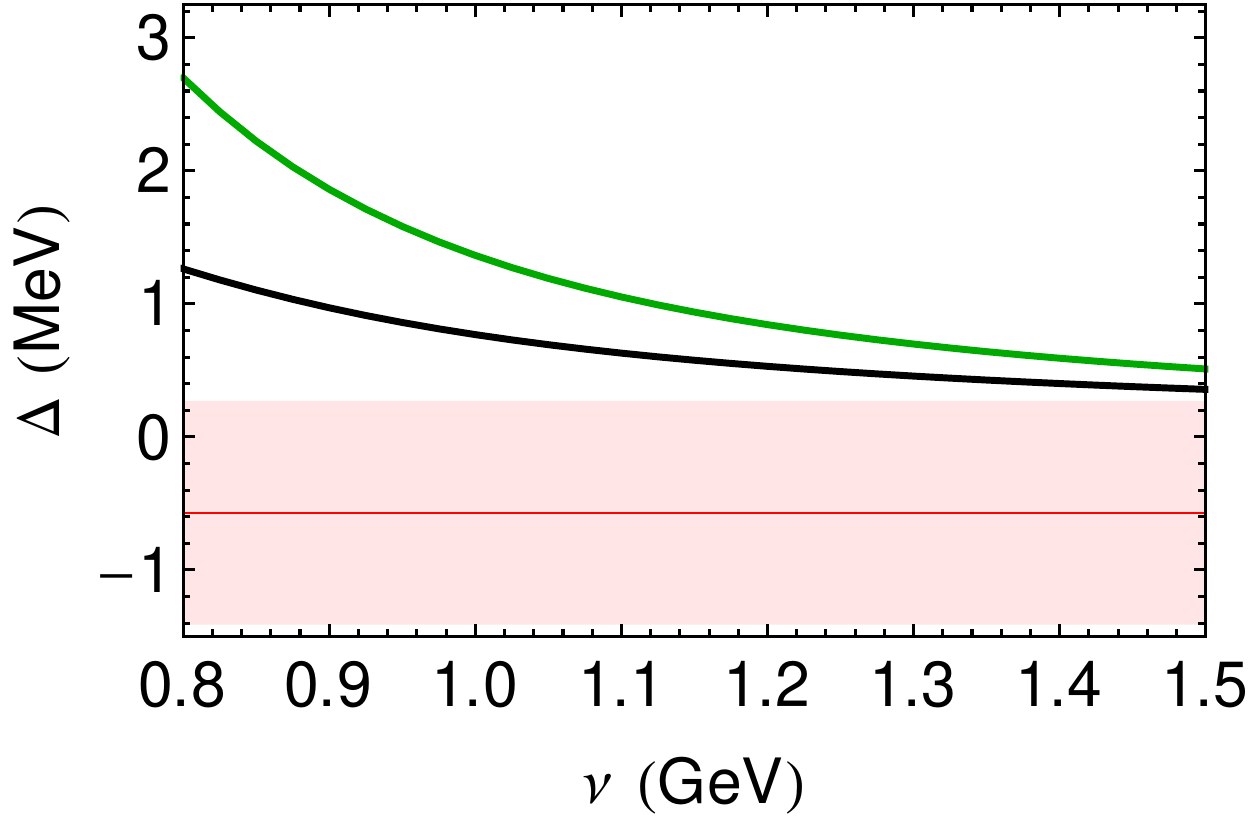}
	%
\caption{
Plots of $\Delta$, the hyperfine splitting for P-wave bottomonium, in the RS' scheme with $\nu_f=1$ GeV and 
$\nuh=m_{b,{\rm RS}'}$. 
Red band is the experimental value. 
{\bf Left panel:} Plot of $\Delta$ with N$^3$LO(N) accuracy with N=0 (dotted red line), 1 (dashed green line), 2 (dash-dotted blue line), 3 (continuous black line). {\bf Right panel:} Plot of $\Delta$ with 
N$^3$LO(3) accuracy (continuous green line) and N$^3$LL(3) accuracy (continuous black line).
\label{Fig:PwaveHFRGI}}   
\end{center}
\end{figure}
We now consider the hyperfine splitting, $\Delta$, defined in \eq{eq:Delta}. It starts giving a nonzero contribution at 
N$^3$LO/N$^3$LL. This observable is sensitive to $\langle {\rm reg} \frac{1}{r^3} \rangle$ (the N$^3$LO and N$^3$LL expression for the fine splitting is also sensitive to the matrix element of this operator). We can then check the convergence in $N$ by computing the N$^3$LO(N) result (i.e. the matrix element) for different $N$'s. We show the outcome in the left panel of \fig{Fig:PwaveHFRGI}. The convergence is similar to the fine case. Corrections are large and so is the $\nu$ scale dependence. The resummation of the hard logarithms improve the agreement with experiment, still the strict weak-coupling result shows a better agreement with experiment. In the above computation, we only have the first term of the perturbative expansion in the strict weak-coupling limit.  We conjecture that higher order terms of the relativistic potential will compensate this behavior. In other words, we do not know the shape of the relativistic corrections with enough accuracy at short distances. This introduces large errors when producing expectation values of them. In this respect, it is interesting to see what lattice simulations can add to this discussion. The hyperfine splitting is specially clean, as it only depends on $V_{S^2}$. Indeed, for P-wave states, any dependence on the delta potential vanishes and only the r-dependence (at nonzero $r$) is relevant. Lattice determinations of $V_{S^2}$ were obtained in \cite{Bali:1997am,Koma:2006fw}. In the first reference the lattice simulations were basically compatible with zero (up to a lattice version of $\delta({\bf r})$, which obviously does not contribute to the hyperfine). The second reference gives a parameterization which has a nontrivial $r$ dependence (with no delta potential). This could give a large contribution to the hyperfine splitting and, thus, making the theoretical prediction incompatible with the experimental figure, which is approximately zero\footnote{Nevertheless, it is not that clear whether the lattice simulations of \cite{Koma:2006fw} at short distances cannot indeed be parameterized by a delta potential. A.P. acknowledges discussions with Gunnar Bali on this point.}.

\begin{figure}[!htb]
	\begin{center}      
	\includegraphics[width=0.49\textwidth]{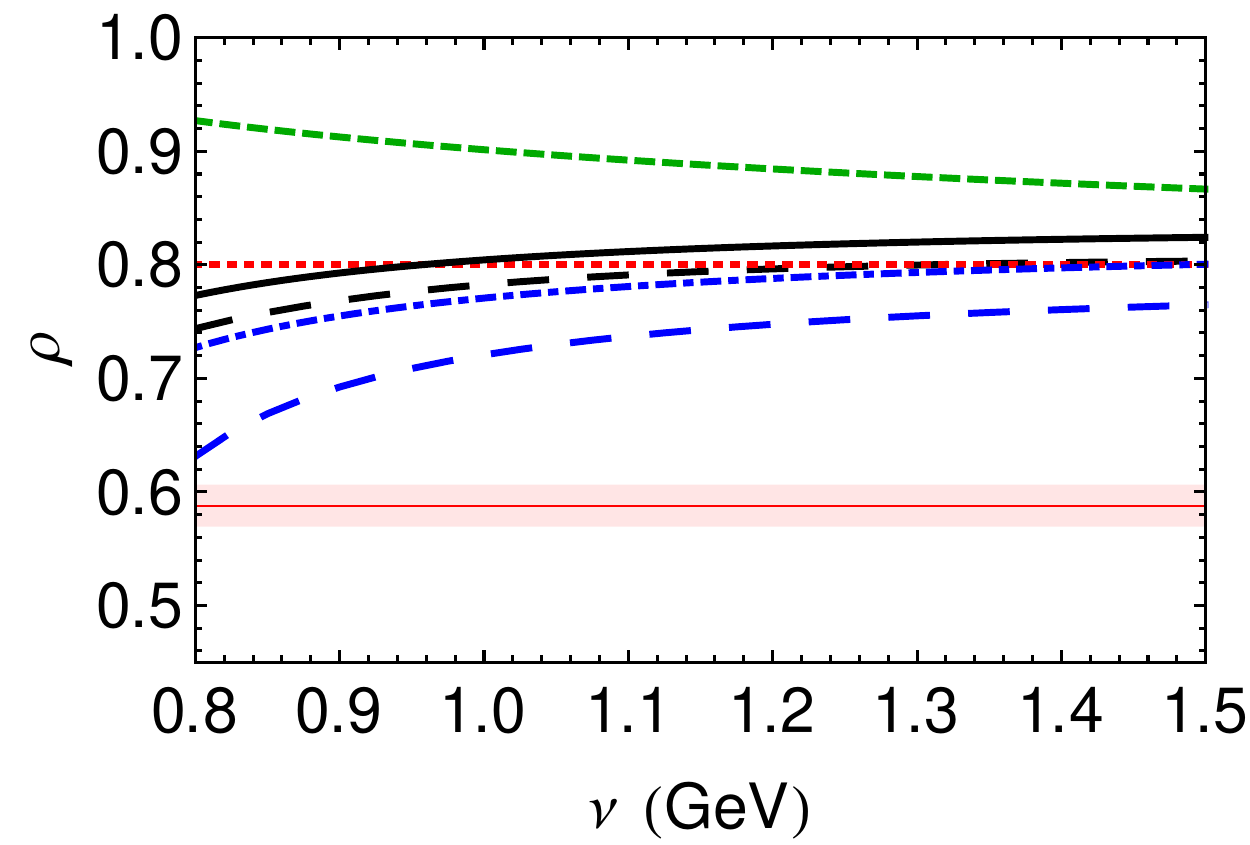}
	%
\caption{Plot of $\rho$ for bottomonium. We plot the LO=LO(N) (dotted red line),  LL=LL(N) (dashed green line), NLO(3) (dot-dashed blue line) and NLL(3) (solid black line) results (the latter two treating the ${\cal O}(\als)$ correction to the relativistic potential perturbatively). The NLO(3) and NLL(3) result from the energy differences are the long-dashed blue and black lines respectively.  \label{Fig:PwaveFineratio}
}   
\end{center}
\end{figure}
Finally we study $\rho$ ratio. We show our results in \fig{Fig:PwaveFineratio}. The LO(N) and LL(N) results are equal (for any $N$) to the strict weak-coupling computation. For the NLO(3) and NLL(3) results, we consider two options: Directly considering the ratio between the
energy differences or treating the ${\cal O}(\als)$ correction to the relativistic potential perturbatively. 
At small scales the difference between both approaches becomes significant, specially for the NLO(3) result, which approaches the experimental result. At present, the spread of values depending on the truncation does not allow to reach definite conclusions. Differences with experiment are of order 20\%, in principle achievable with higher order corrections. 

The issues discussed above deserve further dedicated studies. Indeed, to settle (some of) them, it would be very interesting to compute the next correction in the weak-coupling expansion of the relativistic potential that contributes to this observable. This is a complicated task but within reach. Indeed, for the future, the fine and hyperfine splittings are ideal candidates for dedicated analyses aiming at ${\cal O}(m v^6)$ precision. This is in principle feasible, and may lead to precise predictions with small errors. 

\section{Conclusions}

In this paper, the P-wave heavy quarkonium spectrum  has been obtained for the first time at strict weak coupling with N$^3$LL precision. We have obtained such precision for the equal and non-equal mass cases and 
for the fine and hyperfine splittings as well. We emphasize that these results also give the ${\cal O}(m \als^6 \ln (1/\als))$ correction to the spectrum (for P-wave states) for the first time. Remarkably, the results we obtain are compatible with experiment, for $n=2$, $l=1$ bottomonium, albeit with large uncertainties. For the spin-independent energy combination $\Delta_{SI}$, defined in \eq{MSIPwave}, the convergence is somewhat marginal. For the fine splitting, approximate agreement can be found at scales of around 1 GeV, also for the hyperfine. In any case, the uncertainties are large, to the point that the incorporation of the resummation of logarithms produces energy shifts which are inside the expected uncertainties. For charmonium and $B_c$ we have also performed exploratory studies. 
We found that the scale dependence is larger and the convergence worse. At this stage we refrain of trying quantitative analyses of these states.  

For $\Delta_{SI}$, the N$^3$LL result is the maximal accuracy (in analytic terms) that can be obtained in the foreseeable future. 
For some specific (the fine and hyperfine) energy splittings, it is still within reach (with a quite significant, but finite, amount of effort) to go further analytically, and obtain the complete ${\cal O}(m\als^6)$ result (or its RGI expression, which however could be much more difficult). This implies computing the ${\cal O}(v^2)$ corrections to the leading nonvanishing term. It would give a hint of the size of the relativistic corrections, which is quite compelling. We have already seen that present evaluations of the $\rho$-ratio of the fine splittings are off by around 25\%, even though the description of the individual energy splittings is quite reasonable. It would be interesting to see the impact of the incorporation of the ${\cal O}(v^2)$ corrections to this specific observable. 
For the case of the hyperfine splitting at present only the leading nonvanishing term is known. The present evaluation agrees with experiment. Therefore, it is of great interest to check if such agreement survives the incorporation of higher order corrections.  
Leaving aside these energy differences, going beyond N$^3$LL would require to go to N$^4$LO. For the gross spectrum, $\Delta_{SI}$, this would require the much demanding computation of the static potential with four-loop accuracy, other necessary computations would also be quite difficult. 

In view of the difficulty of obtaining complete higher order corrections, one will have to rely on approximations. The first one, which we have already applied here, explores selective resummations of higher order corrections. We incorporate the static potential (truncated to a given power in $\als$) exactly in the Schr\"odinger equation (see \eq{eq:Schroedinger}). We take the result as the leading ${\cal O}(v^2)$ term. In order for this approach to be sensible, this leading ${\cal O}(v^2)$ term has to be more or less stable when truncating at different orders in the static potential (for large $N$). With $\nu_f=1$ GeV we do not get a convergent pattern, though we get it for $\nu_f=0.7$ GeV, and both results are relatively close for $N=3$ (as long as the scale $\nu$ is not very small). We observe that, compared with the strict weak-coupling computation, we find an almost constant shift downwards of order $\sim 60$ MeV. At this point we do not have a clear explanation for this fact, and only speculate that it may have to do with inefficiencies in the renormalon cancellation in the static potential. Leaving this problem aside, we consider the incorporation of the relativistic corrections. These are renormalon-free quantities but much more sensitive to the hard, and above all, the ultrasoft scales. We can make interesting observations. If we consider the relativistic corrections to the energy without logarithmic resummation, we observe that they yield quite different results in the alternative counting versus the strict weak-coupling computation. The former generates a much bigger correction that deteriorates the agreement with data. Remarkably, the resummation of logarithms fixes this problem. After the resummation of logarithms both the strict weak-coupling computation and the alternative counting scheme yield consistent results for the relativistic corrections of \eq{MSIPwave}.

We also apply this alternative counting scheme to the fine and hyperfine. They are free of renormalon and ultrasoft effects. Therefore, they are potentially rather clean observables. Also interesting observations can be made here. 
The convergence of the static solution is still slow. Nevertheless, a rather reasonable agreement with experiment is obtained for the fine splittings after inclusion of the ${\cal O}(\als)$ corrections to the potential. For the $\rho$ ratio, however, the situation is somewhat inconclusive and so is for the hyperfine splitting. We conjecture that higher order perturbation corrections can be important to obtain precise predictions for these observables. 

\medskip

{\bf Acknowledgments} \\ 
This work was supported in part by  the Spanish grants 
FPA2014-55613-P, FPA2017-86989-P and SEV-2016-0588. J.S. acknowledges the financial support from the European Union's Horizon 2020 research and innovation programme under the Marie Sk\l{}odowska--Curie Grant Agreement No. 665919, and from Spanish MINECO's Juan de la Cierva-Incorporaci\'on programme, Grant Agreement No. IJCI-2016-30028.

\newpage

\appendix

\section{The potential}
\label{app:potential}
We display here the $\MS$ renormalized expressions for the NRQCD potentials necessary to compute the N$^3$LO spectrum.
The static potential reads
\begin{align}
V_N^{(0)}(r;\nu)
&=
 -\frac{C_f\,\als(\nu)}{r}\,
\bigg\{1+\sum_{n=1}^{N}
\left(\frac{\als(\nu)}{4\pi}\right)^n a_n(\nu;r)\bigg\}
\,.
\label{VSDfo}
\end{align}
For the seaked N$^3$LO precision one can truncate at $N=3$ and the coefficients read
\begin{align}
a_1(\nu;r)&=a_1+2\beta_0\,\ln\left(\nu e^{\gamma_E} r\right),\nn\\
a_2(\nu;r)&=a_2 + \frac{\pi^2}{3}\beta_0^{2}+\left(4a_1\beta_0+2\beta_1 \right)\ln\left(\nu e^{\gamma_E} r\right)+4\beta_0^{\,2}\,\ln^2\left(\nu e^{\gamma_E} r\right),\nn\\
a_3(\nu;r)&=a_3+ a_1\beta_0^{2} \pi^2+\frac{5\pi^2}{6}\beta_0\beta_1+16\zeta_3\beta_0^{3}\nn\\
&+\left(2\pi^2\beta_0^{3} + 6a_2\beta_0+4a_1\beta_1+2\beta_2+\frac{16}{3}C_A^{\,3}\pi^2 \right) \ln\left(\nu e^{\gamma_E} r\right)\nn\\
&+\left(12a_1\beta_0^{\,2}+10\beta_0\beta_1\right)  \ln^2\left(\nu e^{\gamma_E} r\right)+8\beta_0^{3}  \ln^3\left(\nu e^{\gamma_E} r\right).
\label{eq:Vr}
\end{align}
The ${\cal O}(\als)$ term was computed in \rcite{Fischler:1977yf}, the ${\cal O}(\als^2)$  in \rcites{Schroder:1998vy,Peter:1996ig}, 
the ${\cal O}(\als^3)$ logarithmic term in \rcite{Brambilla:1999qa}, the light-flavour finite piece in 
\rcite{Smirnov:2008pn}, and the pure gluonic finite piece in \rcites{Anzai:2009tm,Smirnov:2009fh}.

The complete set of relativistic potentials in the on-shell scheme with N$^3$LO accuracy were obtained in the equal mass case in \rcites{Kniehl:2001ju,Kniehl:2002br} (for the NNLO result see \rcite{Gupta:1982qc}). For the unequal mass case (and for the specific renormalization scheme we use in this paper) they were computed in \rcite{Peset:2015vvi}. The resulting expressions read
\begin{align}
&
 \frac{V^{(1,0)}_{\rm on-shell}(r)}{m_1}+\frac{V^{(0,1)}_{\rm on-shell}(r)}{m_2}=
\frac{C_F^2\als^2(e^{-\gamma_E}/r)}{2r^2}\frac{m_r}{m_1 m_2}\left(1+\frac{\als}{2\pi}(a_1-\beta_0)\right)
\\
\nn
&\qquad
-\frac{C_FC_A\als^2(e^{-\gamma_E}/r)}{4m_rr^2}
\left\{1+\frac{\als}{\pi}\left(\frac{89}{36}C_A-\frac{49}{36}T_F n_f-C_F+\frac{4}{3}(C_A+2C_F)\ln\left(\nu r e^{\gamma_E}\right)\right)\right\}
.
\end{align}
\begin{align}
V_{{\bf p}^2,\rm on-shell}^{(2,0)}(r)&=-\frac{C_F\als^2}{3 \pi }\frac{1}{r}C_A\ln\left(\nu re^{\gamma_E}\right),\\
 V_{{\bf p}^2,\rm on-shell}^{(1,1)}(r)&= -\frac{C_F \als (e^{-\gamma_E}/r)}{r}\left\{1+\frac{\als}{4\pi } \left(a_1+\frac{8}{3}C_A\ln \left(\nu  re^{\gamma_E}\right)\right) 
\right\}\label{Vren5},
\end{align}
The spin-dependent and $V_r$ potentials relevant for P-waves states can be found in \eqss{vls20}{vs12}, \eq{V11LS} and \eq{deltaVreg}. For a P-wave state, the equal mass case is trivially recovered by setting $m=m_1=m_2$.

\section{The N$^3$LO heavy quarkonium spectrum}
\label{app:spectrum}
In this appendix we collect the explicit expression for the fixed order N$^3$LO P-wave spectrum. The P-wave spectrum at N$^3$LO, was obtained in \rcite{Kiyo:2013aea,Kiyo:2014uca} for the equal mass case and in \rcite{Peset:2015vvi} for the unequal mass case. It reads
\begin{align}
\label{EnN$^3$LO}
E_{\rm N^3LO}(n,l,s,j)&=E_n^C\left(1+\frac{\als}{\pi}P_1(L_\nu)+\left(\frac{\als}{\pi}\right)^2P_2(L_\nu)
+\left(\frac{\als}{\pi}\right)^3P_3(L_\nu)\right),
\\[2 ex]
P_1(L_\nu)&=\beta_0 L_\nu + \frac{a_1}{2}\,, \\ 
P_2(L_\nu)&=\frac{3}{4} \beta _0^2 L_{\nu }^2+
\left(-\frac{\beta _0^2}{2}+\frac{\beta _1}{4}+\frac{3 \beta _0 a_1}{4}\right) L_{\nu }+c_2 \,, \\ 
P_3(L_\nu)&=\frac{1}{2} \beta _0^3 L_{\nu }^3+
\left(-\frac{7 \beta _0^3}{8}+\frac{7 \beta _0 \beta _1}{16}+\frac{3}{4} \beta _0^2
   a_1\right) L_{\nu }^2\nn\\
&\quad +\left(\frac{\beta _0^3}{4}-\frac{\beta_0 \beta _1 }{4}+\frac{\beta _2}{16}-\frac{3}{8} \beta _0^2 a_1+2 \beta _0 c_2+\frac{3
   \beta _1 a_1}{16}\right) L_{\nu } +c_3  \,,
\end{align}
where $c_i=c_i^c+c_i^{nc}$, $L_{\nu } =\ln\frac{n \nu}{2 C_F m_r \als}+S_1(n+l)$ and $E_n^C=-\frac{m_rC_F^2\als^2}{2n^2}$. Expressions for 
$c_2^c$ and $c_3^c$ can be found in Eqs.~(7.17) and~(7.18) of \rcite{Peset:2015vvi} (see also \rcite{Peset:2015vvi} for definitions and notation);
\begin{align}
c_2^{\rm nc}
&=-\frac{2m_r^2\pi^2C_F^2}{nm_1m_2}\left\{\frac{1-\delta_{l0}}{l(l+1)(2l+1)}\left(D_s+\left(1+\frac{m_1m_2}{2m_r^2}\right)X_{LS}\right)+\frac{8\delta_{l0}}{3}{\cal S}_{12}\right\}\nn\\
&\quad +\frac{ m_r^2\pi^2C_F}{4n^2}\left\{\frac{1}{m_1 m_2}C_F +\frac{1}{m_r^2}\left[-3C_F+\frac{8n}{2l+1}\left(C_F+\frac{C_A}{2}\right)-4nC_F\delta_{l0}\right]\right\}
\end{align}
where, in comparison to \rcite{Peset:2015vvi} we express the results in the spin basis $\{{\bf S},{\bf S^-}\}$:
\begin{align}
{\cal S}_{12} &\equiv\langle {\bf S_1\cdot S_2} \rangle=\frac{1}{2}\left(s(s+1)-s_1(s_1+1)-s_2(s_2+1)\right),\\ 
D_{s} &\equiv \frac{1}{2}\langle S_{12}({\bf r})\rangle=\frac{
2 l (l+1) s (s+1) - 3 X_{LS} - 6 X_{LS}^2
}{
(2l-1)(2l+3)
},\\
X_{LS} &\equiv \langle{\bf L \cdot  S}\rangle= \frac{1}{2}\left[ j(j+1)-l(l+1)-s(s+1) \right],
\end{align}
and finally
\begin{align}
\label{eq:c3nc}
c_3^{\rm nc}&=\pi^2\left(C_F^3\xi_{\rm FFF}^{\rm SD}+C_F^2C_A(\xi_{\rm FFA}^{\rm SD}+\xi_{\rm FFA}^{\rm SI})
+C_F^2T_Fn_f(\xi_{\rm FFnf}^{\rm SD}+\xi_{\rm FFnf}^{\rm SI})-\frac{n}{6}\beta_0c_2^{\rm nc}\right.\nn
\\
&+C_A^3 \,\xi _{\rm AAA}+C_A^2  C_F\,\xi _{\rm AAF}+C_A  C_F T_F n_{f} \xi _{\rm AFn_f}+C_F^2T_F\xi_{\rm FF}+C_F^3\xi_{\rm FFF}^{\rm SI}\Big).
\end{align}
Again, in comparison to \rcite{Peset:2015vvi} we express the results in the spin basis $\{{\bf S},{\bf S^-}\}$:
\be
\xi_{\rm FFF}^{\rm SD}=\frac{1}{3n}
\left\{\frac{-3(1-\delta_{l0})}{l(l+1)(2l+1)}\left(\frac{2m_r^2}{m_1 m_2}D_s+X_{\rm LS}\right)-\frac{8m_r^2}{m_1 m_2}\mathcal S_{12}\delta_{l0}\left[2+3\frac{m_1m_2}{m_2^2-m_1^2}\ln\left(\frac{m_1^2}{m_2^2}\right)\right]\right\}, 
\label{xiFFF}
\ee
\begin{align}
\xi_{\rm FFnf}^{\rm SD}&=\frac{2 m_r^2 }{9 n^2m_1 m_2}\bigg\{\frac{1-\delta_{l0} }{l (l+1) (2 l+1)}\bigg[2 n (4{\cal S}_{12}-D_s) \nn\\
&\quad+ 6 \Big(D_s+\left(1+\frac{m_1m_2}{2m_r^2}\right) X_{\rm LS}\Big)\bigg(\frac{3 n}{2 l+1}+\frac{n}{2 l (l+1) (2 l+1)}+l+\frac{1}{2} \nn\\
&\quad+ 2 n \Big\{S_1(l+n)+S_1(2 l-1)-2 S_1(2 l+1)-l (\Sigma_1^{(k)}+\Sigma_1^{(m)})+n \Sigma_b-\Sigma_1^{(m)}+\frac{1}{6}\Big\}\!\bigg)\bigg]\nn\\
&\quad+ 8 \delta_{l0} {\cal S}_{12} \left[1+4 n \left(\frac{11}{12}-\frac{1}{n}-S_1(n-1)-S_1(n)+n S_2(n)\right)\right]\bigg\},\\
\xi_{\rm FFA}^{\rm SD}&=\frac{m_r^2 }{m_1 m_2}\Bigg\{\frac{1-\delta_{l0}}{l (l+1) (2 l+1) n}\Bigg[\frac{2}{3} \left(D_s+\left(1+\frac{m_1m_2}{2m_r^2}\right) X_{\rm LS}\right) \nn\\
&\qquad \!\times\! \bigg\{22 S_1(2 l+1)-17 S_1(l+n)-5 S_1(2 l-1)+11 \Big[l (\Sigma_1^{(k)}+\Sigma_1^{(m)})-n \Sigma_b+\Sigma_1^{(m)}\Big] \nn\\
&\qquad\quad -\frac{5 (2 l+1)}{4 n}-\frac{15}{2 (2 l+1)}-\frac{5}{4 (l (l+1) (2 l+1))}+\frac{1}{6}+\frac{3}{2} \ln \left(\frac{m_1 m_2}{4 m_r^2}\right)+3 L_H\bigg\} \nn\\
&\qquad -\frac{2}{9} \left(2 D_s+{\cal S}_{12}\right) \nn\\
&\qquad- 2 X_{\rm LS} \bigg(2 (S_1(2 l-1)-S_1(l+n))+\frac{2 l+1}{2 n}+\frac{1}{2 (l (l+1) (2 l+1))}+\frac{3}{2 l+1} \nn\\
&\qquad\quad - 2+\frac{m_1^2-m_2^2}{4m_1m_2}\ln\frac{m_1}{m_2}+\frac{1}{2} \ln \left(\frac{m_1m_2}{4 m_r^2}\right)+L_H\bigg)\Bigg] \nn\\
&\quad- \frac{4 \delta_{l0} {\cal S}_2 }{3 n}\bigg[-\frac{67}{3} S_1(l+n)-7 L_H+\frac{65 S_1(n)}{3}+\frac{44 n \Sigma_2^{(k)}}{3}+\frac{1}{6 n}+\frac{41}{18} \nn\\
&\qquad+ \frac{1}{m_1-m_2}\left((5 m_2-2 m_1) \ln \left(\frac{m_1}{2 m_r}\right)-(5 m_1-2 m_2) \ln \left(\frac{m_2}{2 m_r}\right)\right)\bigg]\Bigg\},
\end{align}
where $L_H=\ln\left(\frac{n}{C_F\als}\right)+S_1(n+l)$. The other color functions  $\xi^{\rm SI}_X$ in \eq{eq:c3nc} are not affected by the change of spin basis, and can be found in Eqs.~(I.15)-(I.21) of \rcite{Peset:2015vvi}.

The case of equal masses is recovered by taking the limit $m=m_1=m_2$ in all the color functions involved in \eq{eq:c3nc}, except for $\xi_{\rm FF}$ (Eq.~(I.18) in \rcite{Peset:2015vvi}), where we shall add the anihilation diagrams and, for equal masses, we obtain instead:
\begin{align}
\xi_{\rm FF}&=\frac{\delta_{l0}}{n}\left(-2 {\cal S}_{12}+\left(2 {\cal S}_{12}-\frac{1}{2}\right) \ln(2)+\frac{19}{30}\right).
\end{align}

Finally, we would like to mention that we can check that the spectrum produced by the potentials obtained in different matching schemes is equal. Indeed, this check gives a nontrivial relation among some finite sums that, even though we did not prove it explicitly, holds true for any arbitrary set of quantum numbers we tried:
\be
S_1(2l)+S_1(1+2l)-2 S_1(l+n)+2l \Sigma_1^{(k)}(n,l)+2(l+1) \Sigma_1^{(m)}(n,l)=1
\,.
\ee 

\section{The N$^3$LL heavy quarkonium spectrum}
\label{sec:NNNLL}
After adding the ultrasoft and soft/hard running to the N$^3$LO result one obtains the N$^3$LL P-wave spectrum. It reads
\begin{align}
\label{EnN$^3$LL}
E_{\rm N^3LL}(n,l,s,j)&=E_n^C\left(1+\frac{\als}{\pi}\tilde P_1(L_\nu)+\left(\frac{\als}{\pi}\right)^2\tilde P_2(L_\nu)
+\left(\frac{\als}{\pi}\right)^3\tilde P_3(L_\nu)\right),
\\[2 ex]
\tilde P_1(L_\nu)&=P_1(L_\nu)=\beta_0 L_\nu + \frac{a_1}{2}\,, \\ 
\tilde P_2(L_\nu)&=\frac{3}{4} \beta _0^2 L_{\nu }^2+
\left(-\frac{\beta _0^2}{2}+\frac{\beta _1}{4}+\frac{3 \beta _0 a_1}{4}\right) L_{\nu }+\tilde c_2 \,, \\ 
\tilde P_3(L_\nu)&=\frac{1}{2} \beta _0^3 L_{\nu }^3+
\left(-\frac{7 \beta _0^3}{8}+\frac{7 \beta _0 \beta _1}{16}+\frac{3}{4} \beta _0^2
    a_1\right) L_{\nu }^2\nn\\
&\quad +\left(\frac{\beta _0^3}{4}-\frac{\beta_0 \beta _1 }{4}+\frac{\beta _2}{16}-\frac{3}{8} \beta _0^2  a_1+2 \beta _0 \tilde c_2+\frac{3
   \beta _1  a_1}{16}\right) L_{\nu } +\tilde c_3 \,.
\end{align}

The coefficients $\tilde c_i=\tilde c_i^{\rm c}+\tilde c_i^{\rm nc}$ are split into contributions from the static potential and those including the relativistic corrections. 

$\tilde c_i^{\rm c}$ only get contributions from the resummation of the ultrasoft logarithms:
\begin{align}
\tilde c_2^c&=c_2^c+\delta c_2^c,\\
\tilde c_3^c&=c_3^c+\delta c_3^c,
\end{align}
where $c_{2,3}^c$ are the coefficients computed for the fixed order spectrum in Eqs.~(147) and~(150) in \rcite{Kiyo:2014uca}  for (un)equal masses, and $\delta c_{2,3}^c$
\begin{align}
\label{deltac2c}
\delta c_2^{\rm c}&=-\frac{ \pi C_A^3}{6}\frac{2\pi}{\beta_0}\ln\frac{\alus}{\als},\\
\label{deltac3c}
\delta c_3^{\rm c}&=\frac{\pi C_A^3}{32}\left[\frac{8}{3}\frac{2\pi}{\beta_0}\ln\frac{\alus}{\als}(\beta_0-2a_1)\right.\nn\\
&\hspace*{1.5cm}\left.-8\pi^2\frac{2\pi}{\beta_0}\frac{\alus-\als}{\als}\left(\frac{8}{3}\frac{\beta_1}{\beta_0}\frac{1}{(4\pi)^2}-\frac{1}{27\pi^2}\left(C_A(47+6\pi^2)-10 T_f n_f\right)\right)\right].
\end{align}

$\tilde c_i^{nc}$ get contributions from the ultrasoft resummation, from the hard resummation, and from the difference of evaluating the ultrasoft energy at the ultrasoft scale (which is included in the ultrasoft part of the coefficients), i.e.:
\begin{align}
\tilde c_2^{\rm nc}&=c_2^{\rm nc}+\delta c_2^{\rm nc},\\
\tilde c_3^{\rm nc}&=c_3^{\rm nc}+\delta c_3^{\rm nc},
\end{align}
where $c_{2,3}^{\rm nc}$ are the coefficients presented in the previous section and computed for the fixed order spectrum in (\rcite{Peset:2015vvi}) \rcite{Kiyo:2014uca}  for (un)equal masses, and
\begin{align}
\delta c_2^{\rm nc}&=\delta c_2^{\rm nc,h}+\delta c_2^{\rm nc,us},\\
\delta c_3^{\rm nc}&=\delta c_3^{\rm nc,h}+\delta c_3^{\rm nc,us}
\end{align}
where 
\begin{align}
\label{c2ncus}
&\delta c_2^{\rm nc,us}=\frac{2 \pi  C_A C_F}{3 n^2}\left(C_F-\frac{2 n}{2 l+1} (C_A+4 C_F)
\right)\frac{2\pi}{\beta_0}\ln\frac{\alus}{\als},\\
&\delta c_3^{\rm nc,us}=\frac{2\pi}{\beta_0}\frac{\alus-\als}{\als}\left[\frac{\pi \beta_0}{3}\left(\frac{C_A^3}{4}\left(L_{\US}-\frac{5}{6}\right) -C_F^3 L_n^E\right)\right.\nn\\
&\hspace*{0.3cm}+\left.\pi ^3 C_AC_F\left(\frac{2 C_A}{(2 l+1) n}+C_F\left(\frac{8}{(2 l+1) n}-\frac{1}{n^2}\right)\right)\right.\nn\\
&\hspace*{0.3cm}\times\left. \left(\frac{8}{3}\frac{\beta_1}{\beta_0}\frac{1}{(4\pi)^2}-\frac{1}{27\pi^2}\left(C_A(47+6\pi^2)-10 T_f n_f\right)+\frac{\beta_0 }{3 \pi ^2}\left(L_{\US}-\frac{5}{6}\right)\right)\right]\nn\\
&\hspace*{0.3cm}+\frac{\pi  C_F}{3n^2}\frac{2\pi}{\beta_0}\ln\frac{\alus}{\als}\left[ C_F(C_A-2C_F)\beta_0\frac{ n(1-\delta_{l0})}{l(l+1)(2l+1)}+2 C_A a_1 \left(C_F-\frac{2 n}{2 l+1}(C_A+4 C_F)\right)\right.\nn\\
&\hspace*{0.3cm}-\left.C_A\beta_0 \left(C_F-\frac{4 n}{(2 l+1)^2}\left(C_A+C_F(7+6l)\right)+\frac{4 n^2 }{2 l+1}(C_A+4 C_F)\left(\Sigma_b(n,l)-\frac{\pi ^2}{6}\right)\right)\right]\nn\\
&-\frac{\pi^2}{3}\left[\frac{C_A^3}{2}+\frac{4 C_A^2 C_F}{(2 l+1) n}+2 C_A C_F^2 \left(\frac{8}{(2 l+1) n}-\frac{1}{n^2}\right)\right]\left(\frac{\alus}{\als}L_{\nuus}-L_\nu\right)
,\label{c3ncus}
\end{align}
where we have omitted the contribution of $\langle{\rm reg}\frac{1}{r^3}\rangle$ to the S-wave spectrum.

The NNLL hard contribution of the spectrum coefficients is known for general quantum numbers:
\begin{align}
\delta c_2^{\rm nc,h}=&-\frac{ 2 \pi ^2 C_F ^2 (1-\delta_{l0} ) }{n(2 l+1) l (l+1)}\frac{m_r^2}{m_1  m_2}\left\{\delta[ c_F^\one c_F^\two] D_s+X_{LS} \left(\delta c_F^\two+\delta c_F^\one+\frac{\delta c_S^\one m_2}{2m_1 }+\frac{\delta c_S^\two m_1 }{2m_2}\right)\right\}\nn\\
&\label{c2ncSDh}=\frac{ 2 \pi ^2 C_F ^2 (1-\delta_{l0} ) }{n(2 l+1) l (l+1)}\left(1-z^{-\frac{\gamma_0}{2}}\right)\left(X_{LS} +\frac{m_r^2}{m_1m_2}\left(1+z^{-\frac{\gamma_0}{2}}\right)D_s\right),
\end{align}
where we have defined: 
\begin{align}
\delta c_F^{(i)}&=c_F^{(i)}(\nuh,\nu)-c_F^{(i)}(\nu,\nu)\;,\\
\delta c_S^{(i)}&=c_S^{(i)}(\nuh,\nu)-c_S^{(i)}(\nu,\nu)=2(c_F^{(i)}(\nuh,\nu)-c_F^{(i)}(\nu,\nu)),\\
\delta [c_F^\one c_F^\two]&=c_F^\one(\nuh,\nu) c_F^\two(\nuh,\nu)-c_F^\one(\nu,\nu) c_F^\two(\nu,\nu)
\end{align}
and truncated to the appropriate order. 

Finally, the third order hard coefficient is obtained from all the relativistic potentials except for $V_r$ and $V_{S^2}$, from which we only obtain the P-wave contribution. We split it into a spin-dependent and a spin-independent piece
\begin{align}
\label{deltac3nch}
\delta c_3^{nc,h}=\delta c_3^{\rm SI,h}+\delta c_3^{\rm SD,h}
\end{align}
where 
\begin{align}
\delta c_3^{\rm SD,h}=&\frac{ 2 \pi ^2 C_F ^2 (1-\delta_{l0} ) }{n(2 l+1) l (l+1)}\frac{m_r^2}{m_1  m_2}\left\{\left[\beta_0  \left(S_1(l+n)+\frac{4 n-(2 l+1)}{4 n}-\frac{1}{2} (S_1(2 l-1)+S_1(2 l+2))+l \Sigma_1^{(k)}\right.\right.\right.\nn\\
&\left.\left.\left.+l \Sigma_1^{(m)}-n \Sigma_2^{(k)}-n \Sigma_2^{(m)}+\frac{\pi ^2 n}{6}+\Sigma_1^{(m)}\right)-\frac{3 a_1}{4}\right]\right.\nn\\
&\left. \times\left[X_{LS} \left( \delta c_F^\one+\delta c_F^\two+\frac{\delta c_S^\one m_2}{2m_1 }+\frac{\delta c_S^\two m_1 }{2m_2}\right)+\delta[ c_F^\one  c_F^\two] D_s\right]\right.\nn\\
&\left. +\frac{1}{12} (\beta_0 +8 C_A) X_{LS} \left(( \delta c_F^\one+ \delta c_F^\two)+\frac{\delta c_S^\one m_2}{2m_1 }+\frac{\delta c_S^\two m_1 }{2m_2}\right)-\frac{\delta [c_F^\one c_F^\two] D_s}{12} (2 C_A-3 \beta_0 )\right.\nn\\
&\left. -C_A \left(2 S_1(l+n)+\frac{2n-(2l+1)}{2n}-S_1(2 l-1)-S_1(2 l+2)\right) \right.\nn\\
&\left.\times \left[\delta [c_F^\one c_F^\two] D_s+X_{LS} \left(\frac{\delta c_F^\one m_2}{2m_r}+\frac{\delta c_F^\two m_1 }{2m_r}\right)\right]+\frac{1}{3} \delta[ c_F^\one c_F^\two] \mathcal{S}_{12} (7 C_A-2 \beta_0 )\right\}\\
&=\frac{2 \pi ^2 C_F^2 (1-\delta_{l0}) m_r^2 }{l (l+1) (2 l+1) n m_1 m_2}\left(1-z^{-C_A}\right) \left\{\left(D_s \left(z^{-C_A}+1\right)+\frac{ m_1 m_2}{m_r^2}X_{LS}\right)\right.\nn\\
&\left.\times \left[\beta_0 \left(-l \Sigma_1^{(k)}(n,l)-(l+1) \Sigma_1^{(m)}(n,l)+n (\Sigma_2^{(k)}(n,l)+\Sigma_2^{(m)}(n,l))-\frac{\pi ^2 n}{6}-1\right)\right.\right.\nn\\
&\left.\left.+a_1+\frac{z^{-\beta_0-\frac{\gamma_0}{2}}-1}{ z^{-\frac{\gamma_0}{2}}-1}\frac{C_A+C_F}{2}-\frac{ z^{-\beta_0-C_A}-1}{ z^{-C_A}-1}\frac{C_A L_H}{2}\right.\right.\nn\\
&\left.\left.+\frac{\beta_0-C_A}{2}  \left(\frac{2 l+1}{2 n}-2 S_1(l+n)+S_1(2 l-1)+S_1(2 l+2)-1\right)\right]\right.\nn\\
&\left.+\frac{1}{2} D_s \left(z^{-C_A}+1\right) \left[-\frac{\beta_0}{3}+\frac{z^{-\beta_0-C_A}+1}{z^{-C_A}+1}(C_A+C_F) -C_A \left(\frac{2 l+1}{2 n}-\frac{8}{3}-2 S_1(l+n)\right.\right.\right.\nn\\
&\left.\left.\left.+S_1(2 l-1)+S_1(2 l+2)+\frac{z^{-\beta_0-C_A}+1}{z^{-C_A}+1}L_H +\frac{\left(z^{-\beta_0-2 C_A}-1\right) }{z^{-2 C_A}-1}\ln \frac{m_1 m_2}{4 m_r^2}\right)\right]\right.\nn\\
&\left.+\frac{z^{-C_A} }{2 \left(z^{-C_A}-1\right)}\frac{\alh-\als }{\als}\left(\frac{\gamma_1}{2 \beta_0}-\frac{\beta_1 C_A}{\beta_0^2}\right)\left(D_s z^{-C_A}+\frac{m_1 m_2}{2 m_r^2}X_{LS} \right)\right.\nn\\
&\left.-\frac{C_A \left(z^{-\beta_0-C_A}-1\right) }{z^{-C_A}-1}X_{LS}\left(\frac{m_2 }{2 m_r}\ln \frac{m_1}{2 m_r}+\frac{m_1 }{2 m_r}\ln \frac{m_2}{2 m_r}\right)\right.\nn\\
&\left.+\frac{4}{3} \mathcal{S}_{12} \left(\frac{\beta_0}{2}-\frac{7 C_A}{4}\right) \left(z^{-C_A}+1\right)\right\}\nn\\
&-\frac{2 \pi ^2 C_A C_F^2 (1-\delta_{l0}) z^{-\frac{\gamma_0}{2}}}{(l (l+1) (2 l+1) n)}\frac{m_r}{m_1m_2} \left(D_s m_r z^{-\frac{\gamma_0}{2}}+\frac{ m_1 m_2}{2 m_r}X_{LS}\right)\left(\frac{\alh}{\als}L_{\nuh}-L_\nu\right),
\label{c3ncSDh}
\end{align}

\begin{align}
\label{c3ncSIh}
\delta c_3^{\rm SI,h}=&\frac{  \pi ^2 C_F ^2 (1-\delta_{l0} ) }{n(2 l+1) l (l+1)}\frac{m_r^2}{m_1  m_2}\left\{\left[\frac{m_1 }{ m_2}+\frac{m_2}{ m_1}\right] \left[-\frac{ 5 }{12}C_A\left(z^{-2CA}-1\right)\right.\right.\\\nn
&\left.\left.+\frac{1}{3}T_fn_f \left(z^{-2C_A}-1+\left(\frac{20 }{ 13}+\frac{32 }{ 13}\frac{C_F }{C_A}\right)\left[1-z^\frac{-13C_A }{ 6}\right]\right)\right]\right\}.
\end{align}


\end{document}